\documentclass[10pt,aps,twocolumn,floatfix,prc,twoside,tightenlines,nofootinbib,showpacs]{revtex4}

\usepackage[sort&compress]{natbib}
\usepackage[dvips]{graphicx}
\usepackage{amsmath,amssymb,amsopn,bm}
\usepackage{color}

\newcommand{\dd}{\text{d}}
\DeclareMathOperator{\im}{Im}

\newcommand{\ii}{\text{i}}
\newcommand{\erw}[1]{\ensuremath {\left \langle {#1} \right \rangle}}
\newcommand{\op}[1]{\ensuremath{\bm{\mathrm{#1}}}}
\newcommand{\beq}{\begin{equation}}
\newcommand{\eeq}{\end{equation}}
\newcommand{\bce}{\begin{center}}
\newcommand{\ece}{\end{center}}
\newcommand{\eg}{\textit{e.g.}}
\newcommand{\ie}{\textit{i.e.}}

\newcommand{\lsim}{\lesssim}
\newcommand{\gsim}{\gtrsim}
\DeclareMathOperator*{\asy}{\cong}

\begin{document}

\title{Dilepton Radiation at the CERN Super Proton Synchrotron} 

\author{Hendrik van Hees and Ralf Rapp} 
\affiliation{Cyclotron Institute and Physics Department, Texas A{\&}M
  University, College Station, Texas 77843-3366, USA }

\date{\today}

\begin{abstract}
  A quantitative evaluation of dilepton sources in heavy-ion reactions
  is performed taking into account both thermal and non-thermal
  production mechanisms. The hadronic thermal emission rate is based on
  an electromagnetic current-correlation function with a low-mass region
  (LMR, $M\lsim 1$~GeV) dominated by vector mesons ($\rho$, $\omega$,
  $\phi$) and an intermediate-mass region (IMR, $1~\text{GeV} \le M \le
  3~\text{GeV}$) characterized by (the onset of) a multi-meson
  continuum. A convolution of the emission rates over a thermal fireball
  expansion results in good agreement with experiment in the low-mass
  spectra, confirming the predicted broadening of the $\rho$ meson in
  hadronic matter in connection with the prevalence of baryon-induced
  medium effects. The absolute magnitude of the LMR excess is mostly
  controlled by the fireball lifetime, which in turn leads to a
  consistent explanation of the dilepton excess in the IMR in terms of
  thermal radiation.  The analysis of experimental transverse-momentum
  ($q_T$) spectra reveals discrepancies with thermal emission for $q_T
  \gsim 1$~GeV in noncentral In-In collisions, which we address by
  extending our calculations by: (i) a refined treatment of $\rho$
  decays at thermal freezeout, (ii) primordially produced $\rho$'s
  subject to energy-loss, (iii) Drell-Yan annihilation, and (iv) thermal
  radiation from $t$-channel meson exchange processes.  We investigate
  the sensitivity of dilepton spectra to the critical temperature and
  hadro-chemical freezeout of the fireball. The $\rho$ broadening in the
  LMR turns out to be robust, while in the IMR Quark-Gluon Plasma
  radiation is moderate unless the critical temperature is rather low.
\end{abstract}

\maketitle

\section{Introduction}  
\label{sec_intro}

Dilepton invariant-mass spectra provide the unique opportunity to
directly probe the electromagnetic (e.m.) spectral function of the hot
and dense medium created in energetic collisions of heavy
nuclei~\cite{Rapp:1999ej,Alam:1999sc,Gale:2003iz}. In the low-mass
region (LMR, $M \le 1$~GeV), this enables the study of modifications of
the light vector mesons (V=$\rho$, $\omega$, $\phi$) caused by their
interactions with the surrounding matter particles and/or changes in the
underlying condensate structure. In the intermediate-mass region (IMR,
$1~\text{GeV} \leq M \leq 3~\text{GeV}$), electromagnetic emission is
expected to become continuum-like with rather well-defined strength
rendering it a suitable tool to infer temperatures of the excited system
(well) before interactions cease (``thermal freezeout'').
  
Dilepton measurements in heavy-ion collisions at the CERN-SPS (at
center-of-mass energies of $\sqrt{s}$=17.3 and 8.8~AGeV) have proven the
presence of substantial excess radiation beyond e.m.~final-state decays
of produced hadrons in both the
LMR~\cite{Agakichiev:2005ai,Adamova:2002kf} and
IMR~\cite{HELIOS3,Abreu:1999jr,Abreu:2000nj}. This, in particular,
corroborated the presence of interacting \emph{matter} over a duration
of $\sim$10-15~fm/$c$ in central Pb-Au collisions.  Moreover, the
spectral shape of the excess in the LMR could only be accounted for if
major medium modifications of the vacuum $\rho$-meson line shape were
incorporated~\cite{Rapp:1999ej,Alam:1999sc,Gale:2003iz}. Both the
implementation of a dropping mass or a strong broadening of its width as
following from hadronic many-body approaches, were compatible with the
large enhancement observed at masses below the free $\rho$ mass. In
addition, with the same underlying fireball model (lifetime and
temperature evolution), reasonable agreement with the enhancement
measured in the IMR~\cite{Abreu:1999jr,Abreu:2000nj} has been
established~\cite{Rapp:1999zw} (as well as with direct photon
spectra~\cite{Aggarwal:2000th,Turbide:2003si}). The excess for $M \ge
1.5$~GeV was largely attributed to four-pion type annihilations in the
hadronic phase, with a strength determined by the vacuum e.m.~spectral
function, while the calculated Quark-Gluon Plasma (QGP) contribution to
the thermal yield amounted to $\sim$$30\%$ ($\sim$$10$-$20\%$ in the
LMR~\cite{Rapp:2002mm}).

A recent, substantial, improvement in statistics and mass resolution in
low-mass dimuon spectra in In-In collisions~\cite{Arnaldi:2006jq} shows
good agreement with predictions for thermal radiation with in-medium
$\rho$-meson broadening as following from hadronic many-body
calculations~\cite{Rapp:2002mm,Rapp:2004zh}. The shape of the excess
radiation is well described from threshold ($M=2m_\mu$) to
$M\simeq0.9$~GeV, while the absolute yield is overpredicted by about
$30\%$. The latter can be accommodated by a minor adjustment in the
thermal fireball evolution (amounting to a $30\%$ lower fireball
lifetime), and after inclusion of in-medium $\omega$ and $\phi$ decays
(whose contribution is mostly localized around their free mass), as well
as four-pion type annihilation (which sets in at masses $M \gsim
0.9$~GeV), a quantitative description of the NA60 invariant-mass spectra
from threshold to $\sim$$1.4$~GeV in central In-In collisions
emerges~\cite{vanHees:2006ng} (see also
Refs.~\cite{Dusling:2006yv,Ruppert:2007cr}). While the inclusive mass
spectra are well described, the comparison of the calculations with
newly released transverse pair-momentum ($q_T$)
spectra~\cite{Damjanovic:2007qm} reveals some discrepancies at
$q_T>1$~GeV in semicentral In-In.

In the present article we reiterate the main points of our previous
study of dilepton invariant-mass spectra at the SPS, and extend the
analysis to $q_T$ spectra. In particular, we conduct a detailed analysis
of sources at high $q_T$, in terms of (i) an improved treatment of
$\rho$ decays at thermal freezeout (which are subject to maximal blue
shift due to transverse flow), (ii) a component of primordial
(hard-produced) $\rho$ mesons subject to energy loss when traversing the
medium (using high-$p_T$ pion spectra as a guideline), (iii) Drell-Yan
(DY) annihilation in primordial $N$-$N$ collisions which we extrapolate
to small mass by imposing constraints from real photon production, and
(iv) meson $t$-channel exchange contributions to the thermal production
rate which are not included in the many-body vector-meson spectral
functions (but which have been found to be a significant source at high
$q_T$ in real photon production~\cite{Turbide:2003si})\footnote{Note
  that contributions (ii)-(iv) have little bearing on the inclusive
  invariant-mass spectra which are predominantly populated by
  low-momentum sources with $q_T \lsim 1$~GeV. Therefore, the inclusion
  of these contributions does not upset our earlier description of
  inclusive $M$ spectra in terms of thermal radiation and freezeout
  $\rho$'s. Initial results of these studies have been reported in
  Refs.~\cite{Rapp:2006cj,vanHees:2007yi}.}.  Another interesting issue
which has received little attention thus far is how uncertainties in the
critical temperature and hadro-chemical evolution of the fireball affect
dilepton spectra. For hadro-chemical freezeout we investigate the
sensitivity to temperatures in the range $T_{\mathrm{ch}} \simeq
160$-$175$~MeV, representative for top SPS energy according to recent
thermal model analyses~\cite{Andronic:2005yp,Becattini:2005xt}. In
connection with updates of lattice-QCD results indicating a critical
temperature up to $T_c\simeq190$-$200$~MeV~\cite{Karsch:2007vw}, this,
in particular, opens the possibility of a chemically equilibrated hot
and dense hadronic phase for, say, $T=160$-$190$~MeV, which we also
consider. We furthermore conduct a quantitative study of effective slope
parameters of the NA60 $q_T$ spectra, where the investigations of the
fireball chemistry are supplemented with variations of the transverse
flow velocity. Finally, we revisit the consequences of our fireball
refinements on our previous evaluations~\cite{Rapp:1999us} of dielectron
spectra as measured by CERES/NA45 in semicentral Pb-Au
collisions~\cite{Agakichiev:2005ai}, as well as recent data in central
Pb-Au~\cite{Adamova:2006nu}.

Our article is organized as follows: in Sec.~\ref{sec_therm} we recall
the main ingredients of our approach to calculate thermal emission
rates, based on the e.m. spectral function in the vacuum
(Sec.~\ref{ssec_rate}), followed by discussing medium effects on
hadronic emission at low mass (due to in-medium $\rho$, $\omega$ and
$\phi$ spectral functions; Sec.~\ref{ssec_lowmass}) and at intermediate
mass (due to finite-$T$ chiral mixing; Sec.~\ref{ssec_intmass}); a new
element not included in previous spectral-function calculations are
$t$-channel meson-exchange reactions which therefore are elaborated in
more detail in Sec.~\ref{ssec_omt}, while we will be brief on partonic
emission from a Quark-Gluon Plasma (Sec.~\ref{ssec_qgp}). In
Sec.~\ref{sec_nontherm} we evaluate nonthermal dilepton sources:
$\rho$-meson decays at thermal freezeout (whose decay kinematics differ
from thermal radiation; Sec.~\ref{ssec_fo}), an estimate of primordial
$\rho$'s at large $q_T$ (which do not thermalize; Sec.~\ref{ssec_hard}),
and Drell-Yan annihilation (with an extrapolation to low mass;
Sec.~\ref{ssec_dy}). In Sec.~\ref{sec_fireball} we recollect the
ingredients to our thermal fireball model for the space-time evolution
of the medium in heavy-ion collisions, including variations in
hadrochemical freezeout and critical temperature. In
Sec.~\ref{sec_spectra} we implement all dilepton sources into the
fireball to compute dimuon invariant-mass (Sec.~\ref{ssec_mspec}) and
transverse-momentum spectra (Sec.~\ref{ssec_qtspec}) in comparison to
NA60 data; the effect of hadrochemistry and $T_c$ on NA60 spectra is
worked out in Sec.~\ref{ssec_hadrochem}, followed by a slope analysis of
$q_T$ spectra (Sec.~\ref{ssec_slope}) in the context of which we also
investigate different radial flow scenarios; the improvements in the
fireball and dilepton source description as deduced from the NA60 data
are confronted with previous and new CERES/NA45 data in
Sec.~\ref{ssec_ceres}.  Sec.~\ref{sec_concl} contains a summary and
conclusions.

\section{Thermal Dilepton Radiation} 
\label{sec_therm}
\subsection{Emission Rate and Electromagnetic Spectral Function}
\label{ssec_rate}
In thermal equilibrium the rate of dilepton emission per four-volume and
four-momentum can be related to the hadronic e.m.~spectral function
as~\cite{MT84}
\begin{equation}
\begin{split}
\frac{\dd N_{ll}}{\dd^4 x \dd^4 q}
=&-\frac{\alpha^2}{3 \pi^3} \frac{L(M^2)}{M^2} \im
\Pi_{\mathrm{em},\mu}^{\mu}(M,q;\mu_B,T) \\
& \times f^B(q_0;T) \ ,  
\end{split}
\label{rate}
\end{equation}
which in this article we will refer to as thermal dileptons (or thermal
radiation).  The retarded e.m.~current-current correlator is given by
\begin{equation}
\label{ret-se}
\Pi^{\mu \nu}_{\mathrm{em}}(q)=
\ii \int \dd^4 x \mathrm{e}^{\ii q_{\sigma} x^{\sigma}}
\Theta(x^0) \erw{[\op{J}^{\mu}_{\mathrm{em}} (x),
\op{J}^{\nu}_{\mathrm{em}}(0)]} \ ,   
\end{equation}
where $\alpha$=$e^2/(4 \pi)$=1/137 denotes the fine structure constant, 
$M^2=q_0^2-q^2$ the dilepton invariant mass squared with energy $q_0$
and three-momentum $q$, and $f^B(q_0;T)$ the thermal Bose
distribution function ($T$: temperature, $\mu_B$: baryon chemical
potential).  The final-state lepton phase space factor,
\begin{equation}
\label{lept-ps}
L(M)=\left (1+\frac{2 m_l^2}{M^2} \right) \sqrt{1-\frac{4 m_l^2}{M^2}} 
\ , 
\end{equation}
depends on the lepton mass, $m_l$=0.511(105.6)~MeV for electrons
(muons; $l=e,\mu$), but quickly approaches one above threshold,
$M=2m_l$ (\eg, $L(M=0.3~{\mathrm{GeV}})=0.89$ for dimuons).

In the vacuum, the e.m. spectral function, $\im \Pi_{\mathrm{em}}(M)$,
is well known from $e^+e^-$ annihilation into hadrons. It is
characterized by the light vector resonances $\rho$(770), $\omega$(782)
and $\phi$(1020) at low mass (vector dominance model (VDM)) and a
perturbative quark-antiquark continuum at higher mass,
\begin{equation}
\im \Pi_{\mathrm{em}} = \left\{
\begin{array}{ll}
 \sum\limits_{V=\rho,\omega,\phi} \left(\frac{m_V^2}{g_V}\right)^2 \
\im D_V  , \ M \lsim M_{\mathrm{dual}}
\vspace{0.3cm}
\\
-\frac{N_c M^2}{12\pi} \ (1+\frac{\alpha_s}{\pi} +\dots) 
\sum\limits_{i} (e_i)^2   , \ M \gsim M_{\mathrm{dual}} \ 
\end{array}  \right.
\label{Piem}
\end{equation}
($N_c=3$: number of colors, $\alpha_s$: strong coupling constant, $e_i$:
the electric quark charge in units of the electron charge, and $i$ is
running over up, down and strange quark flavors for $M<3$~GeV).
$M_{\mathrm{dual}} \simeq1.5$~GeV signifies a ``duality'' scale above
which the total cross section for $e^+e^-\to \text{hadrons}$ (and thus
the strength of $\im \Pi_{\mathrm{em}}$) essentially behaves
perturbatively with little impact from subsequent hadronization. Since
the hadronic final state in $e^+e^-$ annihilation approximately
resembles a thermal medium (except for strangeness), time-reversal
invariance of strong and electromagnetic interactions implies that, to
leading order in temperature, the equilibrium dilepton emission rate of
hadronic matter is determined by the free e.m.~correlator,
$\Pi_{\mathrm{em}}^{\mathrm{vac}}$, cf.~also
Refs.~\cite{Li:1998ma,Rapp:1999zw}.

In the hadronic basis, the e.m.~spectral function is dominated by the
isovector ($\rho$) channel while the isoscalar channels are suppressed.
The SU(3)-flavor quark model, \eg, predicts a weighting of 9:1:2 for the
e.m. couplings of $\rho$, $\omega$ and $\phi$, respectively.  These
values are roughly in line with the electromagnetic decay widths
$\Gamma_{V\to ee}= \frac{4\pi\alpha^2}{3}{m_V/g_V^2}=7.0$, $0.60$ and
$1.27$~keV for $\rho$, $\omega$ and $\phi$, respectively (the empirical
values of the $\phi$ width should be corrected for phase space by the
ratio $m_\omega/m_\phi$=0.77).

\begin{figure}[!tb]
\centerline{\includegraphics[width=0.45 \textwidth]{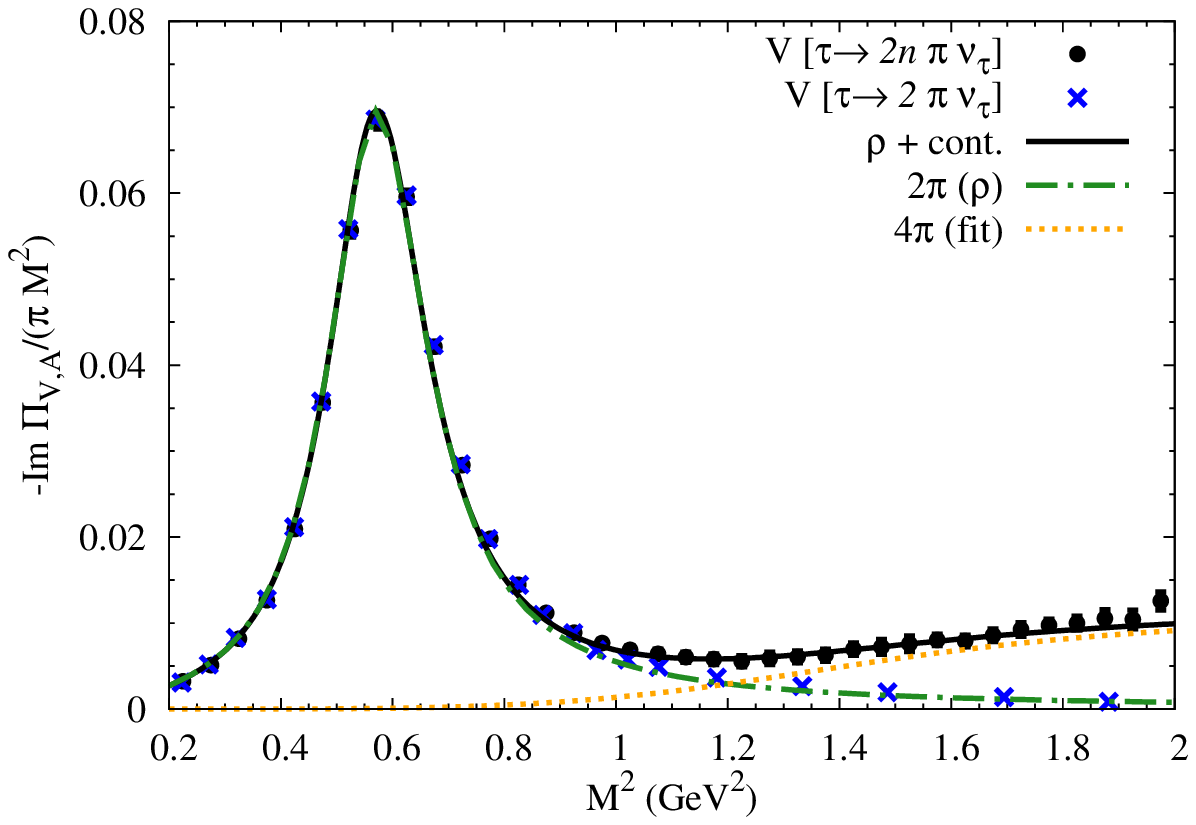}}\vspace*{2mm}
\centerline{\includegraphics[width=0.45 \textwidth]{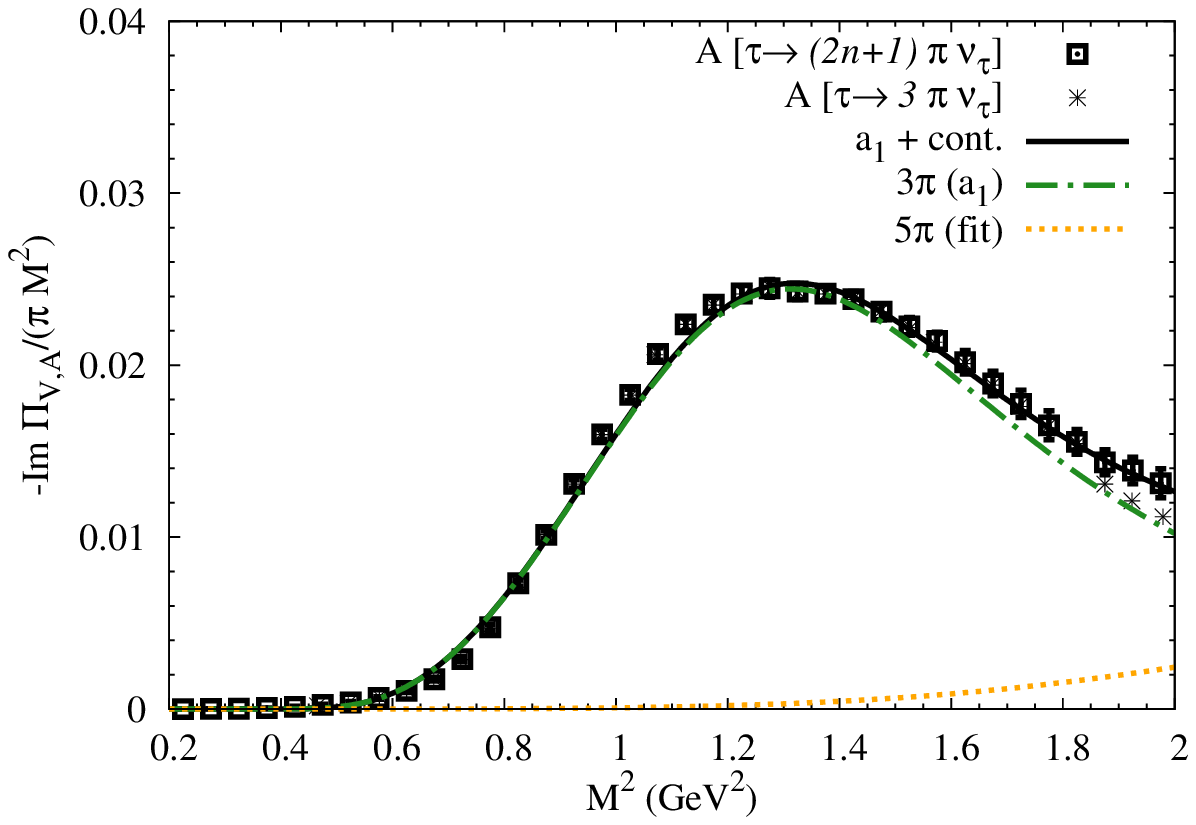}}
\caption{(Color online) Free isovector-vector (upper panel) and
  -axialvector (lower panel) spectral functions as measured in hadronic
  $\tau$-decays~\cite{aleph98}, compared to fits with three- and
  four-pion contributions (the two-pion piece follows from the
  previously calculated $\rho$ propagator~\cite{Urban:1998eg}).}
\label{fig_VAvac}
\end{figure}

The vector-isovector channel furthermore provides the most direct link
to chiral symmetry: under chiral rotations, the $\rho$ channel
transforms into the axialvector-isovector ($a_1$) one. Thus, the $a_1$
is commonly identified as the chiral partner of the $\rho$.\footnote{An
  alternative has recently been suggested in Ref.~\cite{Harada:2003jx}
  in terms of the ``Vector Manifestation'' (VM) of chiral symmetry,
  where the chiral partner of the (longitudinal) $\rho$ is identified
  with the pion.} The pertinent spectral functions in free space have
been accurately measured in hadronic $\tau$
decays~\cite{aleph98,opal99}, cf.~Fig.~\ref{fig_VAvac}, and evaluated in
terms of chiral order parameters ($f_\pi$, four-quark condensates) using
Weinberg sum rules~\cite{Weinberg:1967kj,Das:1967}. To leading order in
temperature (\ie, for a hot pion gas), the mutual in-medium
modifications of vector and axialvector channels can be inferred from
chiral symmetry alone, as will be discussed in Sec.~\ref{ssec_intmass}
below.

The main objective for the remainder of this Section is a realistic
evaluation of the in-medium e.m.~spectral function. In the LMR, we
assume that vector dominance remains valid in the medium. Our
calculations of the $V$-meson spectral functions, $A_V=-2 \im D_V$,
utilize effective hadronic Lagrangians, where the interaction vertices
and coupling constants are constrained by gauge and chiral symmetry, as
well as empirical decay and scattering data. Medium effects are
calculated within hadronic many-body theory \emph{without} introducing
explicit medium dependencies of the bare parameters in the Lagrangian
(which require information beyond the effective hadronic theory). A
careful comparison of our predictions for dilepton spectra in heavy-ion
collisions with experiment can then serve as a basis for identifying
medium effects that go beyond the hadronic many-body framework.

\subsection{Hadronic Emission at Low Mass: In-Medium Vector Mesons} 
\label{ssec_lowmass}

Since the prevalent role in the LMR of the e.m.~correlator is played by
the $\rho$ meson, the latter has been the main focus for studying
in-medium effects in thermal dilepton production. However, with the
current precision of the NA60 dilepton data on the $10$-$20\%$ level,
thermal emission from $\omega$ and $\phi$ decays becomes relevant. Their
contributions are an inevitable consequence of the formation of a
thermal medium, and we therefore incorporate them along with pertinent
medium modifications.

Within the VDM the in-medium e.m.~correlator in the LMR remains directly
proportional to the vector-meson spectral functions as in
Eq.~(\ref{Piem}). The key objective is then to calculate the
vector-meson selfenergies at finite temperature and baryon density
($\varrho_B$), $\Sigma_V$, figuring into the propagator as
\begin{equation}
D_V=\frac{1}{M^2-(m_V^{(0)})^2-\Sigma_{VP}-\Sigma_{VM}-\Sigma_{VB}}
 \ .
\end{equation}
Here, $m_V^{(0)}$ denotes the bare mass, and the selfenergy
contributions are classified into three types: (i) medium modifications
of the pseudoscalar meson cloud, $\Sigma_{VP}$ ($P=\pi\pi$, $3\pi$ or
$K\bar K$ for $V=\rho$, $\omega$ or $\phi$, respectively), and direct
interactions of $V$ with (ii) mesons and (iii) baryons from the
surrounding heat bath, $\Sigma_{VM}$ and $\Sigma_{VB}$.  One should also
note that without the implementation of a phase transition into a
realistic hadronic model (which has not been achieved thus far),
many-body calculations are to be considered as an extrapolation beyond a
certain temperature and/or density. We estimate that, under SPS
conditions, our approach should be reliable up to temperatures of about
$T\simeq 150$~MeV, where the total hadron density amounts to about
$\varrho_h\simeq 2\varrho_0$ (with a net-baryon density of about $0.7
\varrho_0$), while it has increased to about $5 \varrho_0$ at the
expected phase boundary at $T_c\simeq175$~MeV. In the following we
briefly recall the construction of, and constraints on, the underlying
effective Lagrangians.

\subsubsection{$\rho$ Meson}
\label{sssec_rho}

The first step in the calculation of a realistic $\rho$ propagator is a
proper description of its coupling to the pion cloud in vacuum (giving
rise to its two-pion decay width, $\Gamma_{\rho\pi\pi}(m_\rho) \simeq
150$~MeV).  A suitable chiral effective Lagrangian can be obtained by,
\eg, introducing the vector fields as local gauge fields into the chiral
pion Lagrangians and implementing the photon field via vector
dominance~\cite{Kroll:1967it,Sakurai:1969},

\begin{equation}
  \begin{split}
    \mathcal{L}_{\pi\rho\gamma} =& g_\rho \vec \rho_\mu \cdot (\vec \pi
    \times \partial^\mu\vec \pi) \\ 
    & -\frac{1}{2} g_\rho^2
    \left[\vec\rho_\mu\cdot\vec\rho^\mu \vec\pi^2
      -  \vec\rho^\mu\cdot \vec\pi  \vec\rho_\mu\cdot\vec\pi \right] \\
    & +\frac{e m_\rho^2}{g_\rho} A_\mu \rho_3^\mu \, .
\end{split}
\end{equation}

The pertinent two-pion and one-pion-tadpole diagrams specify the
$\rho$'s pion cloud in the vacuum with a transverse selfenergy (using a
proper regularization procedure). The coupling constant, $g_\rho$, a
cutoff parameter, $\Lambda_\rho$ (to render the pion loops finite), and
the bare $\rho$-mass, $m_\rho^{(0)}$, can be readily adjusted to
reproduce $P$-wave $\pi$-$\pi$ scattering phase shifts and the pion
e.m.~formfactor in free space~\cite{Urban:1998eg}. This model also
predicts the two-pion part of the $\tau$-decay data well,
cf.~Fig.~\ref{fig_VAvac}.

Based on the above model in the vacuum, we adopt the in-medium $\rho$
propagator from the hadronic many-body approach developed in
Refs.~\cite{Rapp:1997fs,Urban:1998eg,Rapp:1999us}, to which we refer the
reader for more details. Modifications of the pion cloud are induced by
an in-medium $\pi$ propagator which is dressed with standard $NN^{-1}$
and $\Delta N^{-1}$ excitations in nuclear matter, supplemented with
appropriate vertex corrections to maintain a conserved vector current.
While the calculations of nuclear pion-cloud modifications are often
restricted to the case of vanishing $\rho$ three-momentum
(``back-to-back'' kinematics), we here employ the extension to finite
three-momentum worked out in Ref.~\cite{Urban:1998eg}. This is
particularly important in that it enables (i) to constrain the
corresponding medium effects by nuclear photo-absorption
data~\cite{Rapp:1997ei} and $\pi N\to \rho N$ scattering, and (ii) a
meaningful application to dilepton transverse momentum spectra. At
finite temperature, where a (large) fraction of nucleons is thermally
excited into baryons, only the nucleon density should be employed. To
account for (i) $\pi BN^{-1}$ excitations into higher resonances and
(ii) $\pi B_2B_1^{-1}$ excitations ($B_1\ne N$), an ``effective''
nucleon density $\varrho_N^{\mathrm{eff}}\equiv \varrho_N +
0.5\varrho_{B} \ (B\neq N)$ has been used in the pion 
propagators~\cite{Rapp:1999us} (we estimate the uncertainty implied
by this prescription to be about $\pm$10\%). In
addition, effects of antibaryons are added by multiplying
$\varrho_N^{\mathrm{eff}}$ with a factor $(1+\bar p/p)$ where $\bar p/p$
denotes the measured antiproton-to-proton ratio (this is a small
correction at SPS energies but becomes essential at
RHIC~\cite{Rapp:2000pe}). The modifications of the pion cloud induced by
a pion gas have been found to be rather small in various
analyses~\cite{Schenk:1991xe,Shuryak:1991hb,Rapp:1995fv,vanHees:2000bp,
  GomezNicola:2004gg}. A large part of the pion-gas effect is already
provided by the Bose enhancement factor,
$[1+f^\pi(\omega_1)+f^\pi(\omega_2)]$~\cite{Rapp:1995fv}, while
low-energy $\pi$-$\pi$ interactions are suppressed by chiral symmetry.
The former is included in the present calculations, but the latter are
neglected (being much weaker than baryon-induced effects).

Direct interactions of the (bare) $\rho$ with baryons and mesons have
been approximated by resonance interactions (pion-exchange interactions,
including spacelike pions, are implicit in the nuclear modifications of
the pion-cloud part). For $\rho+N\to B$ (\ie, $\rho BN^{-1}$
excitations), this includes both $S$-wave scattering into
negative-parity baryons and $P$-wave scattering into positive-parity
baryons, with $B=N(1520),\,\Delta(1620),\,\Delta(1700)$ and
$N,\,\Delta(1232),\,N(1440),\,N(1720),\,\Delta(1905),\,N(2000)$,
respectively. An estimate of the involved couplings and cutoff
parameters can be obtained by reproducing the empirical $B\to \rho N,
\gamma N$ decay widths, but more reliable constraints follow from
comprehensive scattering data. For the present model, in addition to
hadronic decay branchings, cold nuclear matter effects induced in both
the pion cloud and via resonant $\rho$-$N$ interactions have been
checked against total photoabsorption cross sections on the nucleon and
nuclei~\cite{Rapp:1997ei}, as well as $\pi N\to \rho N$ scattering
data~\cite{Friman:1997ce,Rapp:1999ej}. Whereas $\gamma N$ and $\rho N$
scattering relates to the low-density limit of the $\rho$-spectral
function, $\gamma A$ data constrain $\im D_\rho(M\to0,q)$ close to
nuclear saturation density, $\varrho_0=0.16$~fm$^{-3}$. For the
finite-temperature case, excitations on hyperons of the type $\rho
Y_2Y_1^{-1}$ are incorporated for $Y_1=\Lambda(1115), \, \Sigma(1192)$
and $Y_2$ resonances with quantum numbers equivalent to the nonstrange
sector and appreciable empirical decay widths into $\rho$'s and/or
photons. $\rho B_2B_1^{-1}$ excitations on thermally excited nonstrange
baryons (\eg, $\Delta(1930)\Delta^{-1}$) are included but turned out to
be small.

In the meson gas, direct $\rho$-$M$ interactions on thermal mesons
($M=\pi$, $K$, $\rho$) have been scrutinized in Ref.~\cite{Rapp:1999qu},
again based on $s$-channel resonance dominance, $\rho M\to R$, and with
pertinent Lagrangians satisfying basic requirements of vector-current
conservation and chiral symmetry. As in the case of excited baryons, the
couplings and cutoffs have been estimated from combined hadronic and
radiative decay widths. The most important contributions to the $\rho$
propagator arise from axialvector mesons, $R=a_1(1260)$ (chiral partner
of the $\rho$), $h_1(1170)$ and $K_1(1270)$, as well as the
$\omega(782)$.

Another valuable consistency check for the in-medium properties of
hadrons are QCD sum rules. The latter relate a dispersion integral over
a spectral function to an operator product expansion (OPE) for the
correlation function at spacelike four-momenta, $q_{\mu} q^{\mu}<0$.
The nonperturbative coefficients of the OPE involve quark and gluon
condensates whose in-medium modifications reflect upon (an energy
integral over) the spectral function.  Generically, a reduction of the
condensates leads to a softening of the spectral function, \ie, a shift
of strength to lower mass. However, as has been elaborated in
Ref.~\cite{Leupold:1997dg} for the case of cold nuclear matter, the
additional low-mass strength in the in-medium $\rho$ spectral function
can be due to both a reduced mass or a broadening of the width, or a
combination thereof. The pertinent ``allowed band'' for width and mass
values at normal nuclear matter density has been quantified in
Ref.~\cite{Leupold:1997dg}. For the $\rho$ spectral function employed
here~\cite{Rapp:1999us}, one finds a mass and width of
$(m_\rho,\Gamma_\rho)|_{\varrho_0}\simeq(820,450)$~MeV at nuclear
saturation density and vanishing three-momentum. These values turn out
to be consistent with the band derived in Ref.~\cite{Leupold:1997dg}
(cf. middle panel of Fig.~2 in there), but more extensive (\ie,
different densities and temperatures) and quantitative comparisons are
certainly desirable (\eg, the low-mass strength (shoulder) in the $\rho$
spectral function which is induced by ``Pisobar'' ($\pi\Delta N^{-1}$)
and ``Rhosobar'' ($\rho BN^{-1}$) excitations, cannot be represented in
Breit-Wigner form as underlying Ref.~\cite{Leupold:1997dg}). However, at
this point, there is no indication for changes in the bare parameters in
the Lagrangian, \ie, at nuclear saturation density hadronic many-body
effects saturate the reduction of quark and gluon condensates as
required by QCD sum rules.

\begin{figure}[!ht]
\centerline{\includegraphics[width=0.45\textwidth,angle=0]{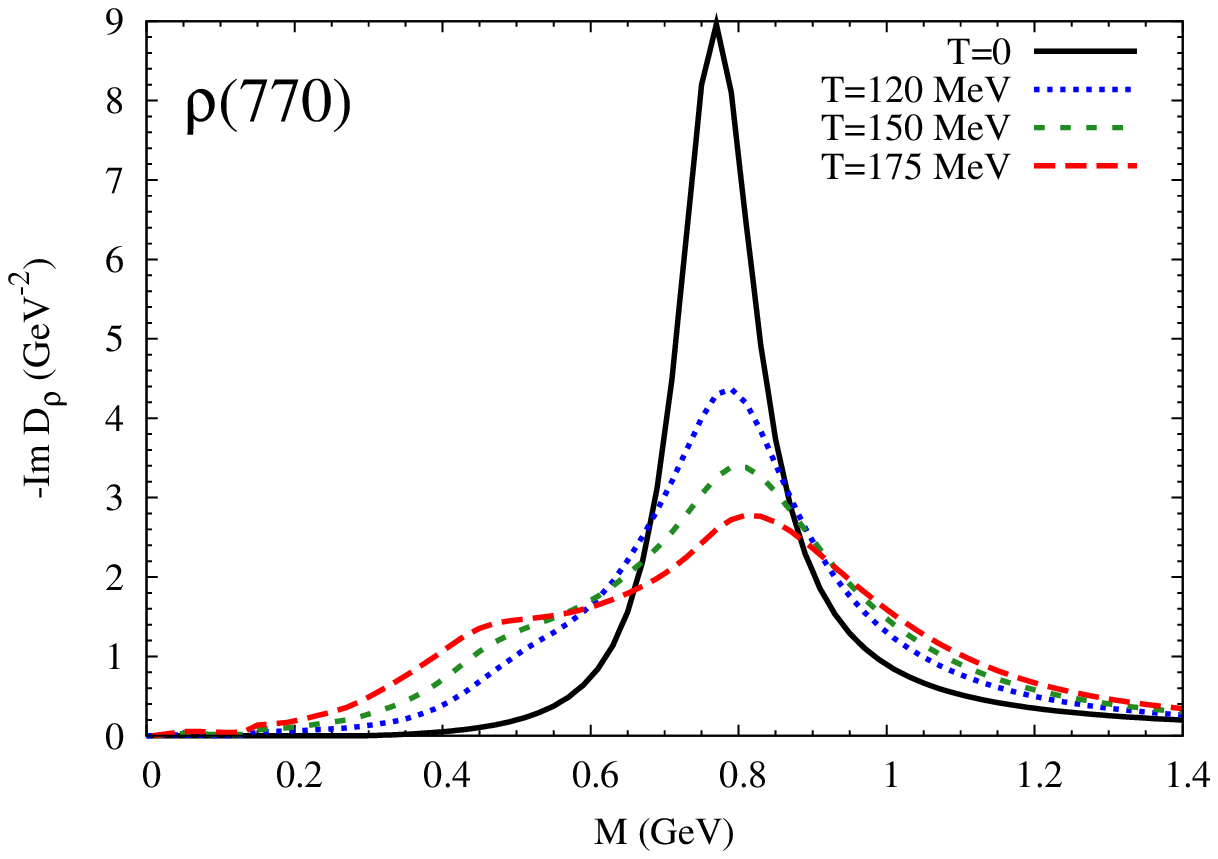}}
\vspace{0.3cm}
\centerline{\includegraphics[width=0.45\textwidth,angle=0]{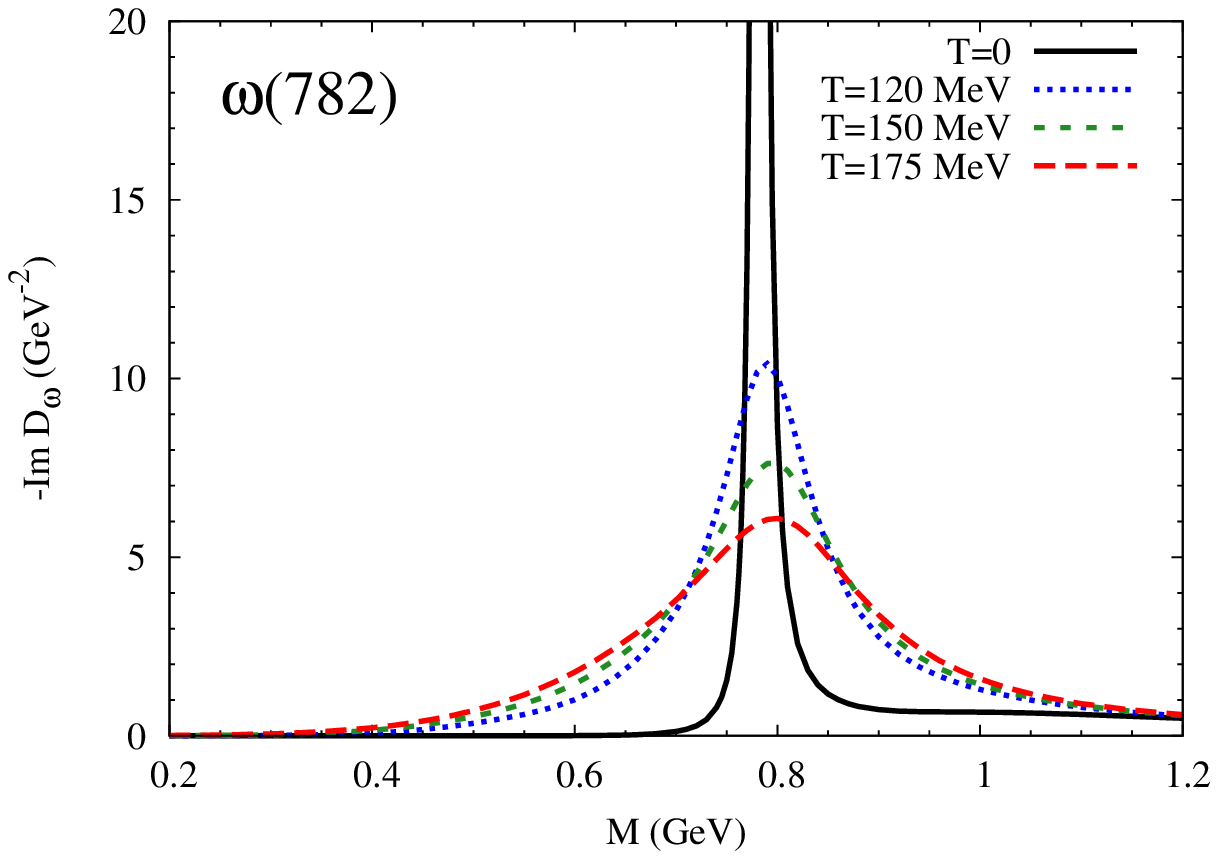}}
\vspace{0.3cm}
\centerline{\includegraphics[width=0.45\textwidth,angle=0]{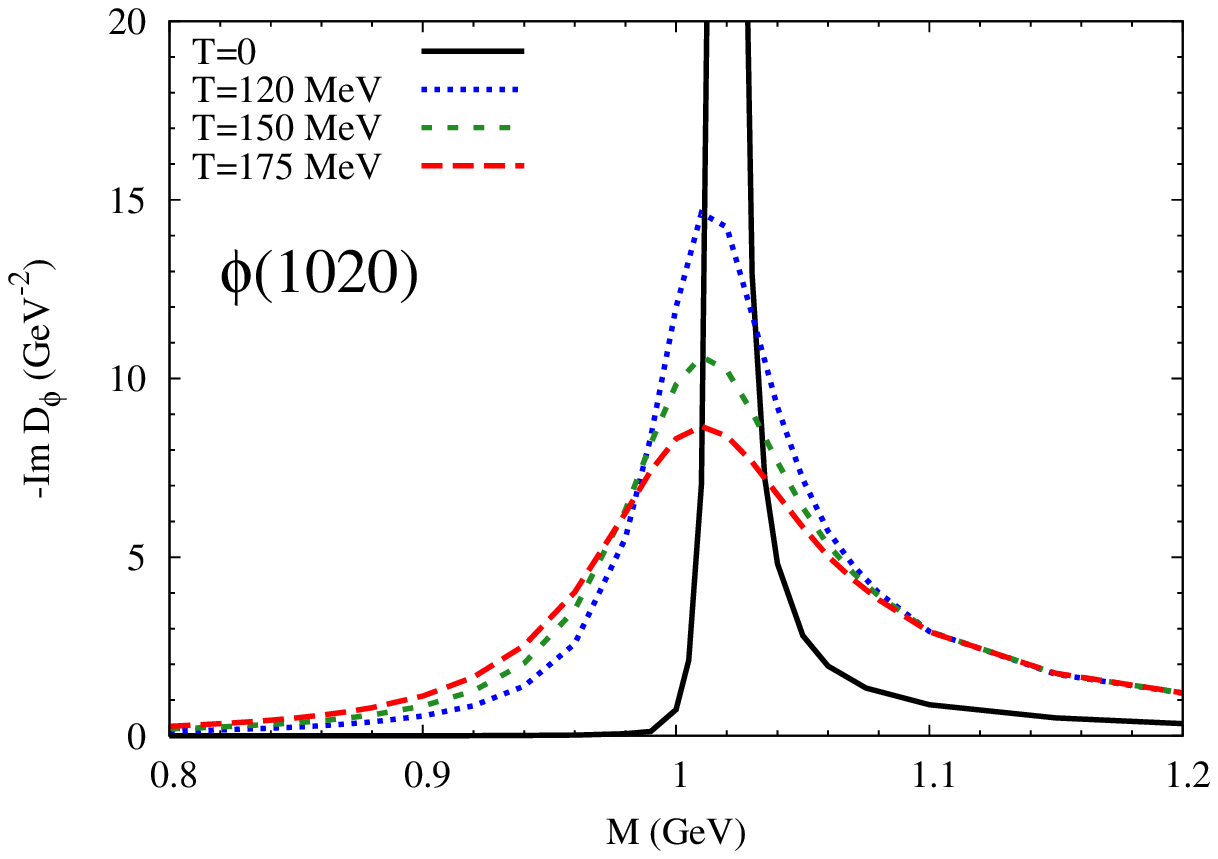}}
\caption{(Color online) In-medium spectral functions of $\rho$ (upper
  panel), $\omega$ (middle panel) and $\phi$ (lower panel) under
  conditions representative for the time evolution of heavy-ion
  collisions at full SPS energy: $(\mu_N, \mu_\pi, \mu_K,
  T)~\text{[MeV]} = (232,0,0,175)$, $(331,39,62,150)$,
  $(449,79,136,120)$ and vacuum for the dotted, short-dashed,
  long-dashed and solid lines, respectively.}
\label{fig_Avec}
\end{figure}
The $\rho$-spectral functions under fireball conditions representing the
evolution of the hadronic phase at full SPS energy
($\sqrt{s}=17.3$~AGeV) are displayed in the upper panel of
Fig.~\ref{fig_Avec}. The resonance strongly broadens with little, if
any, mass shift, essentially leading to its ``melting'' close to the
expected phase boundary. These features are a rather generic consequence
of a multilevel excitation spectrum in which real parts occur with
altering signs while imaginary parts are negative definite and strictly
add up. Also note the development of a pronounced low-energy strength
around $M=0.4$~GeV which leads to a very large enhancement over the
vacuum spectral function (and will be augmented by thermal Bose factors
in the dilepton rate).  At lower temperatures, the presence of
meson-chemical potentials (for $\pi$, $K$, $\eta$) implies higher meson
and baryon densities than in chemical equilibrium at a given
temperature, entailing stronger medium effects. The build-up of the
chemical potentials (necessary to preserve the experimentally observed
hadron ratios throughout the hadronic cooling) therefore facilitates the
observation of in-medium effects in the dilepton spectra. A strong
broadening of the $\rho$ in hot and dense matter has also been reported
in other calculations at finite $T$ and
$\varrho_B$~\cite{Eletsky:2001bb,Riek:2004kx}.

The in-medium $\rho$-spectral function has been rather extensively
employed to calculate dilepton spectra in heavy-ion collisions. At full
SPS energy it has been implemented into fireball
models~\cite{Rapp:1997fs,Rapp:1999us}, transport
simulations~\cite{Cassing:1997jz} and hydrodynamic
evolutions~\cite{Huovinen:1998ze}, with fair success (and consistency)
in describing CERES/NA45 data in central and semicentral Pb-Au
collisions~\cite{Agakichiev:2005ai}. When averaged over a typical
space-time evolution, the in-medium $\rho$ width at SPS amounts to
$\Gamma_\rho^{\mathrm{med}} \simeq 350$-$400$~MeV, almost three times
its vacuum value, implying $\Gamma_\rho(T_c) \simeq m_\rho$ when
extrapolated to the expected QCD-phase boundary. Despite an experimental
pion-to-baryon ratio of $\sim 5:1$, medium effects induced by baryons
have been identified as essential at SPS energies early
on~\cite{Rapp:1997fs}. This has led to the prediction of an even larger
excess at lower beam energies, which was tentatively confirmed in a
$40$~AGeV Pb+Au run of CERES/NA45~\cite{Adamova:2002kf} (albeit with
rather large errors). The lightlike limit of the $\rho$-spectral
function has been applied to calculate thermal photon
spectra~\cite{Turbide:2003si}, supplemented with meson-exchange
reactions which are not included in $\im D_\rho$ (these become
significant at momenta above $1$~GeV and will be discussed in
Sec.~\ref{ssec_omt} below). After convolution over the same fireball
model as used for dileptons, the direct-photon excess observed by WA98
in central Pb-Pb reactions at SPS~\cite{Aggarwal:2000th}, as well as by
PHENIX in central Au-Au collisions at RHIC~\cite{Reygers:2006kh}, can be
reasonably well accounted for (with additional primordial and QGP photon
sources, the latter being subdominant at SPS). The in-medium
$\rho$-spectral function following from hadronic many-body
theory~\cite{Rapp:1999us} has thus passed a rather wide range of both
theoretical and phenomenological tests, in both cold nuclear as well as
hot hadronic matter.

\subsubsection{$\omega$ Meson}
\label{sssec_omega}

Medium effects on the $\omega$ have received less attention so far,
especially in the context of ultrarelativistic heavy-ion collisions
(URHICs) where its dilepton decays are dominated by free decays after
thermal freezeout (commonly referred to as a ``cocktail'' contribution).
We here adopt the same approach as for the $\rho$, following
Ref.~\cite{Rapp:2000pe}. In the vacuum, the hadronic $\omega$ width,
$\Gamma_{\omega \rightarrow 3\pi}$, is composed to approximately equal
parts of a direct $\omega\to 3\pi$ and a $\omega\to\rho\pi$ coupling,
represented by anomalous terms in the interaction Lagrangian,
\begin{alignat}{2} 
\mathcal{L}_{\omega\rho\pi}&= G_{\omega\rho\pi} \,
  \epsilon_{\mu\nu\sigma\tau} \, \omega^\mu \, \partial^\nu
  \vec{\rho}^{\,\sigma} \cdot \partial^\tau \vec\pi \, , \label{rho-om-pi}
  \\
  \mathcal{L}_{\omega 3\pi} &= G_{\omega 3\pi} \
  \epsilon_{\mu\nu\sigma\tau} \, \omega^\mu \, (\partial^\nu \vec{\pi} \times
  \partial^\sigma \vec{\pi}) \cdot
\partial^\tau \vec{\pi} \, .  \label{om-3pi}
\end{alignat} 
With a hadronic formfactor cutoff-parameter, 
$\Lambda_{\omega\rho\pi}=1$~GeV, and using the VDM~\cite{Rapp:1999qu},
one finds for the radiative decay width 
$\Gamma_{\omega\to\pi\gamma}=0.72$~MeV which approximately agrees 
with the updated experimental value of 
$0.76 \pm 0.03$~MeV~\cite{Yao:2006px}.

In cold nuclear matter, the renormalization of the $\pi$-$\rho$ cloud
has been found to be rather sensitive to the medium effects on the
$\rho$~\cite{Wachs00,Rapp:2000pe,Riek:2004kx,Cabrera:2007}, which is due
to the fact that the $\omega\to\pi\rho(\to\pi\pi\pi)$ decay proceeds via
the low-mass tail of the $\rho$-spectral function below its nominal mass
of $770$~MeV. In Ref.~\cite{Wachs00} $80$\% of the estimated in-medium
broadening at nuclear saturation density is attributed to the in-medium
$\rho$ spectral function (taken from Ref.~\cite{Urban:1998eg}),
resulting in $\Gamma_{\omega P}(\varrho_0)\simeq60$~MeV. Likewise, a
large contribution to second order in the nuclear density, ${\cal
  O}(\varrho_N^2)$, to $\Gamma_{\omega P}(\varrho_N)$ arises due to the
simultaneous dressing of both $\pi$ and
$\rho$~\cite{Cabrera:2007}. Following Ref.~\cite{Rapp:2000pe}, we
restrict ourselves to the impact of the in-medium $\rho$ by folding the
phase space in the $\omega\to\pi\rho$-decay width over the spectral
function evaluated in the previous section,
\begin{equation}
\begin{split}
\label{Gamorp}
\Gamma_{\omega\to\rho\pi}(s) = &
\frac{2 G_{\omega\rho\pi}^2}{8\pi} \, \int\limits_0^{M_{\mathrm{max}}} 
\frac{M \dd M}{\pi} \ A_\rho(M) \, 2 q_{\mathrm{cm}}^3
\\
& \times [1+f^\pi(\omega_\pi)+f^\rho(E_\rho)] \, 
F_{\omega\rho\pi}(q_{\mathrm{cm}})^2 \, ,
\end{split}
\end{equation}
where $M_{\text{max}}$$=$$\sqrt{s}-m_\pi$,
$\omega_\pi^2=m_{\pi}^2+q_{\text{cm}}^2$,
$E_\rho^2=M^2+q_{\text{cm}}^2$, and $q_{\text{cm}}$ is the
three-momentum of $\pi$ and $\rho$ in the rest frame of the $\omega$. In
addition we account for cold-nuclear matter effects due to the two most
important direct $\omega$-$N$ excitations, $\omega N(1520)N^{-1}$ and
$\omega N(1650) N^{-1}$, as identified in Ref.~\cite{Lutz:2001mi} in a
coupled-channel analysis of empirical vector-meson nucleon scattering
amplitudes. Recent measurements of $\pi^0\gamma$-invariant mass spectra
in photon-induced production off hydrogen and Nb targets have revealed a
low-mass shoulder~\cite{Trnka:2005ey} that has been interpreted as
reduced $\omega$ mass in nuclear matter. However, this interpretation is
not free of controversy and may also be explained by a large increase of
the $\omega$ width to about $90$~MeV~\cite{Kaskulov:2006zc}. The latter
is compatible with the $\omega$ propagator employed in the present
paper.

Finite-temperature effects on the $\omega$ propagator include
Bose-enhancement factors in the $\pi$-$\rho$ cloud, the in-medium $\rho$
propagator in Eq.~(\ref{Gamorp}), a schematic estimate of the inelastic
$\pi\omega\to\pi\pi$ width based on Ref.~\cite{Haglin:1994xu}, and a
full calculation of the thermal $\omega+\pi\to b_1(1235)$
loop~\cite{Rapp:2000pe}.

The predicted average $\omega$ width in the hadronic phase of URHICs is
$\Gamma_\omega^{\mathrm{med}}$$\simeq$$100$~MeV~\cite{Rapp:2002mm},
cf.~middle panel of Fig.~\ref{fig_Avec}, quite consistent with the
results of Ref.~\cite{Martell:2004gt}.  This has the interesting
consequence~\cite{Rapp:2000pe} that if dilepton spectra in HICs show a
structure of $\sim$$100$~MeV width in the $\rho$-/$\omega$ mass region, it
would be an unambiguous signature of the in-medium $\omega$ spectral
function since it is narrower than the lower limit given by a free
$\rho$ contribution, $\Gamma_\omega^{\rm med} < \Gamma_\rho^{\rm
  vac}\simeq 150$~MeV.

\subsubsection{$\phi$ Meson}
\label{sssec_phi}

For the $\phi$, collision rates in a meson gas have been estimated to
amount to a broadening by $\sim 20$~MeV at
$T=150$~MeV~\cite{Alvarez-Ruso:2002ib}. The dressing of the kaon cloud
is presumably the main effect for $\phi$ modifications in nuclear
matter, increasing its width by $\sim 25$~MeV at saturation
density~\cite{Cabrera:2003wb}. Recent data on $\phi$ absorption in
nuclear photoproduction suggest substantially larger
values~\cite{Ahn:2004id}.  Since a quantitative, empirically constrained
treatment of the $\phi$ in hot and dense matter is not available at
present, we will consider the $\phi$ width as a parameter. Following
Ref.~\cite{vanHees:2006ng} we augment the microscopic calculations of
Ref.~\cite{Rapp:2000pe} by a factor of $4$ to roughly agree with the
phenomenological values extracted from nuclear
photoproduction~\cite{Ahn:2004id}.  The corresponding $\phi$ spectral
functions under SPS conditions are displayed in the lower panel of
Fig.~\ref{fig_Avec}.  The average width over the fireball evolution
amounts to $\Gamma_{\phi}^{\text{med}} \simeq 80 \;\text{MeV}$.

\subsection{Hadronic Emission at Intermediate Mass: Multi-Pion  
Annihilation}
\label{ssec_intmass}

Above the $\phi$ mass, the hadronic structure of the (vacuum)
e.m.~correlator becomes more involved being characterized by overlapping
broad resonances which combine into continuum-like multi-meson states
(most notably isovector four-pion states), recall Fig.~\ref{fig_VAvac}.
To estimate medium effects in this regime, we take recourse to
model-independent predictions following from a low-temperature expansion
evaluated using chiral-reduction techniques for the corresponding matrix
elements involving thermal pions. This method has been first worked out
in Ref.~\cite{Dey:1990ba} in the chiral limit ($m_\pi=0$) and leads to
the chiral mixing formula for the isovector part of vector and
axialvector correlators,
\begin{equation}
\begin{split}
\label{mix}
\Pi_{V}(q) &= (1-\varepsilon) \; \Pi_{V}^{\mathrm{vac}}(q) +
\varepsilon \; \Pi_{A}^{\mathrm{vac}}(q) \, ,\\
\Pi_{A}(q) &= \varepsilon \; \Pi_{V}^{\mathrm{vac}}(q) +
(1-\varepsilon) \; \Pi_{A}^{\mathrm{vac}}(q) \, .
\end{split}
\end{equation}
The mixing coefficient $\varepsilon$ is determined by pion tadpole
diagrams via a loop integral
\begin{equation}
\varepsilon=\frac{2}{f_\pi^2} 
\int \frac{d^3k}{(2\pi)^3 \omega_k^\pi}~f^\pi(\omega_k^\pi;T)
\end{equation}
($\omega_k^\pi$: on-shell pion energy, $f_\pi=93$~MeV: pion decay
constant). For $\varepsilon\to 1/2$, $V$ and $A$ correlators degenerate
signaling chiral-symmetry restoration (at
$T_c^{\mathrm{mix}}$$\simeq$160~MeV). Note that the admixture of the
axialvector part in the $M=1$-$1.5$~GeV region (which corresponds to the
$a_1$ resonance) fills in the ``dip'' structure of the vector correlator
(recall Fig.~\ref{fig_VAvac}); for $\varepsilon\to 1/2$ both spectral
functions merge into the perturbative-QCD continuum for $M\ge1$~GeV
which has been interpreted~\cite{Rapp:1999ej} as a lowering of the
quark-hadron duality scale, $M_{\mathrm{dual}}$,
cf.~Eq.~(\ref{Piem}). The simple form of Eq.~(\ref{mix}) only holds in
the soft-pion limit, \ie, when neglecting the pion four-momenta in
$\Pi_{V,A}$. More elaborate treatments~\cite{Steele:1997tv,Urban:2001uv}
will broaden and somewhat reduce the enhancement in the $M\simeq
1$-$1.5$~GeV region~\cite{vanHees:2006iv}. We here implement the mixing
effect on the (isovector) four-pion part utilizing Eq.~(\ref{mix}) with
a mixing parameter
$\hat\varepsilon=\frac{1}{2}\varepsilon(T,\mu_\pi)/\varepsilon(T_c,0)$
with finite pion mass ($m_\pi=139.6$~MeV), critical temperature, $T_c$,
and pion chemical potentials ($\mu_\pi>0$) implemented as an overall
fugacity factor, $z_\pi={\rm e}^{\mu_\pi/T}$, in Boltzmann approximation
($\mu_\pi$ varies according to the thermal fireball evolution discussed
in Sec.~\ref{sec_fireball} below). The two-pion piece, as well as the
three-pion piece corresponding to $a_1$ decay, $a_1 \rightarrow
\pi+\rho$, have been removed as they are included via the
$\rho$-spectral function as discussed above. A detailed analysis of the
derivation of Eq.~(\ref{mix}) in the presence of a finite pion-chemical
potential leads to the following expression for the mixing effect on the
vector-isovector current correlation function:
\begin{equation}
\begin{split}
\label{mix-fin-mupi}
\Pi_V(q)=&(1-\hat{\varepsilon}) z_{\pi}^4 \Pi_{V,4 \pi}^{\rm vac} +
\frac{\hat{\varepsilon}}{2} z_{\pi}^3 \Pi_{A,3 \pi}^{\rm vac} + 
\frac{\hat{\varepsilon}}{2} (z_{\pi}^4+z_{\pi}^5) \Pi_{A,5 \pi}^{\rm vac} 
\ . 
\end{split}
\end{equation}
 
\subsection{Hadronic Emission at High Momentum: 
Meson $t$-Channel Exchange}
\label{ssec_omt}

As discussed in the Introduction, a current discrepancy in the
theoretical description of the NA60 data concerns the dilepton yield at
high transverse pair-momentum, $q_T>1$~GeV, for masses $M\le1$~GeV.  It
is therefore important to scrutinize the vector-meson spectral function
calculations for high momentum sources. The latter are typically
associated with $t$-channel meson exchanges, rather than $s$-channel
resonances whose decay products are suppressed at momenta far off the
resonance. In the spectral-function approach, $t$-channel exchange
processes are encoded in medium modifications of the pseudoscalar meson
cloud. As elaborated in Sec.~\ref{sssec_rho}, the pion-cloud
modifications of the $\rho$ explicitly include interactions with
baryons, but not with thermal mesons. Consequently, $t$-channel meson
exchanges are present for $\rho B$ interactions, but not for
$\rho\pi$. E.g., the experimental $\pi N\to\rho N$ cross section is
properly reproduced at large $\sqrt{s}$ where $t$-channel $\pi$ exchange
dominates.  Indeed, it has been found in Ref.~\cite{Turbide:2003si} in
the context of thermal photon production that $t$-channel exchanges in
$\rho+\pi \rightarrow \pi+\gamma$ outshine the contributions from the
$\rho$-spectral function at momenta above $\sim$$1$~GeV for $T=200$~MeV
(at smaller temperatures the importance of $t$-channel exchanges is
reduced).  Most notably, $\omega$ exchange emerged as the single most
important hadronic high-$q_T$ thermal photon source.

\begin{figure}[!t]
\centerline{\includegraphics[width=0.3\textwidth]{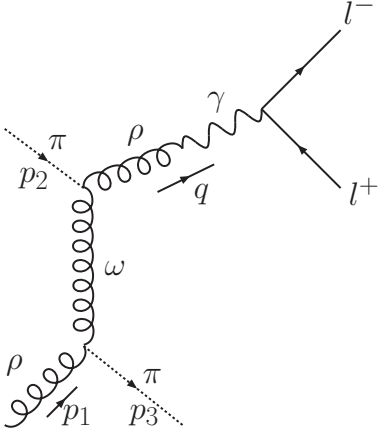}}
\caption{The Feynman diagram corresponding to the annihilation of a
  $\rho$ meson via $\omega$-meson exchange in the $t$ channel.}
\label{fig-om-t}
\end{figure}
To estimate the relevance of $t$-channel exchanges in $\pi\rho$
scattering for thermal dileptons, we adopt the same simplified treatment
as in Ref.~\cite{Turbide:2003si}, properly extended to a finite
invariant mass of the virtual photon (lepton pair): rather than
implementing the pertinent diagram into the $\rho$ selfenergy, we employ
the standard kinetic theory expression for the thermal production rate
\begin{equation}
\begin{split}
\label{rate-om-t}
\frac{\dd N_{\pi\rho\to\pi ll}^{(\text{t-ch})}}{\dd^4 x \dd^4 q} 
= & \int \frac{\dd^3 p_1}{2 \omega^\rho_{p_1} (2 \pi)^3} 
\int \frac{\dd^3 p_2}{2 \omega^\pi_{p_2}  (2 \pi)^3}
\int \frac{\dd^3 p_3}{2 \omega^\pi_{p_3} (2 \pi)^3} \\
& \times (2 \pi)^4 \delta^{(4)}(p_1+p_2-p_3-q)
\sum |\mathcal{M}_{\pi\rho\to\pi ll}|^2 \\
&  \times [1+z_\pi f^B(\omega^\pi_{p_3};T)]
 \frac{z_{\pi}^3 f^B(\omega^\rho_{p_1};T) f^B(\omega^\pi_{p_2};T)}
{4 (2 \pi)^6} \ ,
\end{split}
\end{equation}
which allows to calculate the contribution for $\pi+\rho\to\pi+l^+ +
l^-$ in terms of the corresponding matrix element,
$\mathcal{M}_{\pi\rho\to\pi ll}$, in Born approximation. The latter
substantially facilitates the task of maintaining gauge invariance. For
the $\rho \omega \pi$ vertex, the anomalous-coupling Lagrangian,
Eq.~(\ref{rho-om-pi}), is used to calculate the matrix element,
\begin{widetext}
\begin{equation}
\begin{split}
\label{Msqr}
\sum |\mathcal{M}_{\omega \text{t}}|^2 = & \frac{4 \pi^2 \alpha g_{\rho
    \omega \pi}^4 C^2}{M^2} (m_{\rho}^{(0)})^4 |D_{\rho}(M)|^2
\frac{1}{|t-m_{\omega}^2|^2} L(M) \Theta(M-2 m_l) \\ 
&\times \{
(M^2-m_{\pi}^2)^2(m_{\pi}^2-m_{\rho}^2)^2 -2 t [m_{\pi}^2(M^2-m_{\rho}^2)^2
+(M^2-m_{\pi}^2)(m_{\rho}^2-m_{\pi}^2) s] \\
& \quad 
+ t^2 [(M^2+2 m_{\pi}^2+m_{\rho}^2)^2-2(M^2+2 m_{\pi}^2+m_{\rho}^2)s+2s^2] 
-2 t^3 (M^2+2 m_{\pi}^2+m_{\rho}^2-s) + t^4 \} \ ,
\end{split}
\end{equation}
\end{widetext}
corresponding to the Feynman diagram depicted in Fig.~\ref{fig-om-t};
$M$ denotes the invariant mass of the lepton pair;
$s=(p_1+p_2)^{\mu}(p_1+p_2)_{\mu}$ and $t=k^{\mu} k_{\mu}$ are the usual
Mandelstam variables with $k^{\mu}=(p_3-p_1)^{\mu}$ the exchanged
four-momentum.  The $\rho$-$\gamma$ vertex has been written in the
form~\cite{Rapp:1999qu}
\begin{equation}
\mathcal{L}_{\rho \gamma}=-C m_{\rho}^2 A_{\mu} \rho^{\mu}
\end{equation}
where $C=e/g_\rho=0.052$.  Medium effects on the $\rho$ meson are
implemented by using the full in-medium propagator, $D_\rho(M)$, for the
intermediate $\rho$ meson which couples to the photon (upper $\rho$ line
in Fig.~\ref{fig-om-t}, with $M^2=q^2$) and by folding Eq.~(\ref{Msqr})
over the in-medium spectral function for the incoming $\rho$ (lower
$\rho$ line in Fig.~\ref{fig-om-t}, with $m_\rho^2=p_1^2$ and a weight
$-2 m_{\rho}/\pi \im D_{\rho}(m_\rho)$).  To avoid double counting with
the $\omega$ $s$-channel graph already included in the in-medium
$\rho$-selfenergy, the integral in Eq.~(\ref{rate-om-t}) is restricted
to spacelike $\omega$-exchange momenta, \ie, $t<0$.  The coupling
constant, $g_{\rho \omega \pi}=25.8\;\text{GeV}^{-1}$, has been fixed
in~\cite{Rapp:1999qu} by a simultaneous fit to the hadronic and
radiative $\omega$ decays including a hadronic dipole-form factor,
\begin{equation}
\label{form-fac}
F(t)=\left ( \frac{2 \Lambda^2}{2 \Lambda^2-t} \right)^2,
\end{equation}
with $\Lambda=1$~GeV. To maintain gauge invariance in a simplified way
as in Ref.~\cite{Turbide:2003si}, we pull the formfactor outside the
integral and introduce an average momentum transfer, $\bar{t}$ (adapted
to the finite mass of the virtual photon), according to
\begin{alignat}{2}
  \frac{F^4(\bar{t})}{(m_{\omega}^2-\bar{t})^2} &= \frac{1}{2}
  \int\limits_{-1}^1 \dd x \frac{F^4[t(x)]}{[m_{\omega}^2-t(x)]^2} \, ,  \\
  t(x)&= M^2+m_{\pi}^2-2 (M^2+q^2-q p x) \, .
\end{alignat}
The amplitude, Eq.~(\ref{Msqr}), is then multiplied by $F^4(\bar{t})$.
Eq.~(\ref{rate-om-t}) furthermore accounts for pion fugacity factors,
$z_\pi$ due to finite pion chemical potentials, implemented in Boltzmann
approximation (\eg, for the incoming $\pi$ and $\rho$ this amounts to a
total fugacity of $z_{\pi}^3$).
\begin{figure}[!th]
\centerline{\includegraphics
[width=0.45\textwidth]{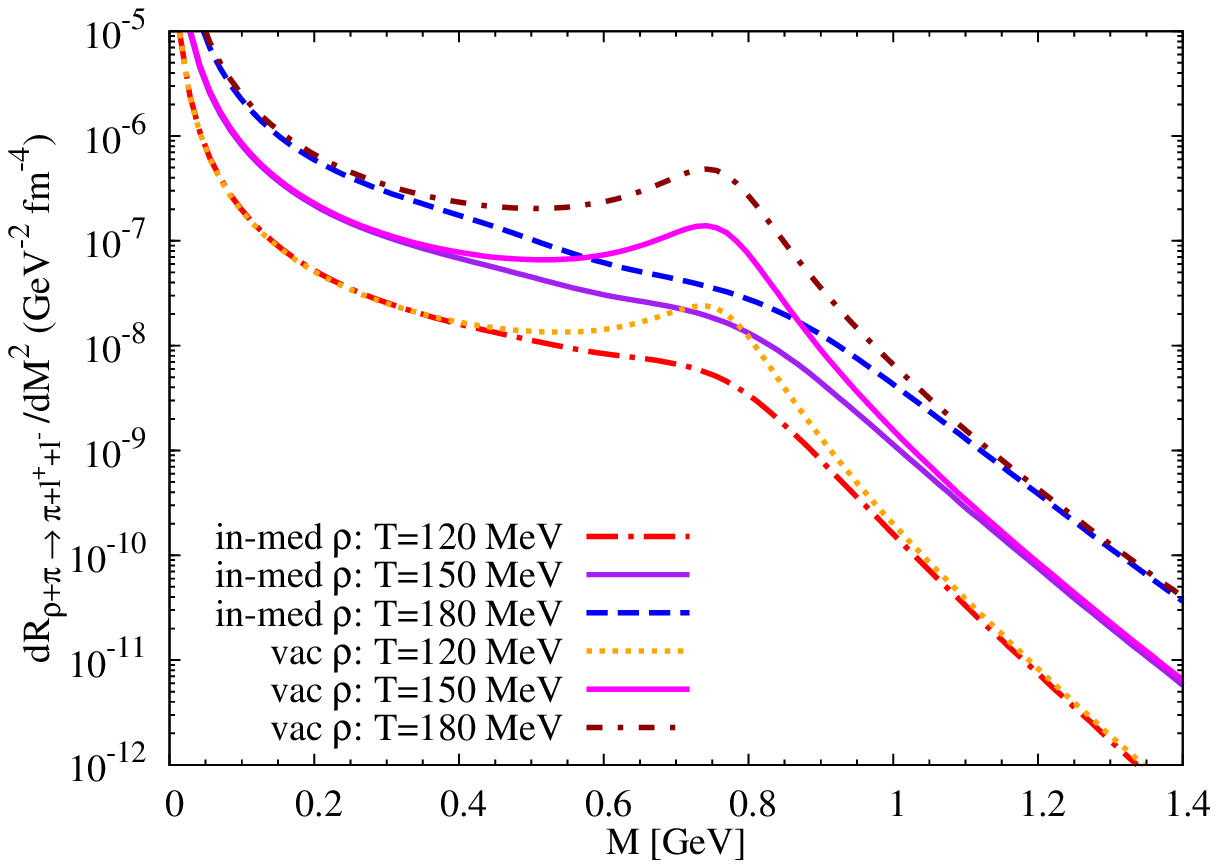}}
\centerline{\includegraphics[width=0.45\textwidth]{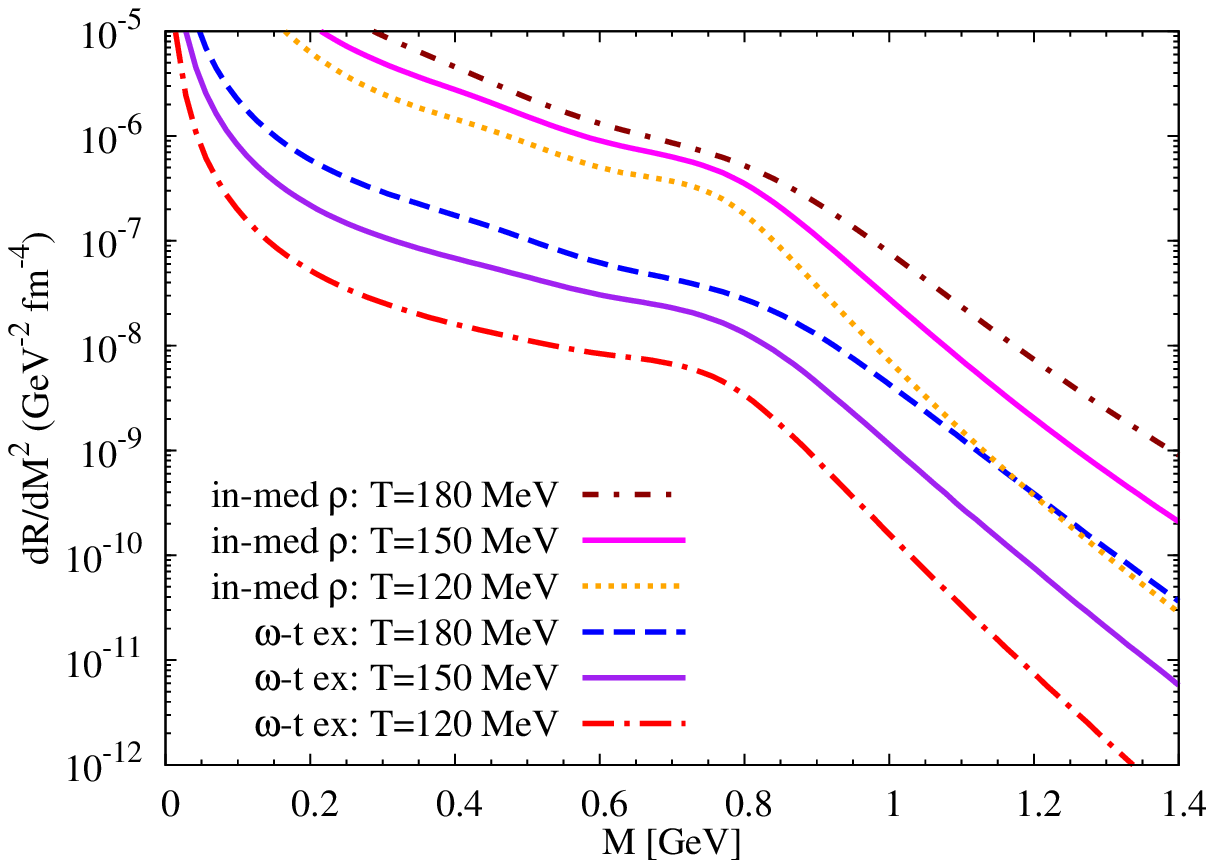}}
\centerline{\includegraphics[width=0.45\textwidth]{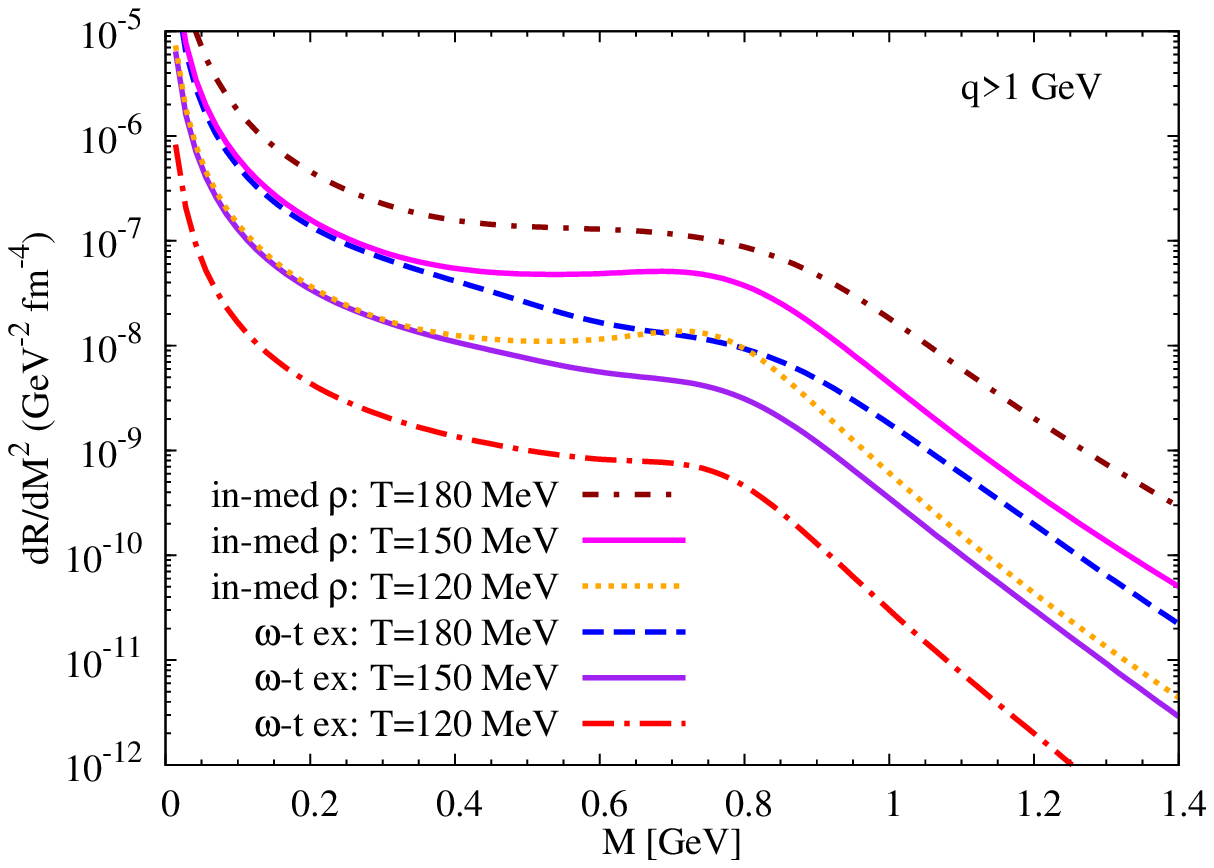}}
\caption{Three-momentum integrated thermal emission rates from $\omega$
  $t$-channel exchange in the $\pi\rho\to\pi e^+e^-$ reaction.  Upper
  panel: comparison of the rates when using either the free or the
  in-medium $\rho$ propagators in both the matrix element and the
  incoming $\rho$-mass distribution.  Middle panel: comparison of the
  in-medium $\omega$ $t$-channel rates with thermal rates from the full
  in-medium $\rho$ spectral function. Lower panel: same as middle panel,
  but restricting the momentum integration to $q>1$~GeV. All curves for
  $T$=150~MeV and 120~MeV include pertinent pion fugacities at
  $\mu_\pi=39$~MeV and 79~MeV, respectively.}
\label{fig_dRdM-om-t}
\end{figure}
In Fig.~\ref{fig_dRdM-om-t} we summarize our results for the $\omega$
$t$-channel emission rate, integrated over three-momentum, for
conditions along our default trajectory for In(158~AGeV)+In collisions
at SPS (including finite $\mu_\pi$). The upper panel illustrates that
the implementation of the in-medium $\rho$ propagator (which was not
done in Ref.~\cite{Turbide:2003si}) leads to a notable reduction in the
emission rate, which is mostly due to the reduction in the intermediate
$\rho$ propagator, $|D_\rho(M)|^2$. We have verified that medium effects
on the $\omega$ propagator are negligible (in the spacelike regime 
finite-width effects are insignificant as long as 
$\Gamma_\omega \ll m_\omega$). 
In the middle panel, the comparison
of the in-medium $\omega$ $t$-channel rates to the ones from the full
in-medium $\rho$ spectral function~\cite{Rapp:1999us}, based on
Eq.~(\ref{rate}), confirms that the former contribution is indeed small
in magnitude. However, if one applies a three-momentum cut of $q>1$~GeV
(lower panel), the relative magnitude of the $t$-channel contribution
increases, as anticipated at the beginning of this section. But even in
this case the relative strength of the $\omega$ $t$-channel exchange is
rather moderate.  The increasing contribution at higher momentum can be
more directly seen when integrating the rate, Eq.~(\ref{rate-om-t}),
over invariant mass bins and plotting it versus $q$, 
cf.~Fig.~\ref{fig_dRdq-om-t}. 
\begin{figure}[!tb]
\centerline{\includegraphics[width=0.45\textwidth]{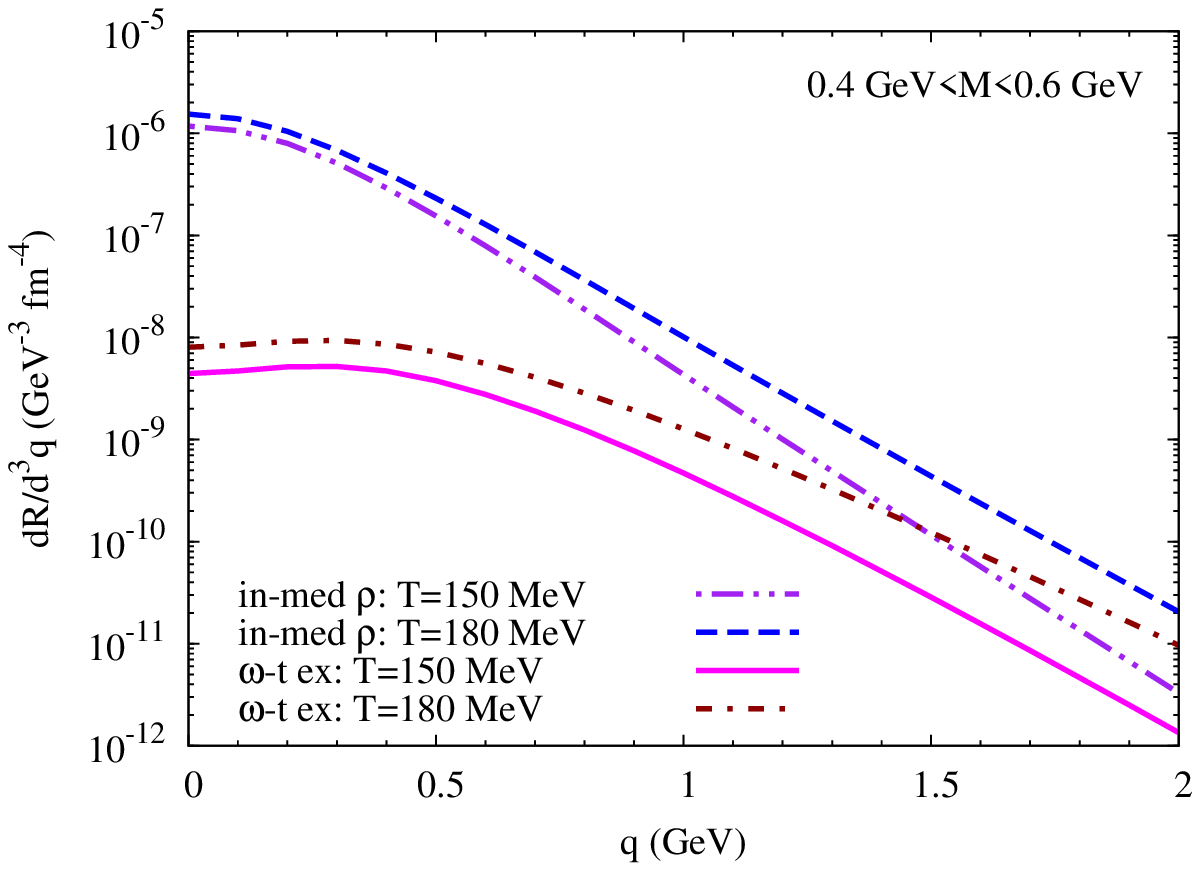}}
\centerline{\includegraphics[width=0.45\textwidth]{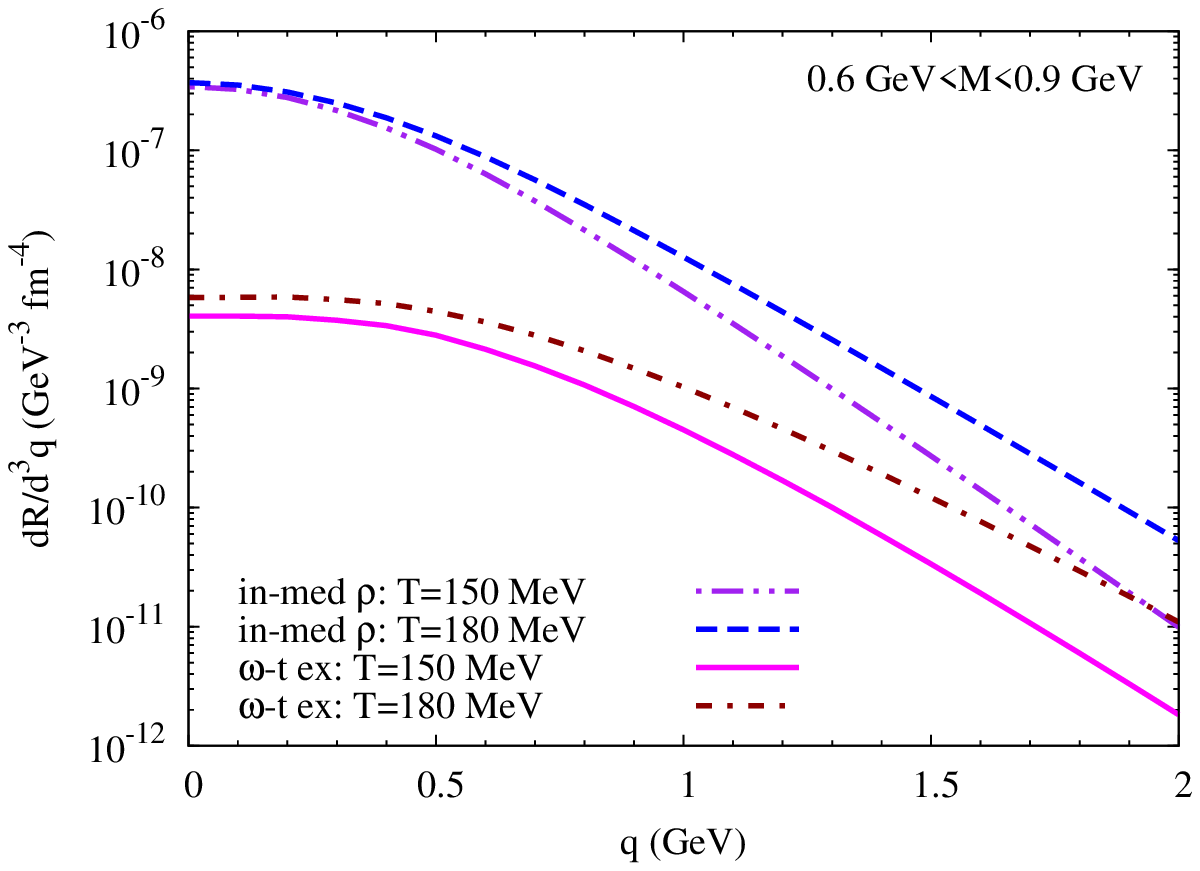}}
\caption{Mass-integrated dilepton rates as a function of three-momentum
  for $\omega$ $t$-channel exchange in $\pi\rho\to\pi e^+e^-$ (computed
  with in-medium $\rho$ propagators) and the leading contribution from
  the full in-medium-$\rho$ spectral function. As in
  Fig.~\ref{fig_dRdM-om-t}, the results for $T$=150~MeV include pion
  fugacity factors.}
\label{fig_dRdq-om-t}
\end{figure}

As indicated above, in our calculations of dilepton spectra in
Sec.~\ref{sec_spectra}, we will implement the $\omega$ $t$-channel
emission rate, Eq.~(\ref{rate-om-t}), by an incoherent addition to the
main contribution, Eq.~(\ref{rate}). This neglects interference terms in
the pertinent selfenergy contribution in the (denominator of the) $\rho$
spectral function, which is justified due to the relative smallness of
the $\omega$ $t$-channel part. To simulate the presence of other
$t$-channel processes (\eg, $\pi$ and $a_1$ exchange), guided by the
photon rate calculations of Ref.~\cite{Turbide:2003si}, we will multiply
the $\omega$ $t$-channel contribution by a factor of 2 in our
calculation of dilepton spectra below.

\subsection{Partonic Emission: Quark-Antiquark Annihilation}
\label{ssec_qgp}

Emission from the QGP is calculated using the hard-thermal-loop improved
rate for $q\bar q$ annihilation~\cite{Braaten:1990wp}, including an
extrapolation to finite three-momentum. It turns out~\cite{Rapp:2002tw}
that this rate has the conceptually attractive feature that it closely
coincides with the rate in hadronic matter when both are extrapolated to
the expected phase-transition region. This is suggestive for a kind of
quark-hadron duality~\cite{Rapp:1999us} and has the additional benefit
that the emission from the expanding fireball can be anticipated to
become rather insensitive to details of how the phase transition is
implemented (\eg, to the values for the critical temperature or ``latent
heat''). This point will be studied explicitly in
Sec.~\ref{ssec_hadrochem} below.

Recent (experimental and theoretical) developments suggest that the QGP
features significant nonperturbative effects for temperatures up to
$\sim$$1.5$-$2 T_c$, \eg, the survival of (hadronic) resonances and/or
bound states, even in the light-quark sector. Possible consequences for
the dilepton emission rate have been estimated in
Ref.~\cite{Casalderrey-Solana:2004dc}. Depending on the width of these
states, a maximal enhancement of up to a factor of $\sim 2$ over the
perturbative QGP emission rate at intermediate masses (\ie, in the mass
range of $1.5$-$2$~GeV) is conceivable. We will not further pursue this
possibility in the present paper.

\section{Dilepton Sources other than Thermal Radiation}
\label{sec_nontherm}
In this section we address dilepton sources other than radiation from a
thermal source as given by Eq.~(\ref{rate}), \ie, $\rho$ decays after
thermal freezeout, decays of primordially produced $\rho$ mesons which
escape the fireball, and primordial Drell-Yan annihilation.  For
simplicity, we generically refer to these sources as ``non-thermal'',
even though the freezeout decays, \eg, are represented by thermal
``blast-wave'' spectra commonly used to characterize light hadron spectra
in heavy-ion collisions.

\subsection{$\rho$ Mesons at Thermal Freezeout}
\label{ssec_fo}
Dilepton decays of long-lived hadrons (most notably Dalitz decays of
$\eta$ and $\omega$, as well as exclusive $l^+l^-$ decays of $\omega$
and $\phi$) can be rather well separated from the freezeout of the
interacting fireball in heavy-ion collisions, which has given rise to
the notion of the ``hadronic decay cocktail'' contribution to dilepton
spectra, computed using the free spectral shape of the decaying mesons.
For the $\rho$ meson, due to its short lifetime, such a separation is
not well defined. Therefore, when adding thermal radiation to the
cocktail, the $\rho$ is commonly removed from the latter and implemented
into the thermal yield. In previous
works~\cite{Rapp:1997fs,Rapp:1999us,vanHees:2006ng} this was done by
running the fireball an extra $\sim$$1$~fm/$c$ using the in-medium
$\rho$ spectral function. It turns
out~\cite{Rapp:2006cj,vanHees:2007yi}, however, that this description of
$\rho$ decays at thermal freezeout carries an extra factor of $1/\gamma$
relative to a standard blast-wave spectrum of hadrons at thermal
freezeout, where $\gamma=q_0/M$ is the usual Lorentz factor. Roughly
speaking, the in-medium radiation given by Eq.~(\ref{rate}) is
proportional to the electromagnetic decay width times the fireball
lifetime, $(\Gamma_{ll}/\gamma)\tau_{\rm FB}$, while decays after
freezeout are proportional to the branching ratio of electromagnetic to
total lifetime, $\Gamma_{ll}/\Gamma_\rho^{\rm fo}$, where the $\gamma$
factor cancels (as usual, we define $\Gamma_X$ as the partial decay
width in the rest system of the particle). The remainder of this section
will give a more detailed discussion of this.

To account for the correct time dilation effects in the calculation of
dilepton decays of $\rho$-mesons after thermal freezeout, we use the
standard Cooper-Frye description~\cite{cf74},
\begin{equation}
\dd N=q_{\mu} \dd \sigma^{\mu} \frac{\dd^3 q}{(2 \pi)^3 q^0} f_B
\left(\frac{q_{\nu} u^{\nu}}{T} \right ),
\end{equation}
for the phase-space distribution of an on-shell particle ($q_{\mu}
q^{\mu}$$=$$m^2$) at thermal freezeout; $T$ and $u^{\nu}=\gamma
(1,\beta_\perp)$ denote the local temperature and four velocity (flow)
of the fluid cells of the medium, and $\dd \sigma^{\mu}$ is the
hypersurface normal vector defined by an appropriate freezeout
condition. In accordance with the homogeneous fireball model described
in Sec.~\ref{sec_fireball}, thermal freezeout at a constant time $t=x^0$
in the laboratory frame is assumed, \ie, $(\dd \sigma^{\mu}) \equiv
(\dd^3 x,0,0,0)$.  The in-medium $\rho$ spectral function at freezeout
is introduced via the substitution
\begin{equation}
\frac{\dd^3 q}{q_0} \rightarrow \dd^4 q \; 2 \delta^+(q_{\mu} q^{\mu}-m^2)
\rightarrow \dd^4 q \frac{A_{\rho}}{\pi} \ , 
\end{equation}
where $A_{\rho}=-2/3 \im (D_{\rho})_{\mu}^{\mu}$ (as before) includes
the average over the three polarizations via the Lorentz trace which
runs over the physical (\ie, four-momentum transverse) components of the
propagator. To properly treat the low-mass tails of the spectral
function, one would have to resolve the individual resonance decays
figuring into the $\rho$ selfenergy. To simplify our task, we circumvent
this problem by employing the vacuum form of the $\rho$ selfenergy
augmented with a width corresponding to the full-width-half-maximum of
the in-medium spectral function at thermal freezeout
($\Gamma_{\rho}^{\text{fo}} \simeq 260$~MeV). The distribution of $\rho$
mesons at thermal freezeout then reads
\begin{equation}
  \frac{\dd N_{\rho}^{\mathrm{fo}}}{\dd^3 x \dd^4 p}=p^0 \frac{3 A_{\rho}}{8 \pi^4} 
 f^B\left(\frac{p_{\nu} u^{\nu}}{T} \right ) \  .
\end{equation}
In accordance with the averaging procedure over the freezeout duration,
we evaluate the transverse four-velocity for the freezeout $\rho$ at a
time $\Delta\tau_\rho^{\rm fo}/2\simeq0.5$~fm/c after the thermal
emission, Eq.(\ref{rate}), has shut off. 
The dilepton rate follows by folding with the
appropriate partial decay width. Within our vector-meson dominance model
this is given by the matrix element for the process $\rho \rightarrow
\gamma^* \rightarrow l^+ + l^-$, and after integration over $t \in
(t_{\mathrm{fo}},\infty)$ this results in
\begin{equation}
\begin{split}
\label{fo-dilep}
\frac{\dd N_{ll}^{(\mathrm{fo})}}{\dd^3 x \dd^4 q} &= \frac{q^0}{M} \;
\frac{\alpha^2 m_{\rho}^4}{g_{\rho}^2 M^2} \frac{A_{\rho}}{2 \pi^3}
L(M) f^B\left(\frac{p_{\nu} u^{\nu}}{T} \right )
\frac{1}{\Gamma_{\rho}^{\text{fo}}} \\
&=\frac{q_0}{M} \frac{1}{\Gamma_{\rho}^{\text{fo}}} \left ( \frac{\dd
    N_{ll}}{\dd^4 x \dd^4 q} \right )_{t=t_{\mathrm{fo}}},
\end{split}
\end{equation}
where $L$ is the dilepton-phase space factor (\ref{lept-ps}). The second
line of Eq.~(\ref{fo-dilep}) shows that the momentum dependence of the
dilepton distribution from $\rho$ decays after thermal freeze-out
deviates from the rate from a thermal source (\ref{rate}) by a Lorentz
factor $\gamma\equiv q_0/M=\sqrt{M^2+q^2}/M$. The physical origin of
this difference is the time dilation of the total lifetime of a
freeze-out $\rho$ meson with three-momentum $q$ which is absent in the
formula for radiation from a thermal source, because its $\rho$-meson
abundance at each instant of time is fixed by the temperature and
pion-chemical potential of the medium (as required by detailed balance
of $\rho$ formation and decay), and thus the total number of thermal
dileptons is determined by the lifetime of the fireball. Note, however,
that the thermal rate is proportional to $\Gamma_{ll}$ with an
associated time dilation factor $1/\gamma$. The freezeout-$\rho$
dilepton spectra are thus equivalent to standard blast-wave descriptions
of stable hadrons~\cite{Schnedermann:1993ws}.

Our default assumption for the radial profile of the flow field is a
linear dependence on the radius according to
\begin{equation}
\label{flow-profile}
\beta_\perp(r,t)= \beta_\perp^s \frac{r}{R(t)},
\end{equation}
where $R(t)=r_0+a_{\perp} t^2/2$ is the radius of the fire-cylinder,
$\beta_\perp^s =a_{\perp} t$ its surface speed, and $r\le R(t)$ the
radial coordinate of the fluid cell related to the volume element
$\dd^3x=2\pi r \dd r \dd z$ (since we neither address azimuthal
asymmetries nor peripheral collisions in this work, we approximate the
cross sectional area of the fire cylinder as a circle).

\subsection{Primordial $\rho$ Mesons}
\label{ssec_hard}

Another source of non-thermal dileptons is the decay of $\rho$ mesons
which originate from primordial hard-scattering processes and traverse
the interaction zone without equilibrating. We evaluate this
contribution within a schematic jet-quenching
model~\cite{vanHees:2007yi}, as follows.

First, we construct the $q_T$ spectrum of primordial $\rho$ mesons
assuming a power law,
\begin{equation}
\label{primord}
\frac{1}{q_T} \frac{\dd N_{\text{prim}}}{\dd q_T}=\frac{A}{\left (1+B q_T^2
  \right)^{a}} \ ,
\end{equation}
with parameters estimated from $p$-$p$ scattering
data~\cite{AguilarBenitez:1991yy}, $B=0.525\text{GeV}^{-2}$ and $a=5.5$.
The total number of primordial $\rho$'s in $A$-$A$ collision is
determined based on the empirical freezeout systematics of light hadron
production, \ie, we calculate the expected total number of $\rho$ mesons
in In-In at $T_c=175$~MeV and match the norm of the primordial spectrum,
Eq.~(\ref{primord}), to it. At the same time, we have to take care of
the correct scaling properties of the spectrum at high momentum: while
the (total) yield at low $q_T$ is proportional to the number of
participant nucleons, the high-$q_T$ yield should scale with the number
of primordial $N$-$N$ collisions.  We implement this transition by a
continuous linear switching between the two regimes over the range
$1~\text{GeV}<q_T<3~\text{GeV}$.

Second, we implement a Cronin effect for A-A collisions by a ``Gaussian
smearing'' of Eq.~(\ref{primord}),
\begin{equation}
  \frac{\dd N_{\text{prim}}^{\text{cron}}}{\dd^2 q_T} =\int \frac{\dd^2
    k_T}{\pi \Delta k_T^2}
  \frac{\dd N_{\text{prim}}}{\dd^2 k_T} \exp
  \left[-\frac{(q_T-k_T)^2}{\Delta k_T^2} \right],
\end{equation}
with a conservative estimate of $\Delta k_T^2=0.2$~GeV$^2$ as extracted
in Ref.~\cite{Turbide:2003si} based on direct photon spectra in $p$-$A$
reactions.

\begin{figure}[!t]
\centerline{\includegraphics[width=0.45\textwidth]{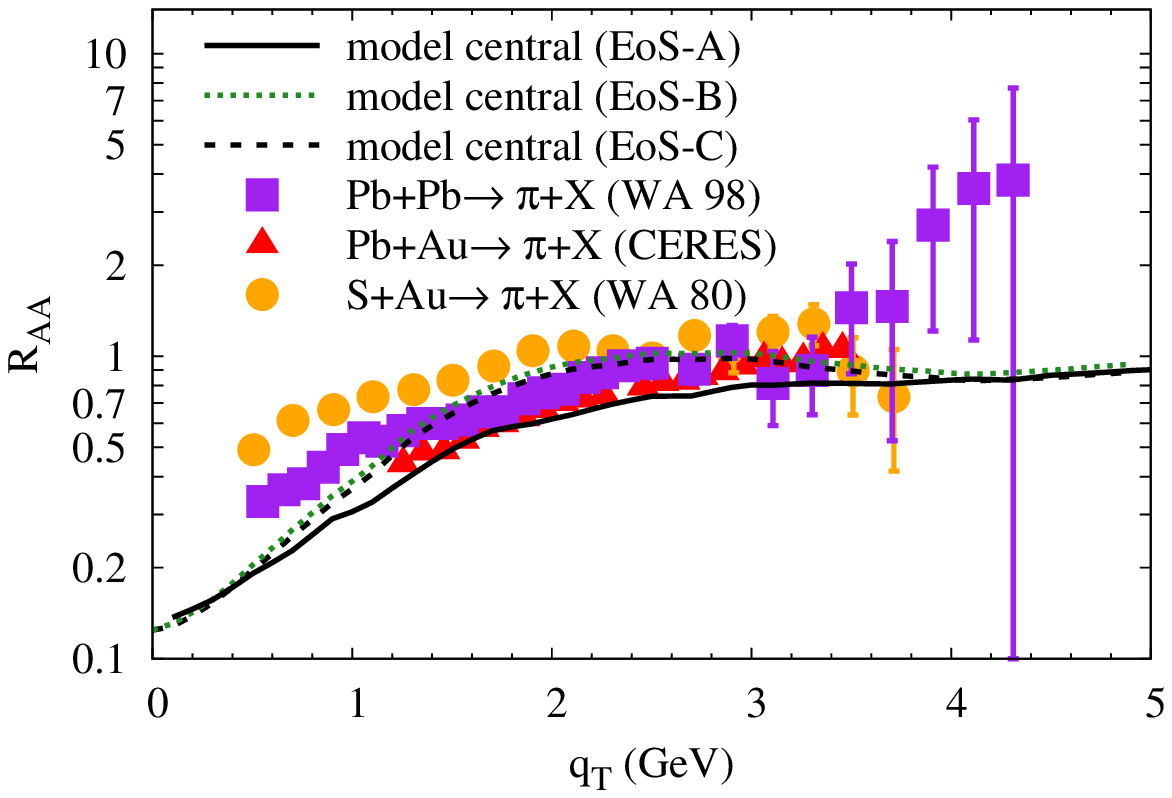}}
\hspace{0.2cm}
\centerline{\includegraphics[width=0.45\textwidth]{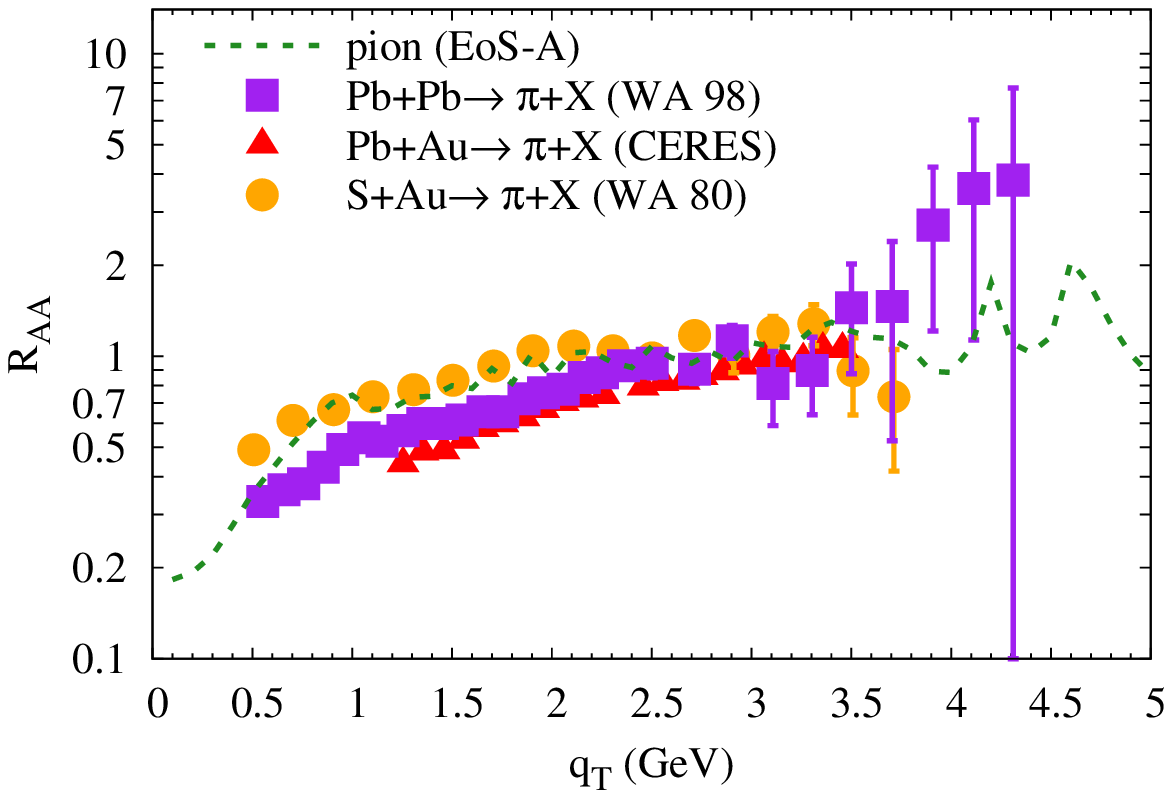}}
\caption{(Color online) Nuclear modification factor for transverse
  momentum spectra of $\rho$ (upper panel) and $\pi$ mesons (lower panel) 
  from primordial (hard) production plus thermal freezeout, 
  compared to a compilation of data for $R_{AA}$ of pions at full SPS
  energy~\cite{dEnterria:2005cs} for central collisions with the
  standard equation of state (EoS-A, full line) and an EoS with
  $T_{\text{ch}}=160\;\text{MeV}$ (EoS-B+C as discussed in
  Sec.~\ref{sec_fireball}, dashed line). The primordial $\rho$ 
  spectra include effects of initial Cronin smearing and jet-quenching in the
  expanding fireball, while the thermal freezeout $\rho$'s are taken
  from Sec.~\ref{ssec_fo} (same for the pions, but without finite-width
  effects).}
\label{fig_raa}
\end{figure}
Third, we calculate a suppression factor representing the probability
for primordial $\rho$ mesons to escape the medium without rescattering,
using Monte Carlo techniques. In line with our fireball model, we start
from a spatially homogeneous distribution in the transverse plane at the
QGP formation time. The escape probability for a $\rho$ with
momentum $q$ is then calculated as
\begin{equation}
P=\exp \left (-\int \dd t~\sigma_{\rho}^{\rm abs}(t)~\varrho(t) 
\right),
\end{equation}
where the absorption cross section
\begin{equation}
\sigma_{\rho}^{\rm abs}(t)=\begin{cases}
\sigma_{\text{ph}}=0.4 \;\text{mb} & \text{for} \quad
t< q_0/m_{\rho}  \, \tau_{\text{f}} \\
\sigma_{\text{had}}=5 \; \text{mb} & \text{for} \quad
t> q_0/m_{\rho} \, \tau_{\text{f}}
\end{cases}
\end{equation}
depends on the $\rho$-meson formation time in its rest frame, which is
assumed to be $\tau_{\text{f}}$$=$$1$~fm, augmented by Lorentz time
dilation in the fireball frame; $\sigma_{\text{ph}}$ and
$\sigma_{\text{had}}$ denote the pre-hadronic and hadronic absorption
cross sections of the $\rho$, respectively, and $\varrho(t)$ the
corresponding particle density of the fireball medium (partonic or
hadronic; if the $\rho$ meson has not formed by the time the system has
hadronized, we use constituent quark scaling to infer the partonic
density as $\varrho_{\rm p}=3\varrho_{\rm had}$, and vice versa).

A rough check of our construction may be obtained by comparing the
$\rho$ spectra to measured pion spectra in $A$-$A$ collisions at the
SPS. Since pion spectra contain both a hard (primordial) and a soft
(thermal + flow) component, a comparison with our $\rho$ spectra should
contain both the primordial part constructed in this section and the
thermal freezeout contribution described in the previous section,
\ref{ssec_fo}. We do this comparison in terms of the nuclear
modification factor, $R_{AA}$, defined as the ratio of the spectrum in
$A$-$A$ collisions over the collision-number scaled spectrum in $p$-$p$. 
The upper panel of Fig.~\ref{fig_raa} shows that the jet-quenching + 
freezeout-$\rho$ model results in fair agreement with a recent
compilation of $R_{AA}$ for pions in S-Au and Pb-Pb systems from 
various experiments at full SPS energy~\cite{dEnterria:2005cs}.
As a check of this procedure, as well as of the fireball model, we
plot in the lower panel of Fig.~\ref{fig_raa} the calculated pion 
$R_{AA}$ for central In-In at SPS (as a sum of the jet-quenching plus 
thermal-freezeout components) for central In-In at SPS; the agreement 
with data is reasonable (note that resonance decays are not included, 
which are mostly concentrated at low $q_T$).

\subsection{Drell-Yan Annihilation and Correlated Charm Decays}
\label{ssec_dy}

In the high-mass region (HMR, $M>3$~GeV), dilepton spectra in nuclear
collisions are expected to be dominated by the Drell-Yan (DY) process,
\ie, primordial annihilation of quarks and antiquarks within the
incoming nucleons. To leading order, this process does not depend on the
strong coupling constant (it is purely electromagnetic, ${\cal
  O}(\alpha_s^0\alpha_{\mathrm{em}}^2)$) and can therefore be rather
reliably evaluated, provided one has a good knowledge of the
parton-distribution functions (PDFs) within the nucleon. In a central
collision ($b=0$) of two equal nuclei with mass number $A$, the
invariant-mass spectrum of Drell-Yan pairs per unit rapidity is given by
\begin{equation}
\left . 
\frac{\dd N_{\text{DY}}^{AA}}{\dd M \dd y} \right |_{b=0}=\frac{3}{4\pi
 R_0^2} \, A^{4/3} \, \frac{\dd \sigma_{\text{DY}}^{NN}}{\dd M \dd y} 
\label{dyaa}
\end{equation}
in terms of the standard DY cross section in an elementary
nucleon-nucleon ($N$-$N$) collision,
\begin{equation}
\frac{\dd \sigma_{\text{DY}}^{NN}}{\dd M \dd y}=K \frac{8\pi\alpha}{9sM}
\sum\limits_{q=u,d,s} e_q^2 \left[ q(x_1) \bar q(x_2) +
\bar q(x_1) q(x_2) \right]  \ .
\label{dynn}
\end{equation}
Here, $q(x_{1,2})$ and $\bar q(x_{1,2})$ denote the (collinear) quark
and anti-quark distribution functions, respectively (neglecting nuclear
effects). Their arguments are related to the center-of-mass (cms)
rapidity, $y$, and the invariant mass of the lepton pair as $x_{1,2}=x
e^{\pm y}$ with $x=M/\sqrt{s}$, where $\sqrt{s}$ is the cms energy of
the $N$-$N$ collision. The root-mean-squared radius parameter in
Eq.~(\ref{dyaa}), $R_0\simeq1.05$~fm, arises from a folding over a
Gaussian thickness function; for simplicity, we will adopt
Eq.~(\ref{dyaa}) also for non-central collisions with an accordingly
reduced number of participants, $A$. For the parton distribution
functions we employ the leading-order parameterization from
Ref.~\cite{GRV95} (GRV94LO), which incorporate isospin asymmetries in
the sea-quark distributions which significantly reduce the need for
additional isospin corrections for nuclei with $N\ne Z$ (\eg, less than
$5\%$ for Pb-Pb collisions), which will be neglected
here~\cite{Spiel98}. In Eq.~(\ref{dynn}) higher-order corrections in
$\alpha_s$ are encoded in an empirical $K$ factor, which turns out to be
$K\simeq1.5$ to reproduce DY production in $p$-$A$
collisions~\cite{Spiel98}.

In addition to generating a $K$ factor, higher-order effects manifest
themselves in a nonzero pair-$q_T$ of the DY dileptons. To obtain a
realistic $q_T$ spectrum, we follow the procedure adopted by the NA50
collaboration~\cite{Abreu:1999jr,Abreu:2000nj}: based on a comprehensive
analysis of $p$-$A$ and $A$-$A$ collisions at the SPS it has been found
that the IMR and HMR dilepton spectra can be fairly well described by
assuming a Gaussian shape for the DY spectrum,
\begin{equation}
\label{dy-qT-spec}
\frac{\dd N_{\text{DY}}}{\dd M \dd y \dd q_T^2} =  \frac{\dd
  N_{\text{DY}}}{\dd M \dd y} \, 
\frac{\exp(-q_T^2/2\sigma_{q_T}^2)}{2\sigma_{q_T}^2}  
\end{equation}  
with $\sigma_{q_T}\simeq0.8-1$~GeV. In Ref.~\cite{Rapp:1999zw}, a value
of $\sigma_{q_T}\simeq0.8$~GeV has resulted in reasonable agreement with 
the NA50 $q_T$ spectra for IMR dimuons in central Pb-Pb; for 
$\sigma_{q_T}\simeq0.9$~GeV, which increases the spectrum at $q_T=2~$GeV
by 50\%, the agreement with NA50 would still be acceptable. Here, we
adopt $\sigma_{q_T}\simeq0.8$~GeV providing a conservative estimate
of the DY contribution. 

The extrapolation of the DY spectrum to the regions where both mass and
momentum are small (say, below $M,q_T\simeq1.5$~GeV) is problematic, but
fortunately its contribution in this regime is small compared to, \eg,
thermal emission in $A$-$A$ collisions. There is, however, an additional
constraint provided by the photon point ($M \to 0$) which allows to
extrapolate the DY spectra to low mass, at least for reasonably large
momenta, $q_T>1$~GeV (which is the regime were the DY yield is noticeable).  
Formally, the photon-production rate follows from
the dilepton one by taking the limit $M\to 0$. More specifically, this
is encoded in the relation
\begin{equation}
q_0 \frac{\dd R_\gamma}{\dd^3 q} = -\frac{\alpha}{2 \pi^2}
\im \Pi_{\text{em},\mu}^{\mu}(M=0,q) f^B(q_0;T)
\label{phrate}
\end{equation} 
with the same e.m.~current-current correlation function,
$\Pi_{\text{em}}^{\mu \nu}$ of Eq.~(\ref{ret-se}), as in the dilepton rate,
Eq.~(\ref{rate}). For $q\gg M$, the $M$-dependence of $\Pi_{\text{em}}$
is weak, and the difference between $q_0 \frac{dN_{ll}}{dM d^3q}$ 
and $q_0 \frac{\dd R_\gamma}{\dd^3 q}$ amounts to a factor of 
$\frac{2\alpha}{3\pi M}$. Thus we can evaluate the DY $q_T$ spectrum 
at a mass $M_{cut}$ and extrapolate it down in mass to the photon point 
using the factor $M_{cut}/M$. For $M_{cut}$=0.8-1~GeV, in connection with
$\sigma_{q_T}\simeq0.8$~GeV, reasonable agreement with the primordial
photon spectrum of Ref.~\cite{Turbide:2003si} is found.     

In addition, there could be ``pre-equilibrium''
contributions from secondary Drell-Yan processes~\cite{Spiel98} (\eg,
$\pi N\to \mu\mu X$ involving primordially produced pions), which turn
out to be rather sensitive to the pion formation time. \textit{E.g.},
for $\tau^\pi_{\rm form}=1$~fm/$c$, the enhancement over primordial
Drell-Yan annihilation in central S-U was found to be $\sim$$10\%$ at
$M=2$~GeV. In a thermal emission description, which we employ here after
a rather early thermalization time of $\tau_0=1$~fm/$c$, it is difficult 
to separate pre-equilibrium radiation from thermal emission (e.g., secondary 
Drell-Yan in $\pi$-$N$ interactions might overlap with thermal 
$q$-$\bar q$ annihilation). A rough estimate of pre-equilibrium effects
may be obtained by varying the thermalization time; decreasing, \eg, 
$\tau_0$ to 0.8~fm/$c$ (which is close to the overlap time of the 2 colliding
nuclei at SPS energies) increases the QGP contribution at $q_T=2$~GeV   
by $\sim$50\%, which is less than 10\% of the Drell-Yan contibution at 
all masses considered. 

For the dilepton contribution from correlated decays of $D$ and
$\bar{D}$ mesons we use the experimental result from $p$-$p$ collisions
extrapolated to In-In, as provided by the NA60
collaboration~\cite{Arnaldi:2006jq}. As a note of caution, we remark
that recent measurements at the Relativistic Heavy-Ion Collider (RHIC)
report substantial modifications of heavy-quark spectra in Au-Au
collisions, relative to $p$-$p$ (as inferred from a suppression and
elliptic flow of ``non-photonic'' single-electron spectra associated
with semileptonic decays of open-charm (and -bottom)
hadrons)~\cite{Adare:2006nq,Abelev:2006db}. Such (possibly
nonperturbative~\cite{vanHees:2005wb}) medium modifications of the charm
momentum spectra presumably translate into a softening of the
invariant-mass spectra of $l^+ l^-$ pairs as well. At the SPS, the
shorter QGP lifetime is likely to lead to smaller effects, but an
explicit measurement of the (delayed) charm decays has been presented
recently~\cite{Shahoyan:2007zz}.

\section{Thermal Fireball Evolution}  
\label{sec_fireball}

Our description of the space-time evolution of central and semicentral
$A$-$A$ collisions is approximated by an expanding thermal fireball
characterized by a time dependent cylindrical volume
as~\cite{Rapp:1999us},
\begin{equation}
  V_{\mathrm{FB}}(t) = \pi \left (r_{\perp,0}+\frac{1}{2} a_\perp t^2
  \right)^2 \left (z_0+v_{z,0} t + \frac{1}{2} a_z t^2 \right )
\label{VFB}  \ ,   
\end{equation}
where we neglect effects due to a finite ellipticity.  The initial
transverse radius $r_{\perp,0}$ is determined by the centrality of the
collision (\eg, $r_{\perp,0}=5.15(4.6)\;\text{fm}$ for central
(semicentral) In-In collisions). The initial longitudinal size, $z_0$,
is equivalent to the formation time, $\tau_0$, of the thermal medium,
which we fix at the standard value of $\tau_0=1$~fm/$c$ (translating
into $z_0\simeq \tau_0\Delta y=1.8$~fm, where $\Delta y=1.8$ represents
the rapidity width of a thermal fireball). For the longitudinal
expansion we employ a moderate acceleration, $a_z=0.045c^2$/fm, together
with an initial velocity of $v_{z,0}=0.6c$ (reminiscent to
Ref.~\cite{Rapp:1999us}), but the dilepton invariant-mass spectra are
essentially unaffected if we use $(v_{z,0},a_z)=(c,0)$ as in
Ref.~\cite{vanHees:2006ng}. The most important parameter is the
transverse acceleration. More recent applications of the fireball model,
both in the dilepton~\cite{vanHees:2006ng} and charm
diffusion~\cite{vanHees:2005wb} context, have used larger values than in
previous work~\cite{Rapp:1999us,Rapp:2000pe}, in the range
$a_\perp=0.08-0.1c^2$/fm. Here, we employ $a_\perp=0.085 c^2/\text{fm}$
as in Ref.~\cite{vanHees:2007yi}.
  
The time evolution of the temperature is determined assuming entropy
conservation. At a given collision energy, the hadro-chemistry of the
fireball is characterized by thermal-model fits to the observed hadron
ratios. In our default scenario we assume the chemical freezeout
temperature to coincide with the critical temperature for QGP formation,
at $T_{\rm ch}=T_c=175$~MeV. For central and semicentral In(158A~GeV)+In
collisions we fix the entropy per (net) baryon at $s/\varrho_B^{\rm
  net}=27$, which, using a hadronic resonance gas (HG) equation of state
(EoS), translates into an associated baryon chemical potential of
$\mu_B^{\rm ch}=232$~MeV, well within the uncertainties of recent
thermal model fits at SPS~\cite{Andronic:2005yp,Becattini:2005xt} (all
other chemical potentials being zero). The subsequent hadronic
trajectory in the phase diagram is then constructed at fixed
$s/\varrho_B^{\rm net}=27$ (isentropic expansion) with the additional
constraints of pion, kaon, $\eta$ and antibaryon number conservation,
requiring the build-up of corresponding chemical potentials. For a given
number of nucleon participants (\ie, collision centrality), the fireball
entropy amounts to $S=2630(1890)$ for central (semicentral) In-In
collisions, translating into a charged particle multiplicity of $\dd
N_{\text{ch}}/\dd y \simeq 195(140) \simeq N_{\text{part}}$ (for a
chemical freezeout temperature of $T_{\rm ch}=160$~MeV, as considered
below, the multiplicities increase by less than 5\%). Using the relation
between the total entropy and volume, $S= s(t) \cdot V(t)$, the entropy
density, $s(t)$, can be used to determine the time evolution of
temperature and baryon density (along the hadronic trajectory) using the
HG EoS, and similarly for the QGP phase using a quasiparticle EoS. The
volume partition in the HG-QGP mixed phase is calculated from the
standard mixed-phase construction, where the fraction of matter in the
hadronic phase is given by
\begin{equation}
f_{\rm HG}(t)=\frac{s_c^{\rm{QGP}}-s(t)}{s_c^{\text{QGP}}-s_c^{\text{HG}}}
 \ , 
\end{equation}
where $s_c^{\rm{HG,QGP}}$ denote the critical entropy densities at
$T_c$. With a formation time of $\tau_0=1$~fm/$c$, the evolution for
central (semicentral) In-In starts in the QGP at $T_0=197(190)$~MeV,
passes through a mixed phase at $T^{\text{ch}}=T_c=175\; \text{MeV}$,
and terminates at thermal freezeout at around $T_{\text{fo}} \simeq
120$-$135$~MeV.

The main uncertainties associated with the fireball evolution are the
transverse acceleration as well as the overall lifetime (which is
somewhat correlated to the longitudinal expansion). It turns out that
the latter is rather sensitive to the absolute magnitude of the
experimentally observed dilepton excess radiation, resulting in
$\tau_{\rm fo} \simeq 6 \; \text{fm}/c$ for central In-In collisions. The
remaining uncertainty consists of the interplay between longitudinal and
transverse expansion. As mentioned above, the variations considered in
our previous works~\cite{vanHees:2006ng,vanHees:2007yi},
$a_\perp=0.08$-$0.085 \, c^2/\text{fm}$ and
$(v_z^0,a_z)=(c,0)$-$(0.6c,0.045\,c^2/{\rm fm})$, have negligible impact
on the invariant mass spectra, while the longitudinal-acceleration
scenario, which we will focus on here, allows for a slightly larger
transverse expansion which appears to be favored by the rather hard $q_T$ 
spectra. In principle, more accurate information on the final state of 
the expansion can be obtained once hadronic spectra for In-In are 
available (our pion spectra are actually in reasonable agreement with 
S-Au data, recall lower panel of Fig.~\ref{fig_raa}). For central Pb-Pb 
collisions at SPS energies, the typical results for transverse 
surface velocity and thermal freezeout temperature are 
$(\beta_\perp^s/c,T_{\rm fo}[{\rm MeV}])=(0.65\pm0.1,120\pm10)$
\cite{Bearden:1996dd,Appelshauser:1997rr,Antinori:2001yi,Adamova:2002wi}, 
which is quite comparable to the
values used here for In-In. We emphasize that all
contributions to the dilepton spectrum (QGP, $\rho$, $\omega$, $\phi$,
and four-pion) are tied to the \emph{same} evolution, thus fixing their
relative weights.

Finally, to illustrate the uncertainties associated with the underlying
equation of state we will investigate three combinations of critical and
chemical freezeout temperatures roughly covering the current theoretical
and experimental ranges:
\begin{itemize}

\item[(A)] Our default scenario, employed in most of our calculations
  thus far, consisting of identical $T_c$ and $T_{\rm ch}$ at an
  ``intermediate'' value of $175$~MeV; thermal freezeout is fixed at
  $(\mu_\pi^{\rm fo},T_{\rm fo})=(79,120)\; \text{MeV}$ (semicentral
  In-In).

\item[(B)] A scenario with a relatively small and identical critical and
  chemical freezeout temperature at $(\mu_B^{\rm ch}, T_{\rm
    ch})=(240,160)$~MeV, compatible with recent thermal model fits in
  Refs.~\cite{Andronic:2005yp,Becattini:2005xt}. For dilepton spectra
  the most important consequences of this scenario are a significantly
  extended QGP phase which will increase its thermal emission
  contribution and reduce the hadronic one, in particular at
  intermediate masses. In addition, due to smaller pion chemical
  potentials in the subsequent hadronic phase, the freezeout temperature
  (at fixed fireball lifetime) will be larger than with EoS-A, at
  $(\mu_\pi^{\rm fo},T_{\rm fo})=(37,136)\;\text{MeV}$.

\item[(C)] A scenario with a large critical temperature $T_c=190$~MeV
  (as suggested by recent lattice QCD computations~\cite{Karsch:2007vw},
  which maximizes (minimizes) the space-time volume occupied by the
  hadronic (QGP) phase. Since chemical freezeout at such a temperature
  is questionable, we allow for a chemically equilibrated hadronic phase
  until chemical-freeze out sets in under the same conditions as in
  EOS-B, at $(\mu_B^{\rm ch}, T_{\rm ch})=(240,160)$~MeV.
\end{itemize}

\section{Comparison to Dilepton Spectra at SPS}
\label{sec_spectra}

We now turn to a systematic analysis of experimental dilepton spectra as
measured at the SPS by the NA60 and CERES/NA45 collaborations in In-In
and Pb-Au collisions, respectively. Based on the various ingredients
developed in the previous sections, we first address the NA60
invariant-mass and transverse-momentum spectra, followed by a
consistency check with earlier and updated CERES data.

\subsection{Invariant Mass Spectra}
\label{ssec_mspec}
\begin{figure}[!t]
\centerline{\includegraphics[width=0.46\textwidth]{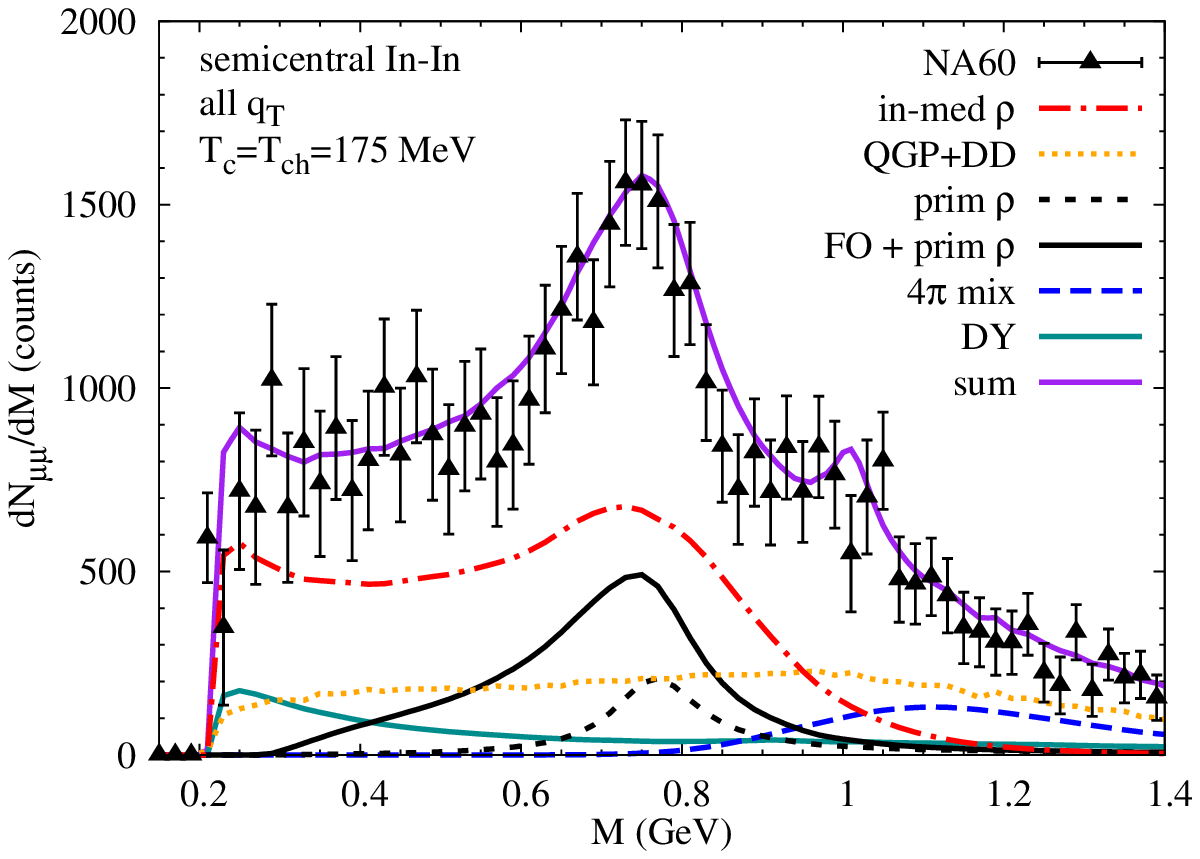}}
\centerline{\includegraphics[width=0.46\textwidth]{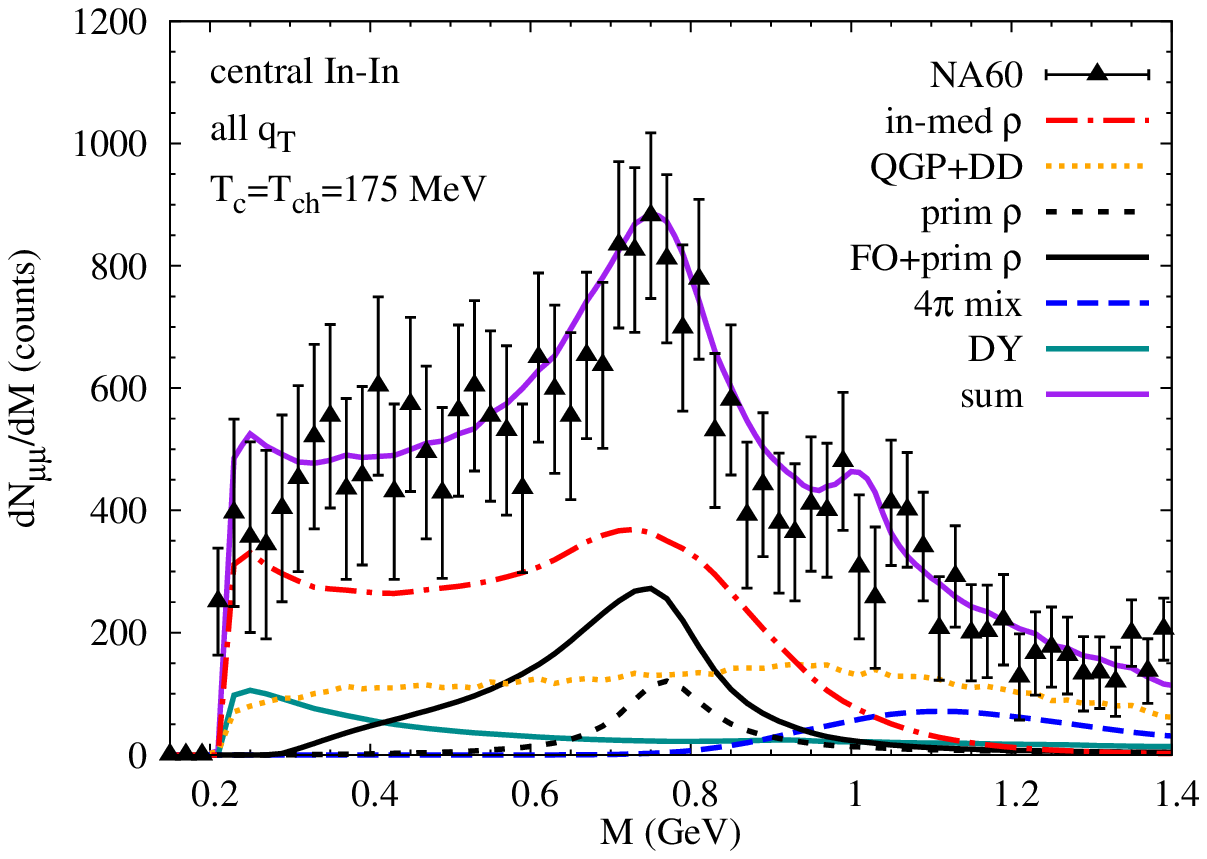}}
\caption{(Color online) NA60 excess dimuon spectra~\cite{Arnaldi:2006jq}
  in semicentral (upper panel) and central (lower panel) In-In
  collisions at SPS compared to theoretical calculations using an
  in-medium e.m.~spectral function. The individual contributions arise
  from in-medium $\rho$-mesons~\cite{Rapp:1999us} (dash-dotted red
  line), 4$\pi$ annihilation with chiral $V$-$A$ mixing (dashed blue
  line), QGP plus correlated open charm decays (dotted orange line) and
  Drell-Yan annihilation (solid turquoise line); the upper dashed brown
  line is the sum of the above, while the solid purple line additionally
  includes in-medium $\omega$ and $\phi$ decays as well as freezeout
  plus primordial $\rho$s (solid black line). In the semi-central data,
  the uncertainty due to the $\eta$ cocktail subtraction is indicated by
  the open and filled data points (the former are based on an estimated
  $\eta$ yield at high $q_T$ while the latter represent an upper limit
  on the $\eta$ by subtracting the dimuon spectrum to zero at
  threshold).}
\label{fig_med}
\end{figure}
\begin{figure}[!t]
\centerline{\includegraphics[width=0.46\textwidth]{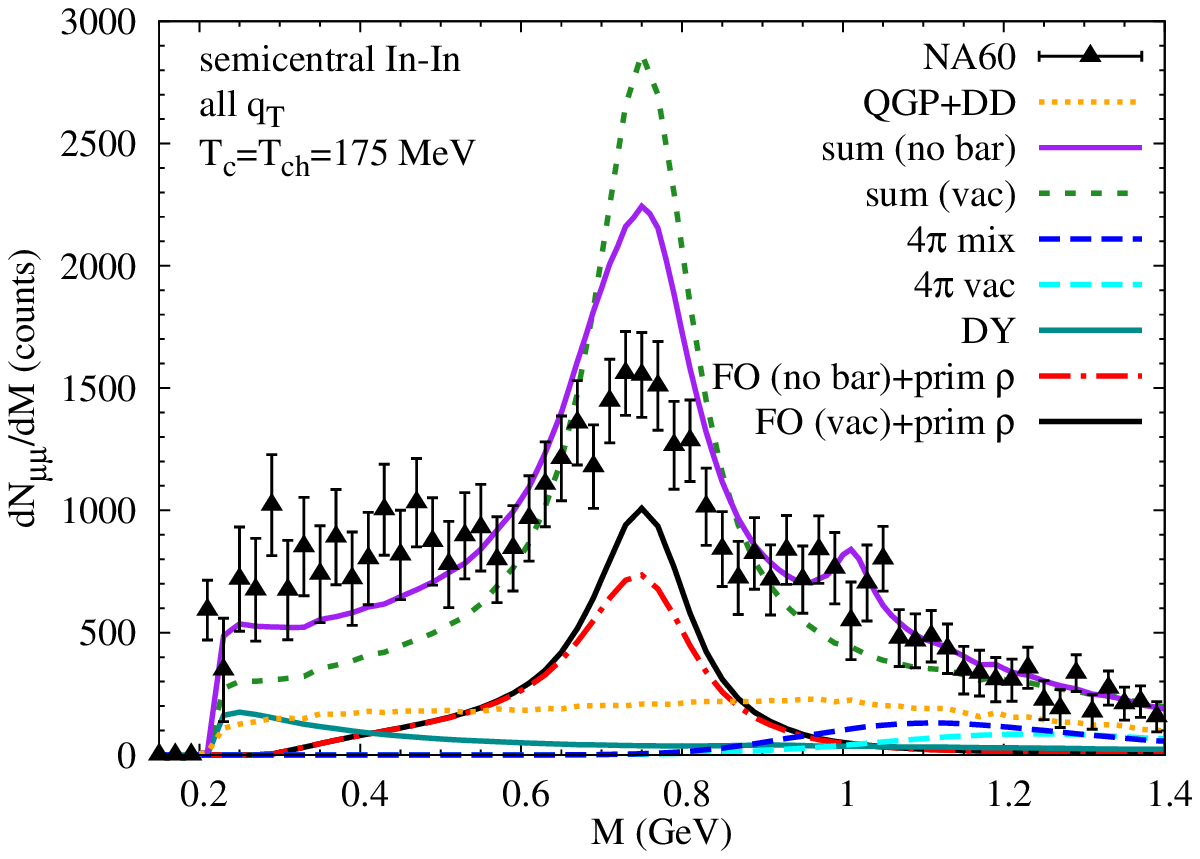}}
\centerline{\includegraphics[width=0.46\textwidth]{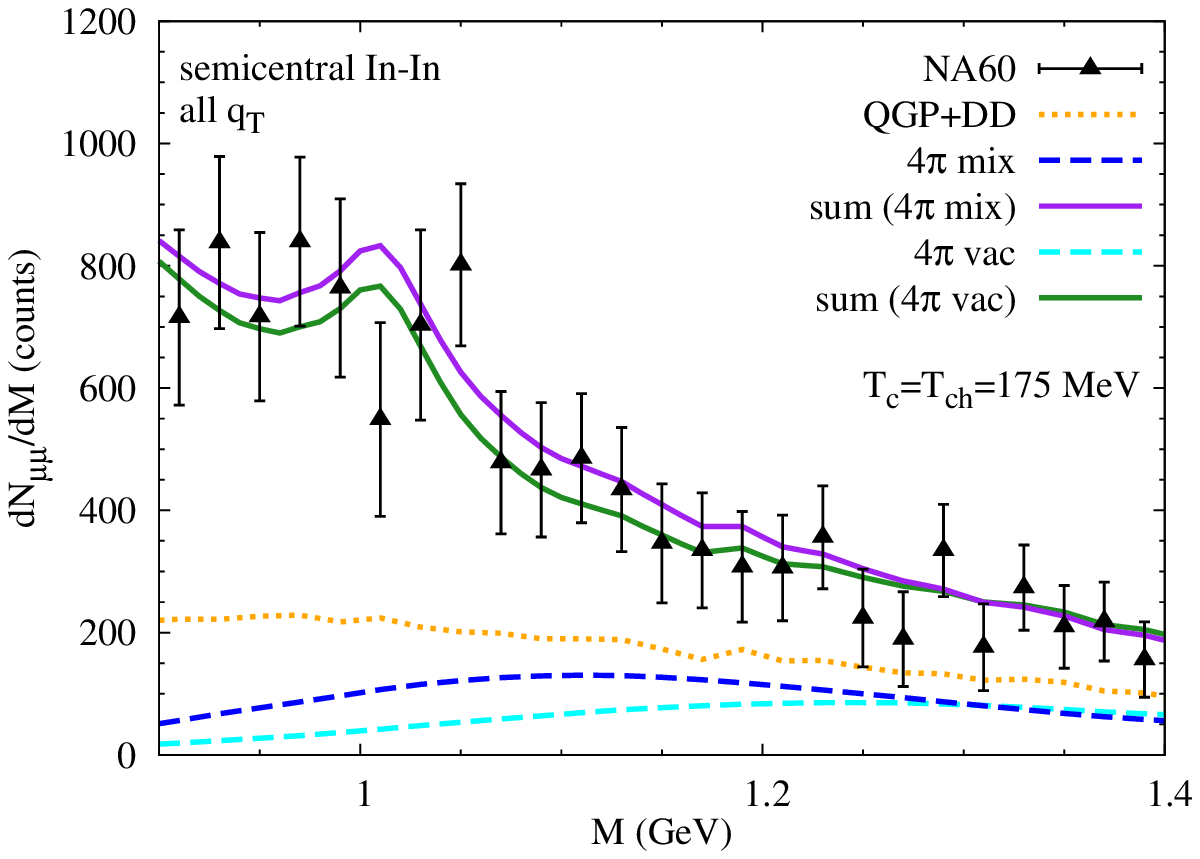}}
\caption{(Color online) Upper panel: NA60 data~\cite{Arnaldi:2006jq}
  compared to thermal dimuon spectra using (i) in-medium $\rho$-,
  $\omega$- and $\phi$-mesons without baryon effects
  (+QGP+charm+in-medium four-pion; solid purple line), and (ii) free
  $\rho$ (+QGP+charm+free four-pion; dashed green line). Lower panel:
  NA60 data~\cite{Arnaldi:2006jq} in the IMR compared to thermal dimuon
  spectra with different implementations of the four-pion contribution,
  using either its vacuum form (lower dashed line) or including chiral
  mixing (upper dashed line), and corresponding total spectra (lower and
  upper solid line, respectively).}
\label{fig_vac}
\end{figure}

Thermal $\mu^+\mu^-$ invariant-mass spectra for $A$-$A$ collisions are
computed by integrating the emission rate, Eq.~(\ref{rate}), over the
fireball evolution (as well as three-momentum),
\begin{equation}
\label{spect}
\frac{\dd N_{ll}}{\dd M}=\frac{M}{\Delta y}
\int\limits_0^{t_{\text{fo}}} \dd t~V_{\text{FB}}(t) 
\int \frac{d^3q}{q_0}~\frac{\dd N_{ll}}{\dd^4 x \, \dd^4 q}
 z_{P}^n \text{Acc}(M,q_T,y) \ ,
\end{equation}
where $\text{Acc}$ denotes the detector acceptance which has been
carefully tuned to NA60 simulations~\cite{na60acc}. The fugacity factor,
$z_{P}^n$=$\mathrm{e}^{n\mu_P/T}$, arises due to chemical
off-equilibrium in the hadronic phase for $T<T_{\rm ch}$; it depends on
the thermal emission source under consideration, cf.~Eq.~(\ref{Piem}):
for the $\rho$, $\omega$ and four-pion contributions one has $z_\pi^n$
with $n=2,3,4$, respectively, while for the $\phi$ one has $z_K^2\cdot
\gamma_s^2$, where $\gamma_s\simeq0.75$ accounts for strangeness
undersaturation for medium-size nuclear collision systems at the
SPS~\cite{Becattini:2003wp}. In addition, appropriate fugacities figure
into the various in-medium selfenergy contributions.

\begin{figure}[!t]
\centerline{\includegraphics[width=0.45\textwidth]{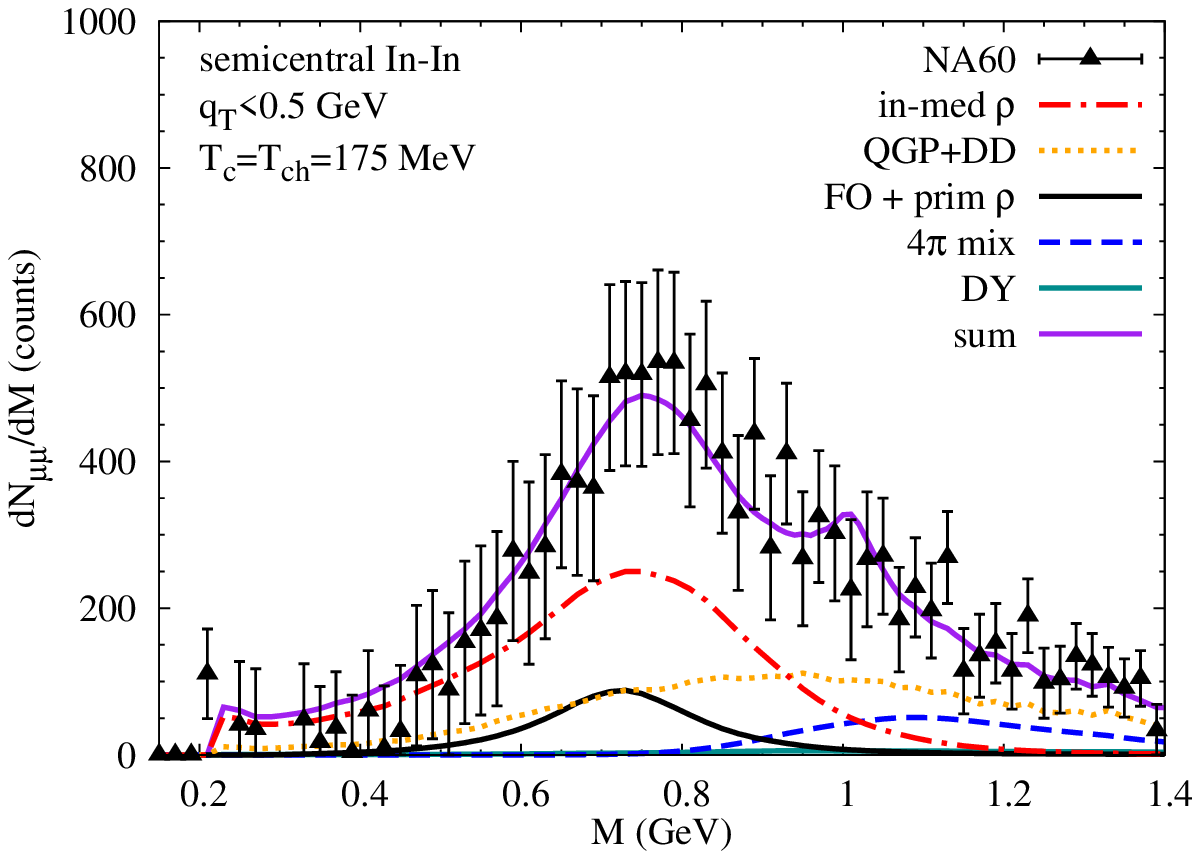}}
\vspace{0.3cm}
\centerline{\includegraphics[width=0.45\textwidth]{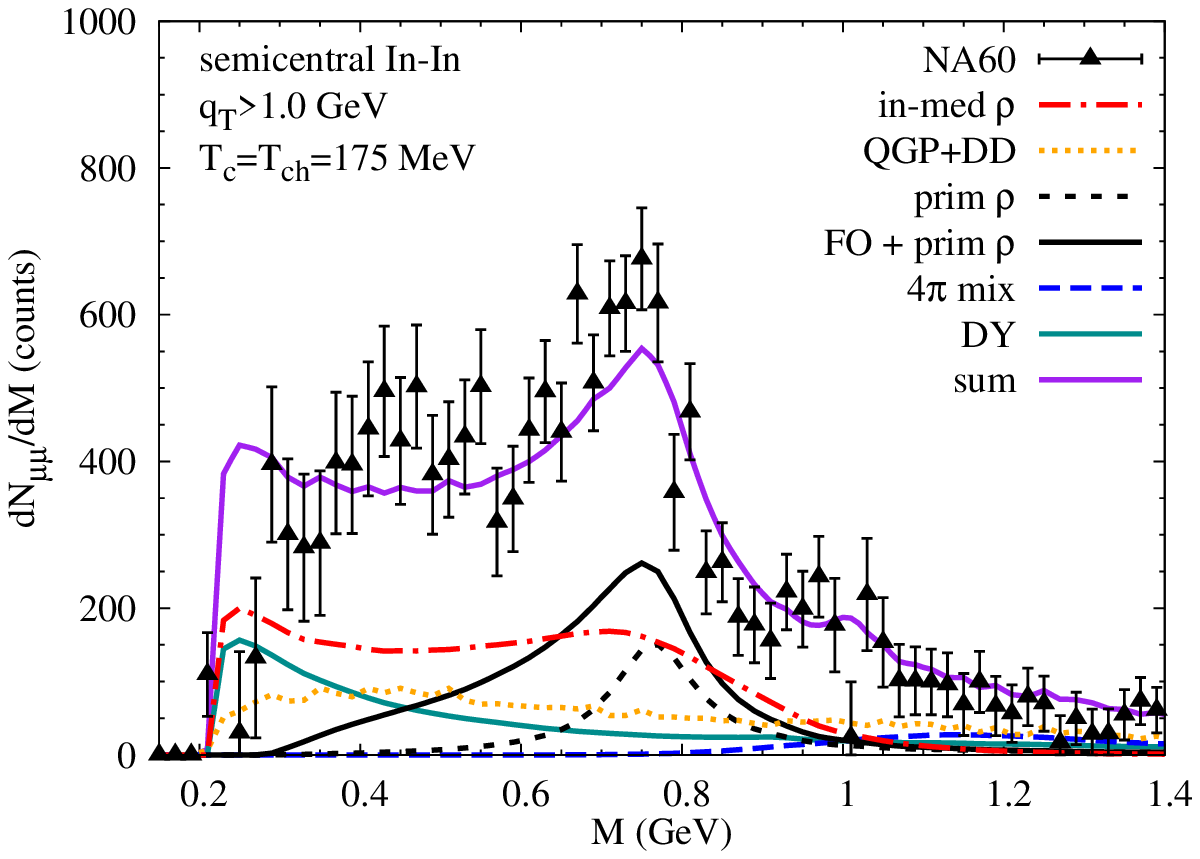}}
\caption{(Color online) Dimuon invariant mass spectra in two different
  bins of transverse pair momentum, $q_T$~\cite{Arnaldi:2006jq}.  Upper
  panel: $q_T\le0.5$~GeV; lower panel: $q_T\ge1.0$~GeV.}
\label{fig_Mspecpt}
\end{figure}
\begin{figure}[!t]
\centerline{\includegraphics[width=0.45\textwidth]{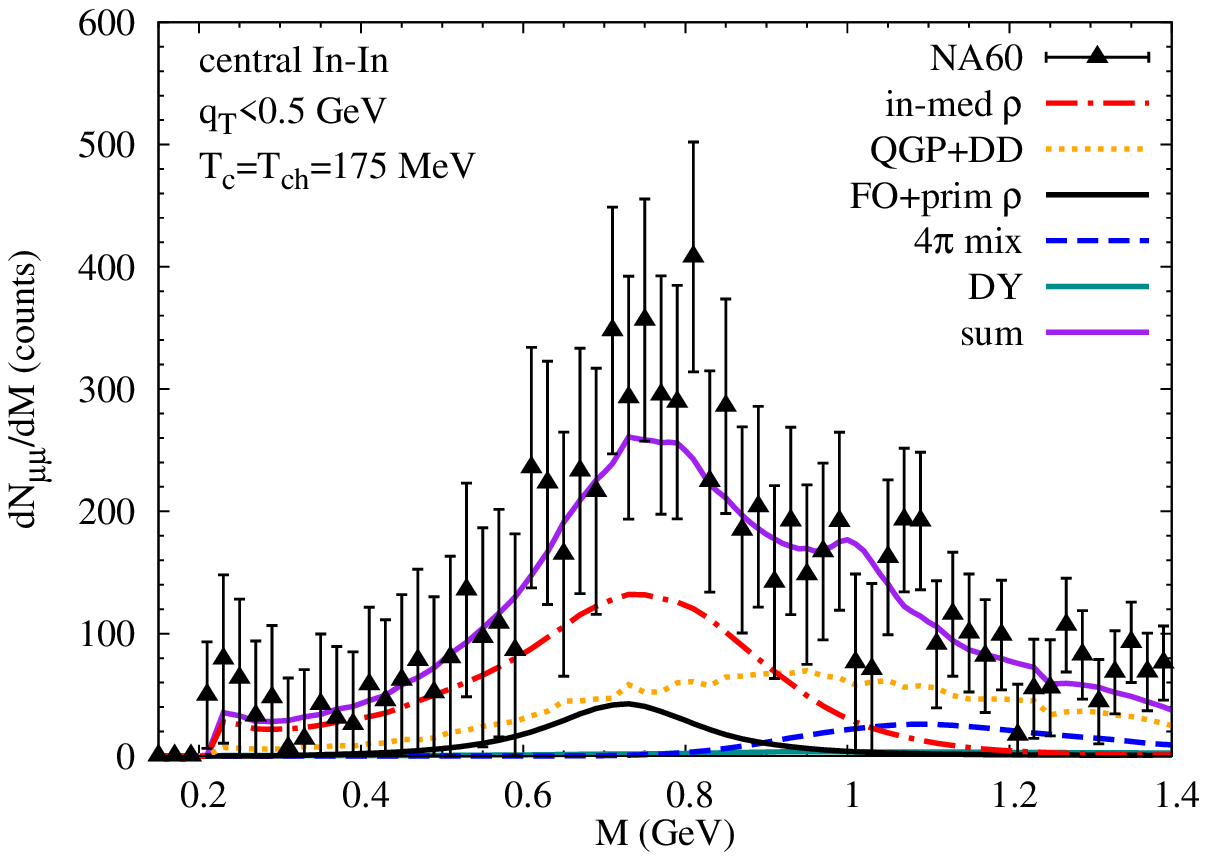}}
\vspace{0.3cm}
\centerline{\includegraphics[width=0.45\textwidth]{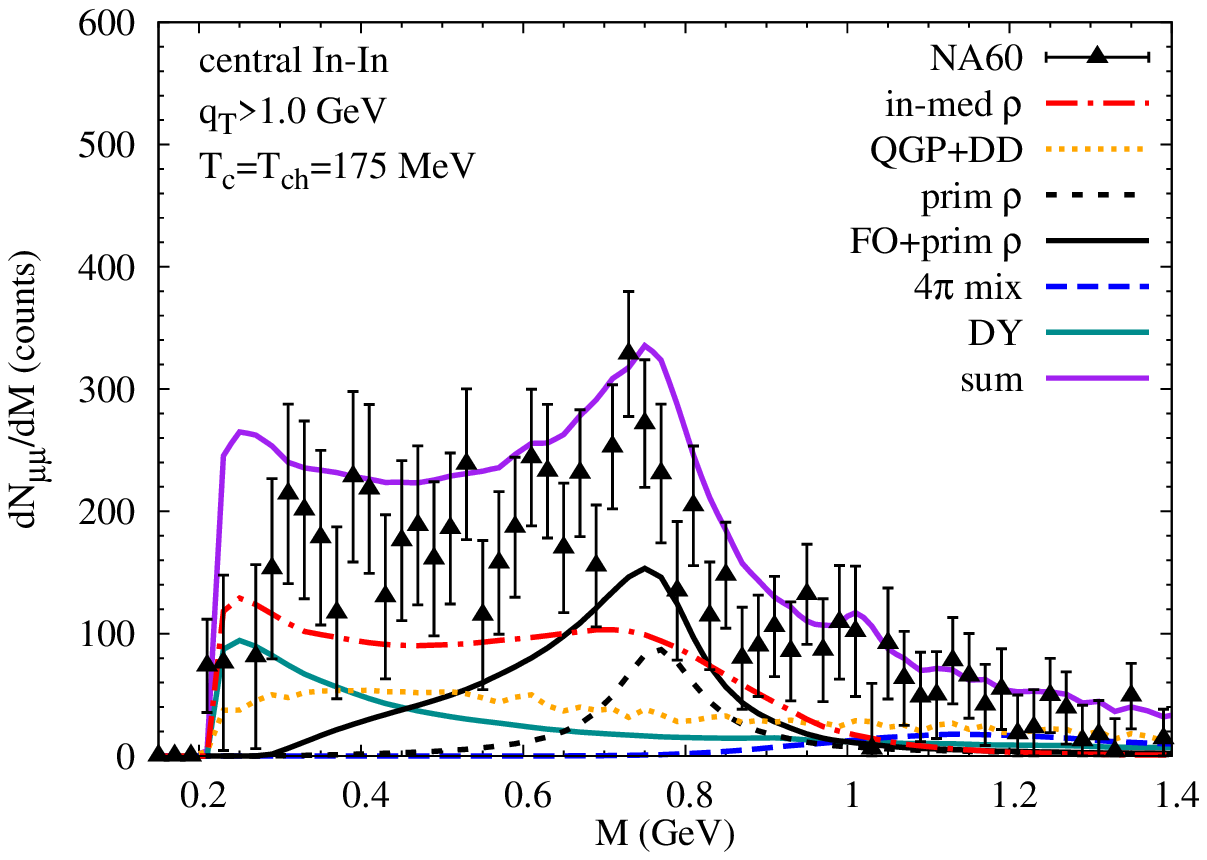}}
\caption{(Color online) The same as Fig.~\ref{fig_Mspecpt} but for
  central collisions.}
\label{fig_Mspecpt_central}
\end{figure}


Initial comparisons~\cite{Arnaldi:2006jq} of NA60 data to theoretical
predictions~\cite{Rapp:2004zh} have focused on the contribution from the
$\rho$ meson which dominates in the LMR. The shape of the in-medium
$\rho$-spectral function describes the experimental spectra well, but
the absolute yields have been overestimated by $\sim 30\%$.  This
discrepancy has been resolved~\cite{vanHees:2006ng} by increasing the
transverse fireball expansion ($a_{\perp}$), reducing the fireball
lifetime to about $6$-$7$~fm/$c$, cf.~Sec.~\ref{sec_fireball} above.  In
addition, the larger transverse expansion leads to harder emission
spectra in $q_T$, which will be helpful in understanding the $q_T$
spectra, as discussed in the following sections.

Fig.~\ref{fig_med} summarizes our results for the mass spectra in
semicentral and central In(158~AGeV)-In collisions, computed for the
EoS-A scenario ($T_c=T_{\rm ch}=175$~MeV). The modifications relative to
our previous work~\cite{vanHees:2006ng} are: (i) the freezeout $\rho$
has been separated from the in-medium $\rho$ contribution, (ii)
primordial $\rho$ and (iii) Drell-Yan contributions have been added. As
a result of the separation (i), the thermal emission lifetime is now
slightly smaller, 6.5(6.2)~fm/$c$ for central (semicentral)
collisions. The inclusive mass spectra (and pertinent conclusions) are
essentially identical to those in Ref.~\cite{vanHees:2006ng}, \ie, the
predicted in-medium effects on $\rho$ spectral function lead to good
agreement with the data in the LMR, while the same fireball evolution
also reproduces the observed excess in the IMR well. The largest source
here is four-pion annihilation, together with smaller contributions from
open-charm decays and QGP emission. In-medium $\phi$ decays are
noticeable but not very significant relative to the uncertainty in the
data.

The sensitivity to the medium effects in the e.m.~correlator is further
illustrated in Fig.~\ref{fig_vac}. The upper panel demonstrates that a
free $\rho$ spectral function is ruled out, but also one which only
includes modifications due to a meson gas clearly does not reproduce the
data, due to both a too narrow peak and a lack of enhancement below the
free $\rho$ mass, especially when approaching the dimuon threshold. In
the lower panel one sees that the effect of chiral mixing on the 4$\pi$
contribution amounts to up to a factor of $\sim$2 enhancement in the
$a_1$ resonance region, but the effect on the total is rather
moderate. Thus, no strong case on the chiral mixing can be made at
present. The prevalence of the four-pion contribution in the IMR is
reminiscent to the hydrodynamic calculations in
Ref.~\cite{Dusling:2006yv} where hadronic rates calculated in the chiral
reduction formalism have been employed. On the contrary, in the fireball
calculations of Ref.~\cite{Ruppert:2007cr} QGP emission dominates in the
IMR. We elucidate on this discrepancy in Sec.~\ref{ssec_hadrochem}
below.

In Figs.~\ref{fig_Mspecpt} and \ref{fig_Mspecpt_central} we compare our
theoretical calculations to NA60 $M$-spectra binned into regions of low
($q_T<0.5$~GeV) and high ($q_T>1.0$~GeV) transverse pair momentum, for
both semicentral and central collisions. Also in this representation the
agreement is fair. There is possibly an indication of a slight over-
(under-) estimate in the high-$q_T$ bin for (semi-) central
collisions. To further scrutinize this issue we now turn to $q_T$
spectra, binned in invariant mass.

\subsection{Transverse Momentum Spectra}
\label{ssec_qtspec}

In analogy to Eq.~(\ref{spect}) for $M$ spectra, $q_T$ spectra are
computed by integrating the dilepton rate, Eq.~(\ref{rate}), over the
space-time evolution of the fireball and a suitable interval in
invariant mass, $[M_{\rm min},M_{\rm max}]$. However, since the
available experimental spectra have been corrected for the detector
acceptance, the Monte-Carlo simulation of the acceptance function (with
radial and angular dependencies due to directed (radial flow) and random
(thermal) motion of the virtual photon) is now replaced by explicit
integrations over the radial coordinate, $r$, of the fireball, the
azimuthal angle, $\phi_q$, of the virtual photon momentum and its
rapidity, $y$,
\begin{equation}
\begin{split}
\label{qT-spect}
\frac{\dd N_{ll}}{q_T \dd q_T}=\int\limits_0^{t_{\text{fo}}} \dd t
\int\limits_{0}^{R(t)} \dd r \int\limits_{M_{\rm min}}^{M_{\rm max}}
\dd M \int\limits_{0}^{2 \pi} \dd \phi_q \int\limits_{y_{<}}^{y_{>}} \dd y \; 
2 \pi r \; z(t)  
\\ 
\times M \ z_{P}^n \ \frac{\dd N_{ll}}{\dd^4 x \dd^4 q} \ .  
\end{split}
\end{equation}
As before, $R(t)$ denotes the radius of the expanding fire-cylinder
(corresponding to the first parenthesis in Eq.~(\ref{VFB})), $z(t)$ its
longitudinal length (second parenthesis in Eq.~(\ref{VFB})). Note that
the rate, Eq.~(\ref{rate}), is calculated in the thermal rest frame
while the integrations in Eq.~(\ref{qT-spect}) are in the laboratory
frame. The relation between the four-momentum, $q^{\mu}$, in the
laboratory frame and the one in the local rest frame of the fluid cells,
$\bar{q}^{\mu}$, is determined by a boost with the radial flow velocity,
Eq.~(\ref{flow-profile}).  Due to rotational invariance the thermal
rate, Eq.~(\ref{rate}), depends only on the magnitude of the
three-momentum in the thermal rest frame and therefore one of the
integrations over the spatial and momentum azimuthal angles in
Eq.~(\ref{qT-spect}) becomes trivial yielding a factor of $2 \pi$.  To
gain qualitative insights into the behavior of dilepton $q_T$ spectra
from thermal sources, let us assume that the three-momentum dependence
of the hadronic e.m.~current correlator is weak (it would be absent in
the absence of medium effects), so that for not too large invariant-mass
intervals, the integrand in Eq.~(\ref{qT-spect}) can be considered
constant except for the Bose distribution.  For $M \gg T$, it is
possible to approximate the qualitative behavior of the $q_T$ spectra in
analytic form~\cite{Schnedermann:1993ws,Gorenstein:2001ti}.  For a
linear flow profile, Eq.~(\ref{flow-profile}), and in Boltzmann
approximation one has
\begin{equation}
\begin{split}
\frac{\dd N_{ll}}{q_T \dd q_T \dd M \dd t}= & C \int_0^{R(t)} \dd r \int_0^{2
  \pi} \dd \phi \int_{-\infty}^{\infty} \dd y \\ & \times r \exp \left
  (-\frac{q_0-|\beta_{\perp}| q_T \cos \phi}{T \sqrt{1-\beta_{\perp}^2}}
\right),
\end{split}
\end{equation}
with $C$ weakly $q_T$ dependent. Substituting $q_0=m_{\perp} \cosh y$
($m_T=\sqrt{M^2+q_T^2}$) the integrals over $y$ and $\phi$ are given by
modified Bessel functions. For the $y$-integral we can use their
asymptotic form for $m_T/T \geq M/T \gg 1$:
\begin{equation}
\begin{split}
\label{spect-besselI0}
  \frac{\dd N_{ll}}{q_T \dd q_T \dd M \dd t}= & C \int_0^{R(t)} \dd r \,
  r
  \sqrt{\frac{2 \pi T \sqrt{1-\beta_{\perp}^2}}{m_T}} \\
  & \times \exp \left(-
    \frac{m_T}{T \sqrt{1-\beta_{\perp}^2}} \right) 
   I_0 \left (\frac{|q_T| |\beta_{\perp}|}{T
      \sqrt{1-\beta_{\perp}^2}} \right).
\end{split}
\end{equation}
In the non-relativistic limit, $q_T \ll M$, we use the
asymptotic form $I_0(x) \asy_{x \rightarrow 0} 1+x^2/4 \simeq
\exp(x^2/4)$ to obtain
\begin{equation}
\begin{split}
  \frac{\dd N_{ll}}{q_T \dd q_T \dd M \dd t} & \asy_{q_T \ll M} C
  \int_0^{R(t)} \dd r \, r \sqrt{\frac{2 \pi T \sqrt{1-\beta_{\perp}^2}}{m_T}}
    \\
& \times \exp \left(-\frac{m_T}{T \sqrt{1-\beta_{\perp}^2}}+\frac{q_T^2
    \beta_{\perp}^2}{4 T^2(1-\beta_{\perp}^2)} \right). 
\end{split}
\end{equation}
We have numerically verified that this expression can be further
simplified by the following approximate treatment of the $r$ integral:
In the pre-factor of the exponential substitute $\beta_{\perp}$ by its
average over $r$,
\begin{equation}
\erw{\beta_{\perp}} =\frac{2}{R^2} \int_0^R \dd r r \beta_{\perp}^s
\frac{r}{R}= \frac{2}{3} \beta_{\perp}^s \ ; 
\end{equation}
use the $r$ average of the argument in the exponential, which, together
with the non-relativistic approximation $q_T \asy_{q_T \ll M}=\sqrt{2
  M(m_T-M)}$, results in
\begin{equation}
\begin{split}
\label{fancy-pocket}
\frac{\dd N_{ll}}{q_T \dd q_T \dd M \dd t} \asy_{q_T \ll M} & C
\sqrt{2 \pi T \sqrt{1-\erw{\beta_{\perp}}_r^2}} \\ 
& \times \exp \left(-\frac{M^2
    (\beta_{\perp}^s)}{2T^2} \right )\\ 
& \times \sqrt{\frac{1}{m_T}} \exp \left
  (-\frac{m_T}{T_{\text{eff}}} \right)
\end{split}
\end{equation}
where
\begin{equation}
\begin{split}
\label{fancy-Teff}
T_{\text{eff}} &=\frac{T}{1-(M-T) (\beta_{\perp}^s)^2/(4 T)} \\
& \asy_{\beta_{\perp}^s \rightarrow 0} T+\frac{M}{2}
\erw{\beta_{\perp}^2}_r,
\end{split}
\end{equation}
and $\erw{\beta_{\perp}^2}_r=(\beta_{\perp}^s)^2/2$. The form for
$\beta_{\perp}^s \ll 1$ is the known ``pocket formula'' for the
parametric dependence of the effective slopes of hadronic $q_T$ spectra
on the fireball temperature, the particle mass and the surface-flow
velocity~\cite{Gorenstein:2001ti}.  We recall that in the
ultrarelativistic limit, $q_T \gg M$, the dependence of the effective
temperature on the radial velocity is given by the Doppler blue-shift
expression for a thermalized gas of massless particles,
\begin{equation}
T_{\text{eff}}(\beta_{\perp})=T
\sqrt{\frac{1+\beta_{\perp}}{1-\beta_{\perp}}} \ , 
\end{equation}
which follows from the large-$q_T$ limit of $I_0$ in
Eq.~(\ref{spect-besselI0})~\cite{Gorenstein:2001ti} which also provides
an additional factor $1/\sqrt{q_T}$. For our $r$-dependent flow profile
the effective temperature is given by the blue-shift value for an
average value $\erw{\beta_{\perp}}_r=\xi \beta_{\perp}^s$ where $\xi \in
(0,1)$ is not easily determined by simple approximations of the radial
integral. Numerical studies show that in the case of a linear
radial-flow profile, Eq.~(\ref{flow-profile}), and typical parameters in
the region of the $\rho$ peak ($M=0.75\;\text{GeV}$,
$\beta_{\perp}^s\simeq0.5$), $\xi \simeq 0.8$-$0.85$ leads to a good
estimate for $T_{\text{eff}}$ at high $q_T$.

It is important to note that for vector-meson decays after thermal
freezeout, cf.~Sec.~\ref{ssec_fo}), the $q_T$ spectra are harder by an
additional factor $m_T/M$, \ie, applying the same approximations to
Eq.~(\ref{fo-dilep}) as used to derive Eq.~(\ref{fancy-pocket}), one
obtains
\begin{equation}
\begin{split}
\label{fancy-pocket-fo}
\frac{\dd N_{ll}^{(\text{fo})}}{q_T \dd q_T \dd M} \asy_{q_T \ll
  M} & C \sqrt{2 \pi T \sqrt{1-\erw{\beta_{\perp}}_r^2}} \\
& \times \exp \left(-\frac{M^2
    (\beta_{\perp}^s)^2}{2T^2} \right ) \\
& \times \frac{\sqrt{m_T}}{M} \exp \left (-\frac{m_T}{T_{\text{eff}}}
\right) \frac{1}{\Gamma_{\rho}^{(\text{fo})}}
\end{split}
\end{equation}
with $T_{\text{eff}}$ again given by Eq.~(\ref{fancy-Teff}).

The additional factor $m_T/M$ in (\ref{fancy-pocket-fo}) compared to
(\ref{fancy-pocket}) originates from the additional factor $q^0/M$ in
the Cooper-Frye (CF) formula~(\ref{fo-dilep}) compared to the
McLerran-Toimela (MT) emission formula~(\ref{rate}), as discussed in
detail in Sec.~\ref{ssec_fo}.
\begin{figure}[!t]
\centerline{\includegraphics[width=0.47\textwidth]{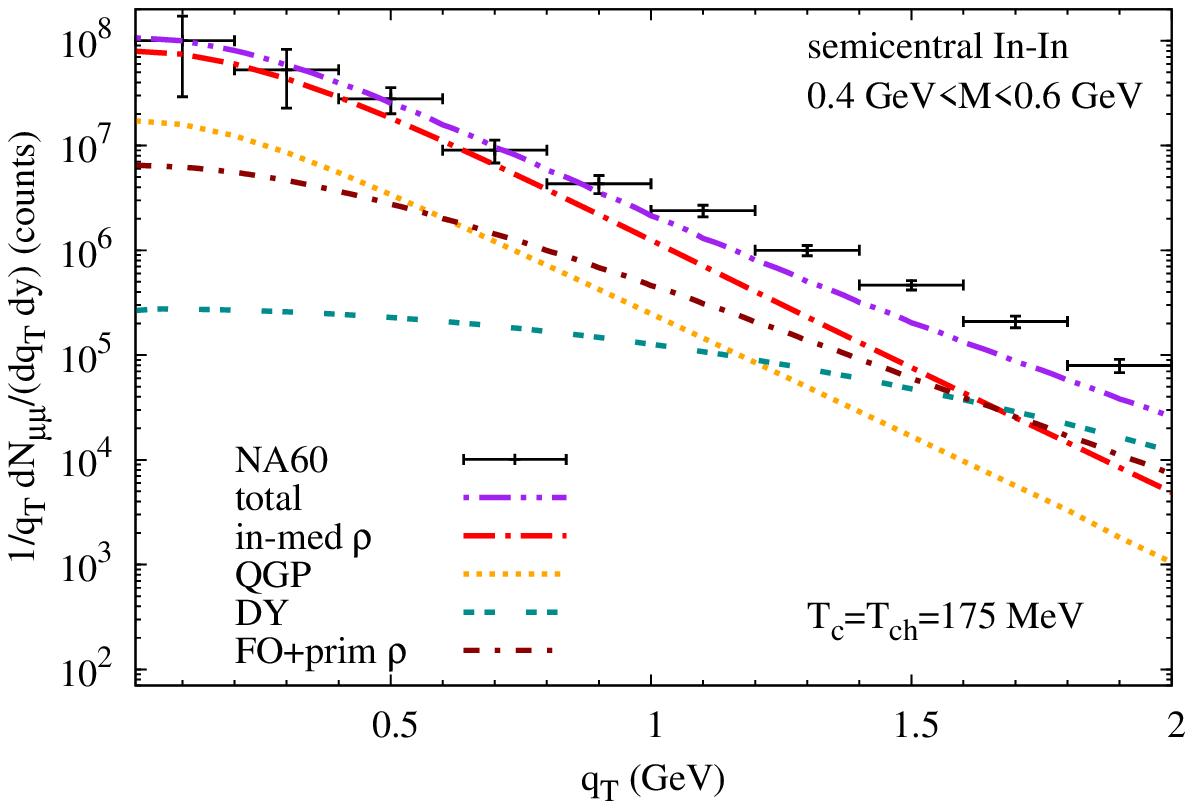}}
\centerline{\includegraphics[width=0.47\textwidth]{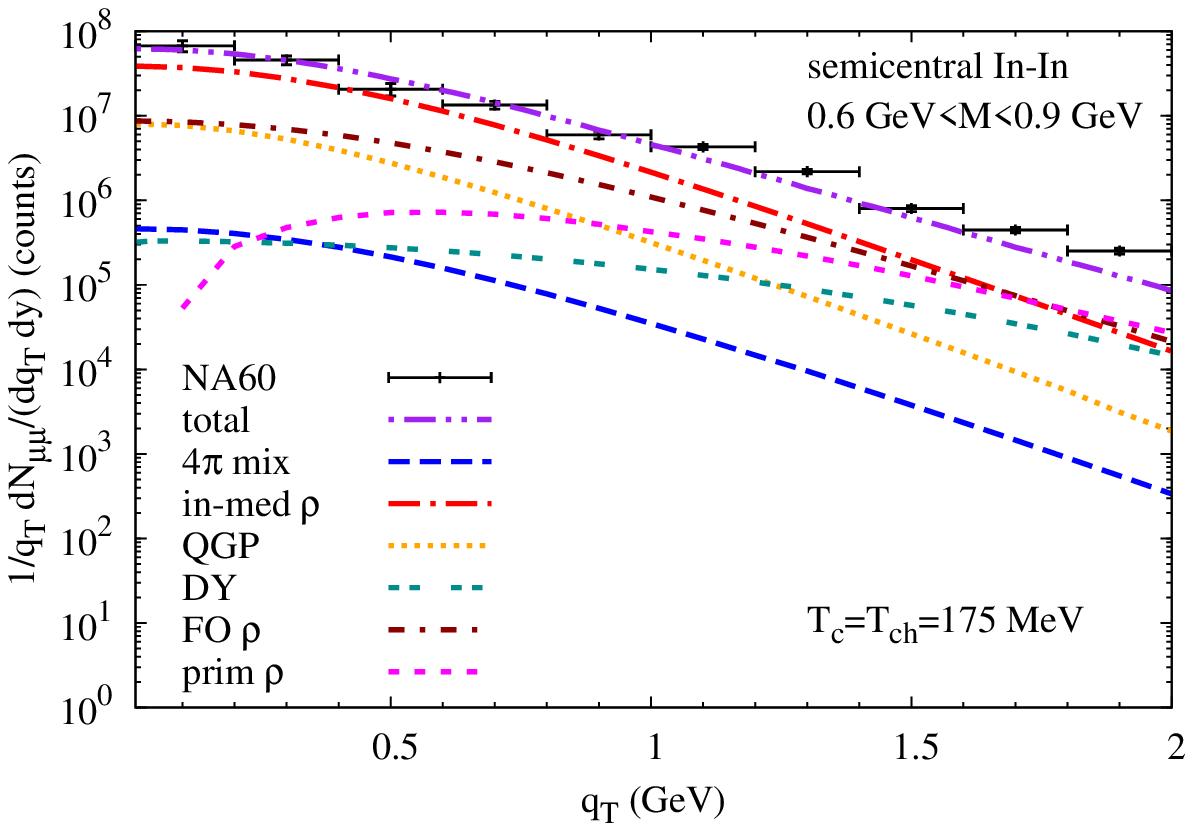}}
\centerline{\includegraphics[width=0.47\textwidth]{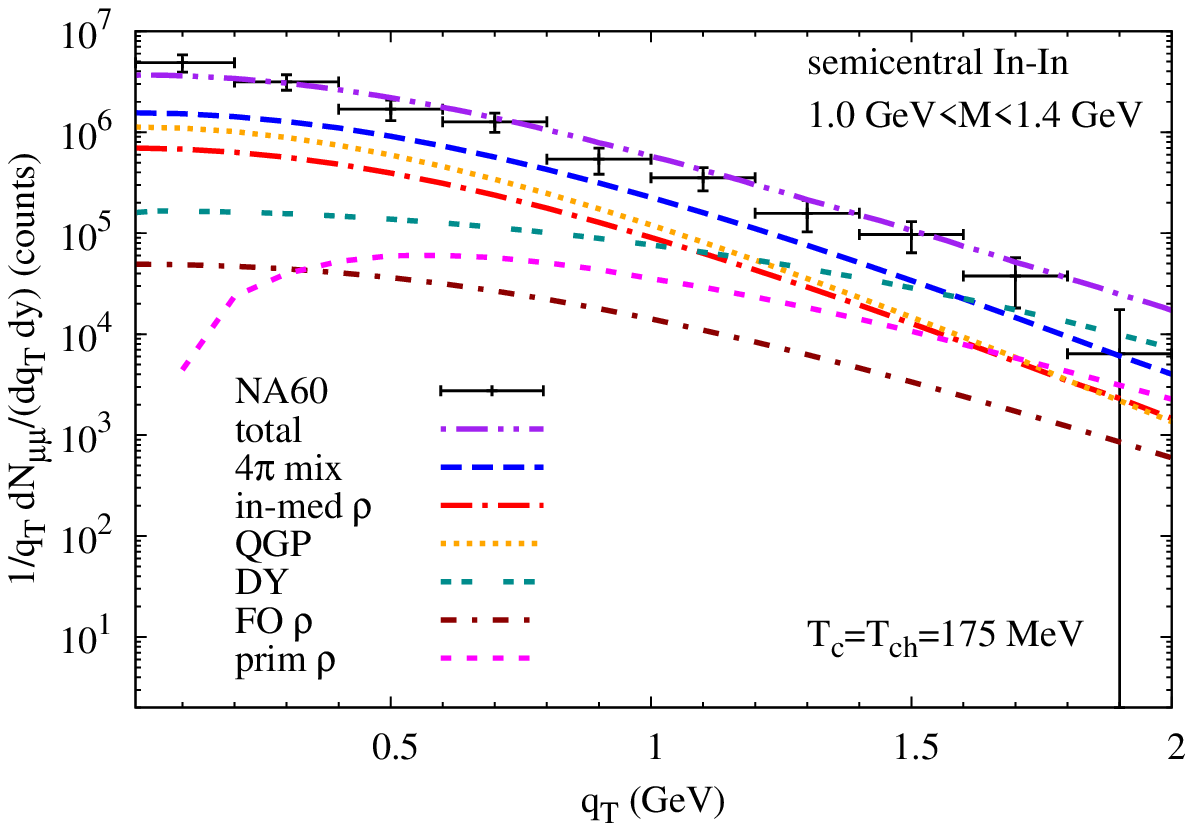}}
\caption{(Color online) Dimuon transverse pair-momentum spectra in
  semicentral In(158~AGeV)-In collisions for three bins of invariant
  mass~\cite{Damjanovic:2007qm}. Upper panel: $0.4~\mathrm{GeV}\le M \le
  0.6$~GeV; middle panel: $0.6~\mathrm{GeV}\le M \le 0.9$~GeV; lower
  panel: $1.0~\mathrm{GeV}\le M \le 1.4$~GeV.  }
\label{fig_ptspec}
\end{figure}
\begin{figure}[!t]
\centerline{\includegraphics[width=0.47\textwidth]{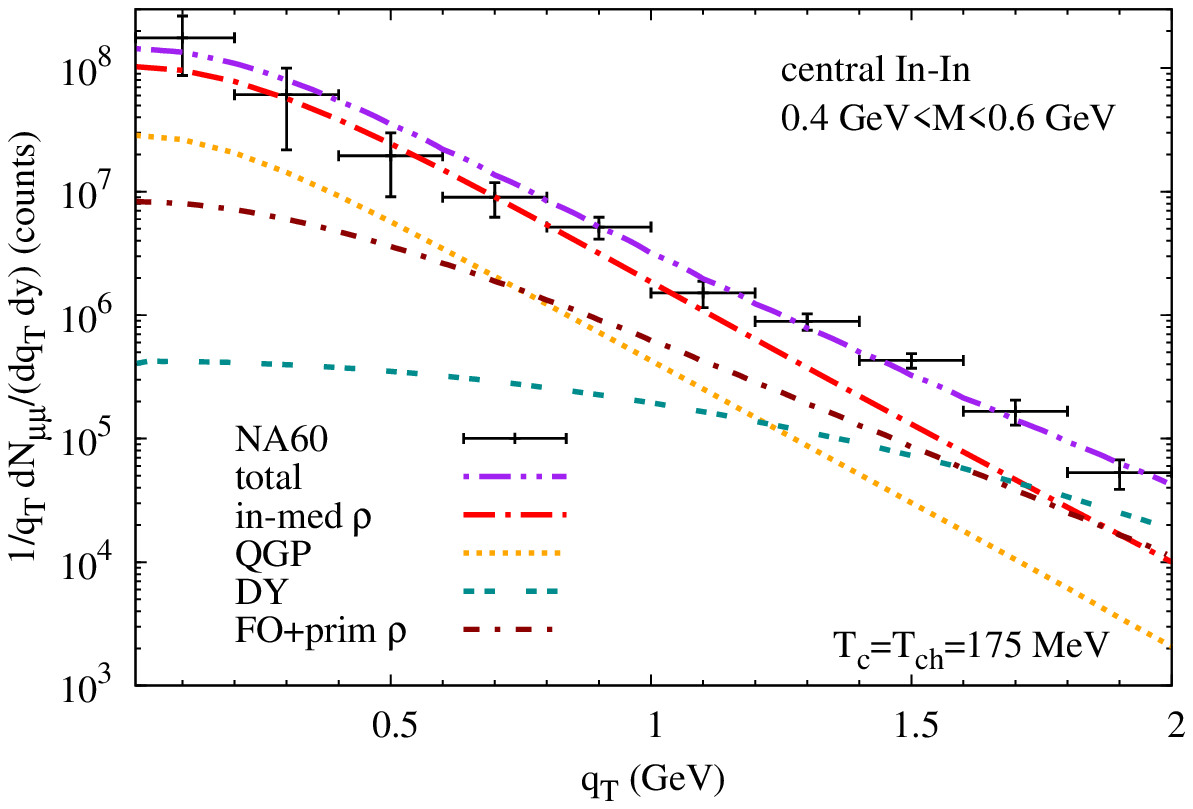}}
\centerline{\includegraphics[width=0.47\textwidth]{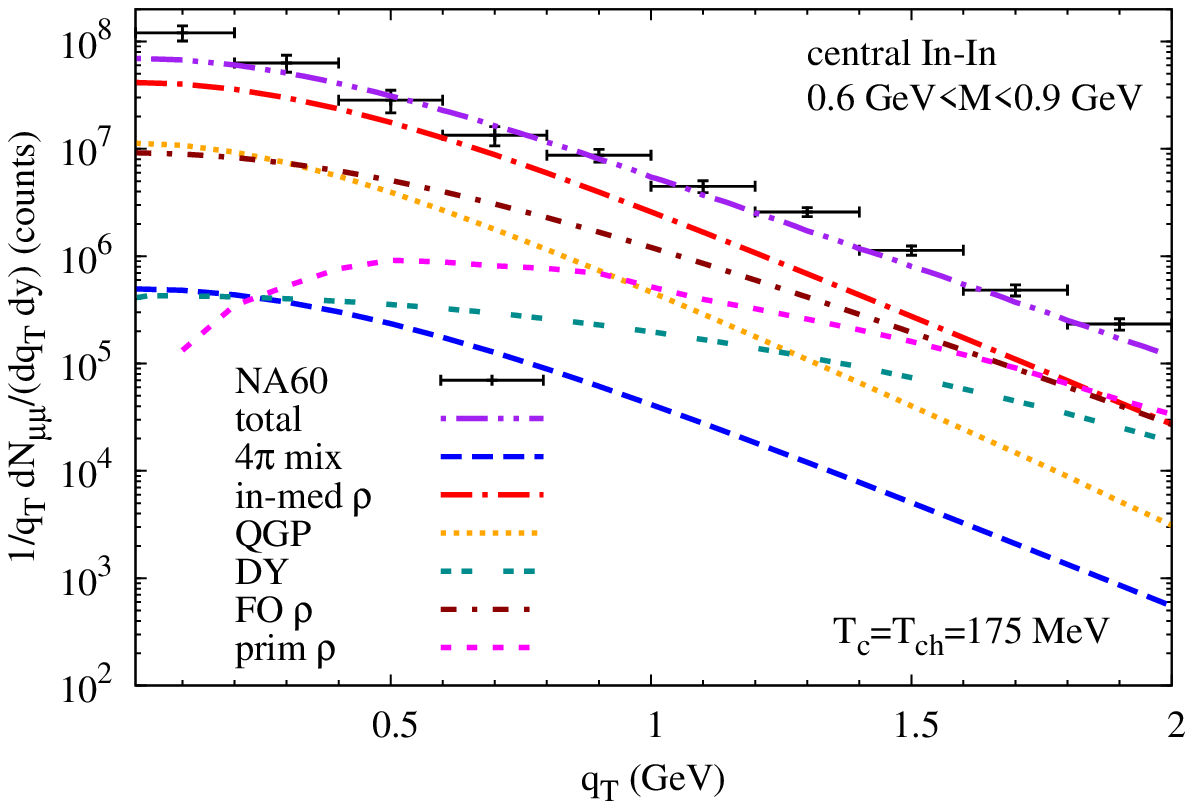}}
\centerline{\includegraphics[width=0.47\textwidth]{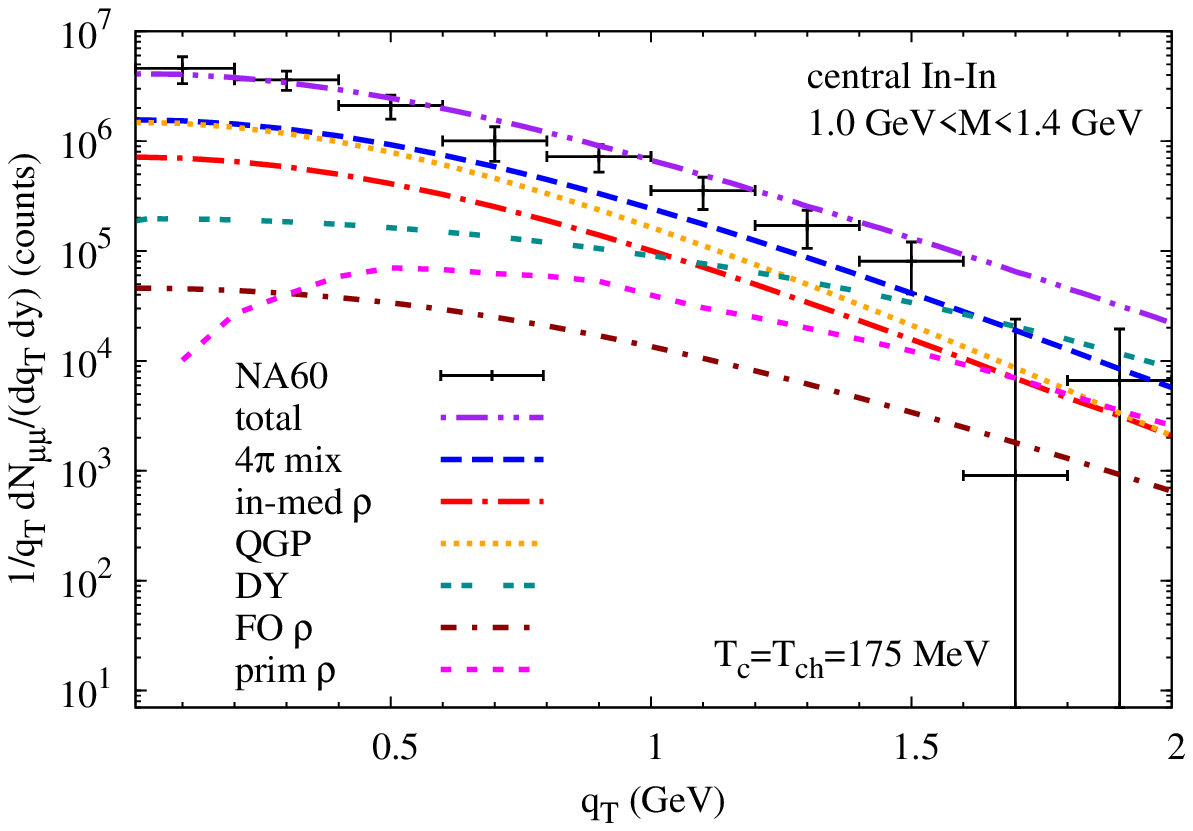}}
\caption{(Color online) Same as Fig.~\ref{fig_ptspec}, but for central
  In(158~AGeV)-In.}
\label{fig_ptspec_central}
\end{figure}

The above discussion of effective slopes in thermal emission spectra
obviously does not apply to non-thermal dilepton sources, \ie, decays of
``primordial $\rho$ mesons'' (Sec.~\ref{ssec_hard}) and Drell-Yan
annihilation (Sec~\ref{ssec_dy}). Both are characterized by a power-law
behavior at high $q_T$, where their contribution becomes potentially
important. Toward lower $q_T$, the primordial $\rho$ contribution is
much suppressed due to $\rho$ absorption (``jet quenching''), while the
Drell-Yan process is no longer well-defined.  However, the Drell-Yan
dileptons carry the hardest slope of all sources considered, which
renders even a naive extrapolation of their spectra to low $q_T$ very
small (no more than a few percent for $q_T\le 1$~GeV in all mass bins
below $M=1.4$~GeV). At $q_T\simeq 2$~GeV, in turn, the Drell-Yan
contribution is quite appreciable.

Figs.~\ref{fig_ptspec} and \ref{fig_ptspec_central} summarize our
calculations for semicentral and central In-In collisions, respectively,
in three different mass bins in comparison to the NA60
data~\cite{Damjanovic:2007qm}. Open-charm decays have been removed from
the experimental spectra and are consequently not included in the theory
curves either. The parameters of our fireball evolution are as described
in Sec.~\ref{sec_fireball} within our default EoS-A scenario
($T_c=T_{\text{ch}}=175$~MeV). Since the overall normalization of the
experimental spectra is not known at present, we have fixed it in each
mass bin using the invariant mass spectra in the low-$q_T$ bin ($0 \leq
q_T \leq 0.5 \; \text{GeV}$). This easily translates into a $10$-$20\%$
uncertainty in the absolute normalization.  For the central collisions
the agreement between theory and data is quite satisfactory in all mass
bins. In the high-mass bin $1$~GeV$\le$$M$$\le$$1.4$~GeV, lower panel in
Fig.~\ref{fig_ptspec_central}), this procedure appears to overestimate
somewhat the $q_T$ spectra at $q_T \ge 1$~GeV, which, however, seems not
to be reflected in the corresponding $q_T$ bin in the mass spectra
(lower panel in Fig.~\ref{fig_Mspecpt_central}). This could very well be
due to the uncertainty in the underlying normalization procedure. Note
again that the compared to Ref.~\cite{vanHees:2006ng} additionally
implemented hard components (primordial $\rho$'s and Drell-Yan) are
insignificant at $q_T\le 1$~GeV, \ie, for the total yield and the
understanding of the spectral shape of the inclusive $M$-spectra. On the
other hand, at higher $q_T$, these contributions are essential (even
dominant) for a proper description of the spectra.  In fact, in
semicentral collisions, the experimental transverse momentum spectra in
the two mass bins below ($0.4$~GeV$\le q_T \le 0.6$~GeV) and around
($0.6$~GeV$\le q_T \le 0.9$~GeV) the free $\rho$-meson mass turn out to
be underestimated (and too soft) for $q_T \gtrsim 1$~GeV. Especially in
the low-mass bin, the discrepancy again appears to be larger than one
could anticipate from the mass spectrum in the $0.4$-$0.6$~GeV regime in
the $q_T\ge1$~GeV bin (lower panel in Fig.~\ref{fig_Mspecpt}).  The fact
remains, however, that the theoretical $q_T$ spectra in the low and
intermediate mass bin are somewhat too soft. Even though the final
transverse flow velocity of the fireball model is about $7\%$ smaller
for semicentral relative to central In-In collisions, this difference
would not be able to account for the discrepancy (it amounts to a change
in slope by about $10$~MeV for the freeze-out $\rho$, and even less for
thermal radiation). The nuclear suppression factor, $R_{AA}(q_T)$, in
Fig.~\ref{fig_raa} (solid line) suggests that there might be room for a
30\% increase of the primordial plus freezeout $\rho$ contribution
(possibly more if one could compare to the $R_{AA}$ of $\rho$ mesons;
due to their larger mass a more pronounced flow effect could enhance
their $R_{AA}$). Whether this would suffice in the low-mass bin, is
questionable.

\begin{figure}[!t]
\centerline{\includegraphics[width=0.47\textwidth]{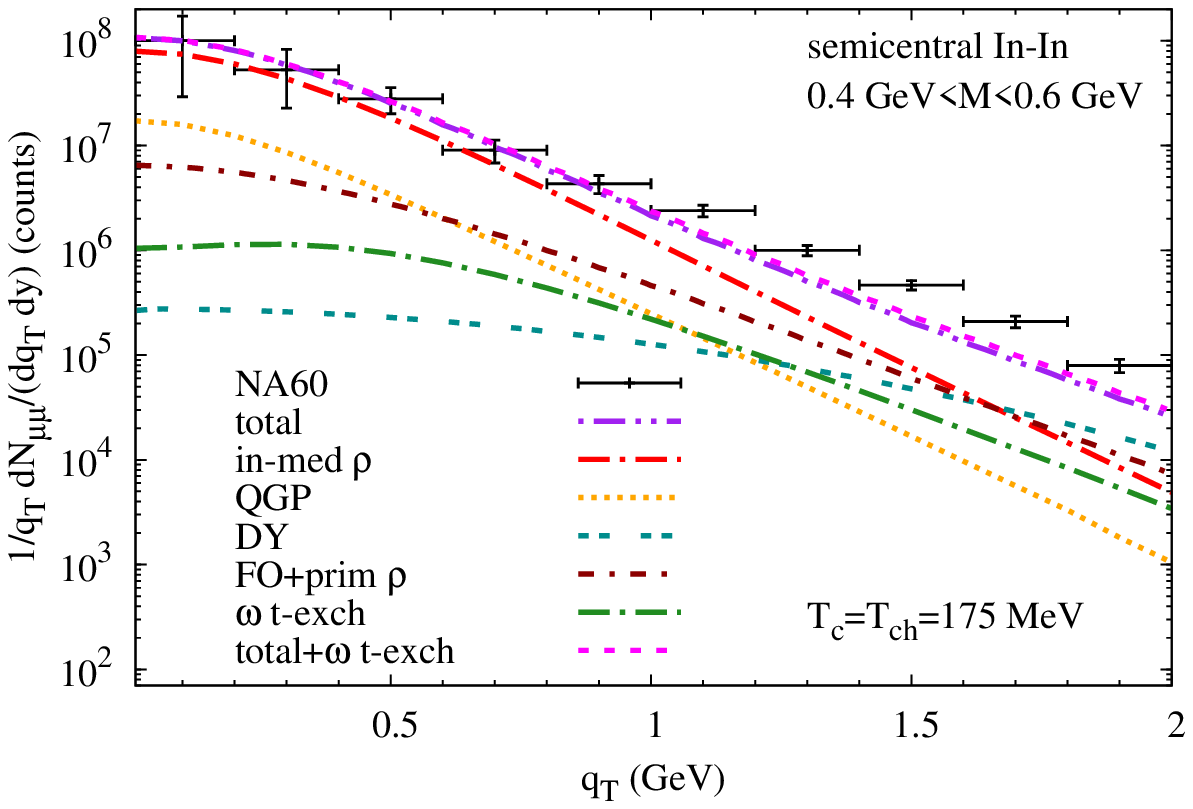}}
\centerline{\includegraphics[width=0.47\textwidth]{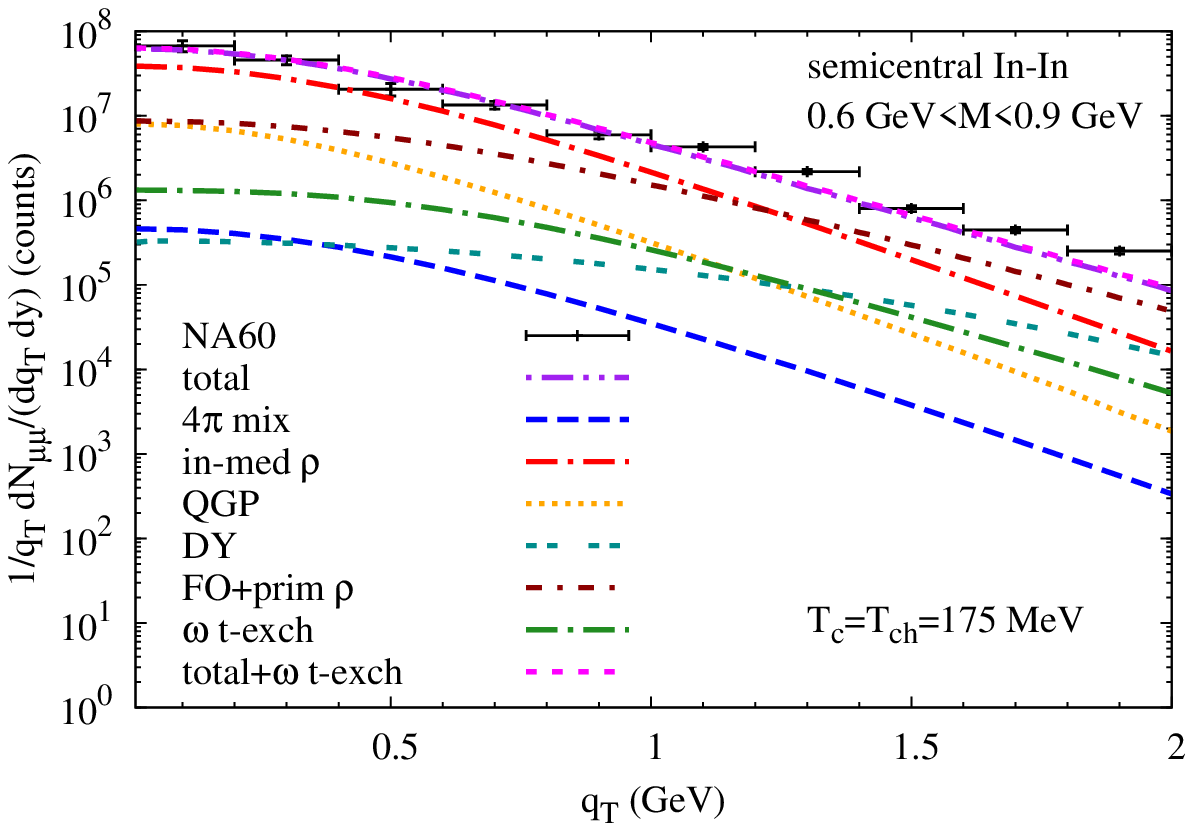}}
\centerline{\includegraphics[width=0.47\textwidth]{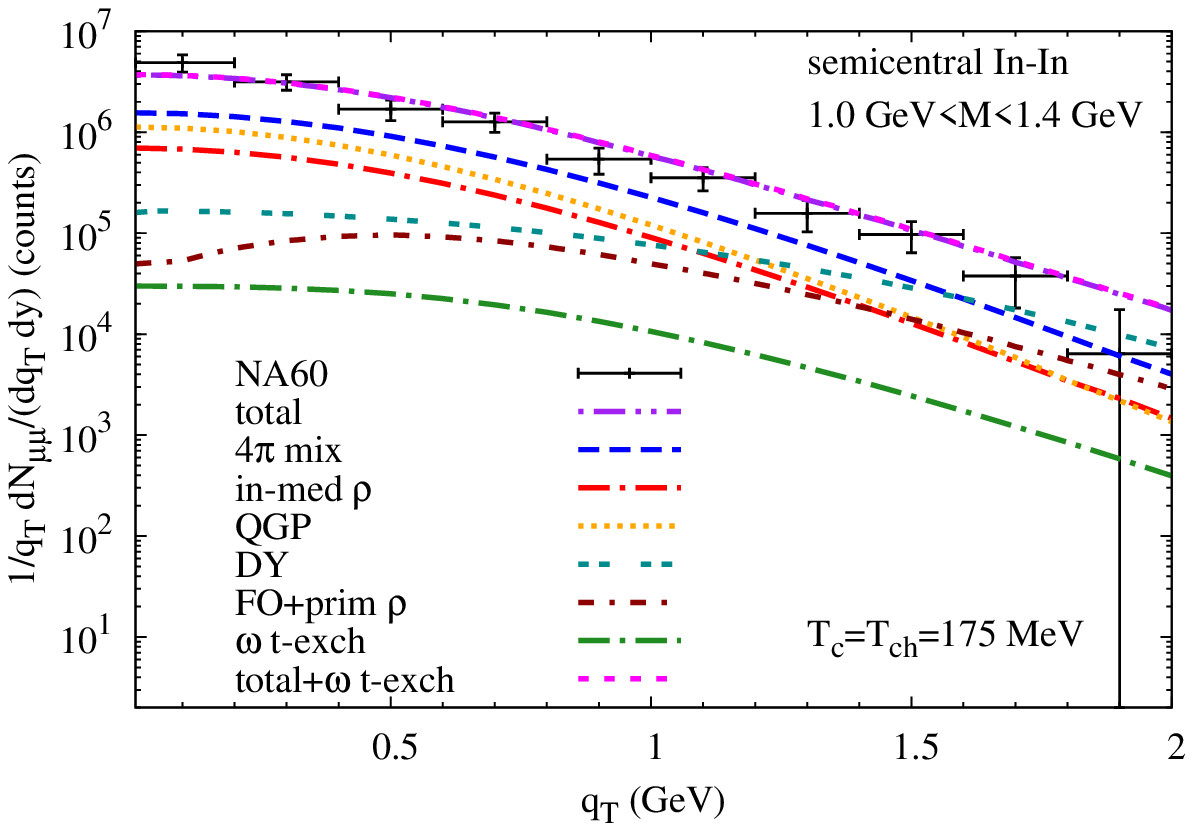}}
\caption{(Color online) Same as Fig.~\ref{fig_ptspec}, but additionally
  including (factor 2 augmented) contributions from $\omega$ $t$-channel
  exchange in $\pi\rho\to\pi \mu\mu$.}
\label{fig_ptspec_om_t}
\end{figure}
Another possibility consists of meson $t$-channel exchange in elastic
scattering of $\rho$ mesons off pions in the heat bath (with subsequent
conversion into a lepton pair).  These processes have been calculated
previously for real photon production~\cite{Turbide:2003si} (which are
given by the same diagrams except that the final-state photon is
on-shell), where $\omega$ exchange has been found to be the most
important process at high $q_T$; however, contributions from other
exchange processes (\eg, $\pi$ and $a_1$ exchange in
$\pi\rho\to\pi\gamma$, or reactions involving strange
mesons~\cite{Turbide:2003si}) are also appreciable. To roughly account
for the latter, we augment the rates computed in Sec.~\ref{ssec_omt} by
a factor of two (in addition, we recall that $\pi\rho$ initial states in
the hadronic fireball emission carry a pion fugacity factor to the third
power, $z_\pi^3$).  The convolution of these rates over the fireball
evolution leads to a contribution to dilepton-$q_T$ spectra as shown by
the lower dash-dotted line in each of the panels in
Fig.~\ref{fig_ptspec_om_t}; it indeed provides the hardest spectrum
among all thermal sources (QGP, in-medium vector mesons and four-pion
annihilation), in line with the rates displayed in
Fig.~\ref{fig_dRdq-om-t}. Consequently, the relative importance of the
$t$-channel processes grows with $q_T$ but remains rather moderate even
at $q_T\simeq 2 \; \text{GeV}$, up to $15\%$ and $5\%$ of the total
theoretical yield in the $M=0.4$-$0.6 \;\text{GeV}$ and $0.6$-$0.9
\;\text{GeV}$ mass bins, respectively (negligible for $M>1$~GeV). Note
that the slopes of the $t$-channel emission spectra resemble the data
quite well, but our present estimate of their strength is insufficient
to resolve the discrepancies at high $q_T$. However, their impact on the
slope of the total spectra is not insignificant, as we will see in
Sec.~\ref{ssec_slope} below.

\subsection{Hadro-Chemical Freezeout and Critical Temperature}
\label{ssec_hadrochem}

In all our calculations of thermal dilepton spectra thus far, the medium
evolution was based on the notion that the critical ($T_c$) and chemical
freezeout ($T_{\rm ch}$) temperature coincide, at $T_c=T_{\rm
  ch}=175$~MeV. More recent theoretical (lattice QCD) and
phenomenological (thermal model fits to hadron ratios) studies, however,
allow for the possibilities that $T_c$ could be significantly larger
($190$-$200$~MeV) and $T_{\rm ch}$ significantly smaller
($150$-$160$~MeV). While a smaller $T_c\simeq160$~MeV is still viable, a
larger $T_{\rm ch}$ appears unlikely since in a high-density hadronic
phase number-changing reactions affecting the chemistry are to be
expected (in addition, thermal model fits start to become unstable at
temperatures above $\sim$$180$~MeV due to uncertainties in the
high-lying, high degeneracy hadron resonance spectrum). Therefore, in
this section we study the sensitivity of the invariant-mass and $q_T$
spectra to the hadro-chemistry of the fireball, keeping its geometry and
flow parameterization as well as the total lifetime the same as before
(including all normalizations, where applicable). As quoted in
Sec~\ref{sec_fireball}, in addition to our standard equation of state
(EoS-A: $T_c=T_{\text{ch}}=175\;\text{MeV}$), we investigate two
alternative scenarios: EoS-B with $T_c=T_{\text{ch}}=160\;\text{MeV}$
and EoS-C with $T_c=190\;\text{MeV}$,~$T_{\text{ch}}=160\;\text{MeV}$.
In general, a lower chemical freezeout temperature (where all meson
chemical potentials are by definition zero) entails a larger volume at
chemical freezeout, and thus also smaller baryon chemical potentials
(\ie, smaller baryon densities), as well as smaller meson chemical
potentials in the subsequent evolution. This furthermore implies that,
to obtain roughly the same overall dilepton yield, thermal freezeout
occurs at a \emph{larger} temperature (and smaller $\mu_{\pi,K}$),
which, in principle, leads to (somewhat) harder $q_T$ spectra, both at
thermal freezeout and for thermal emission throughout the hadronic
evolution. In the remainder of this Section, we focus on semicentral
collisions.

\begin{figure}[!t]
\centerline{\includegraphics[width=0.45\textwidth]{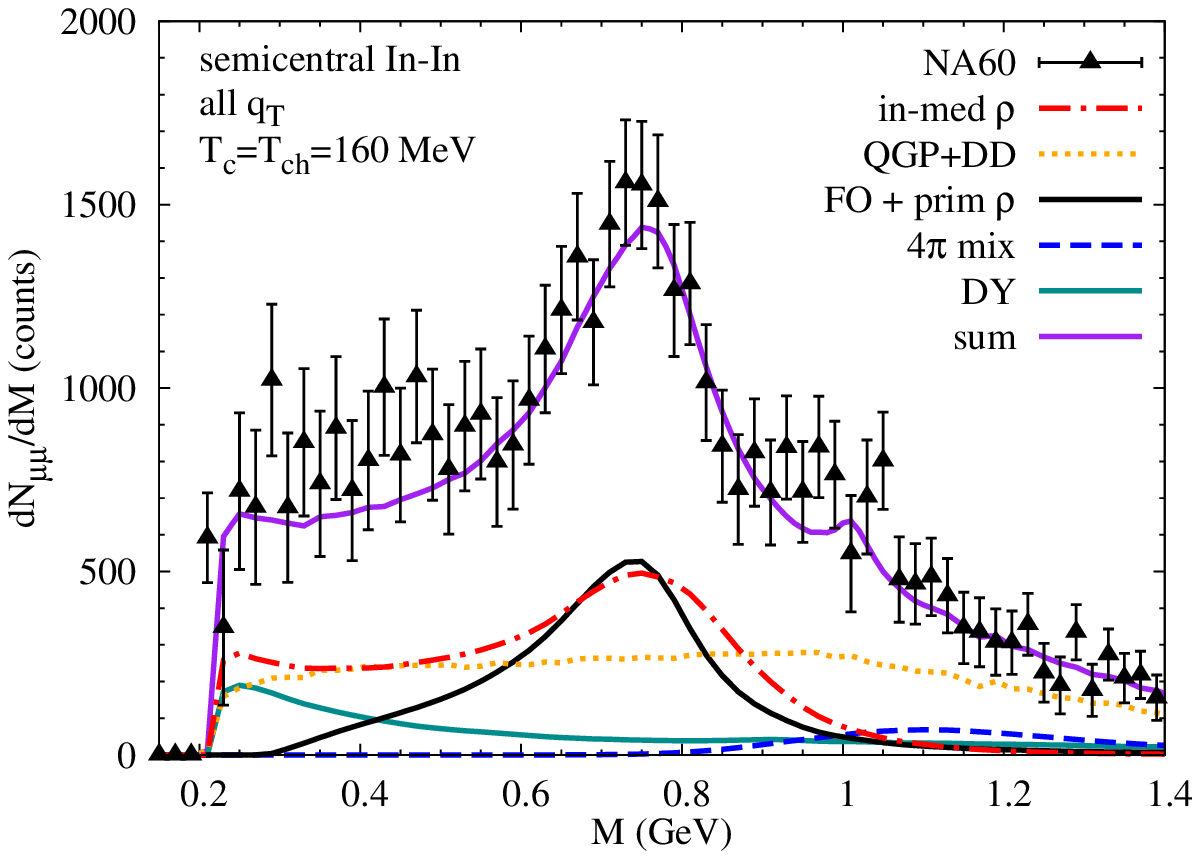}}
\centerline{\includegraphics[width=0.45\textwidth]{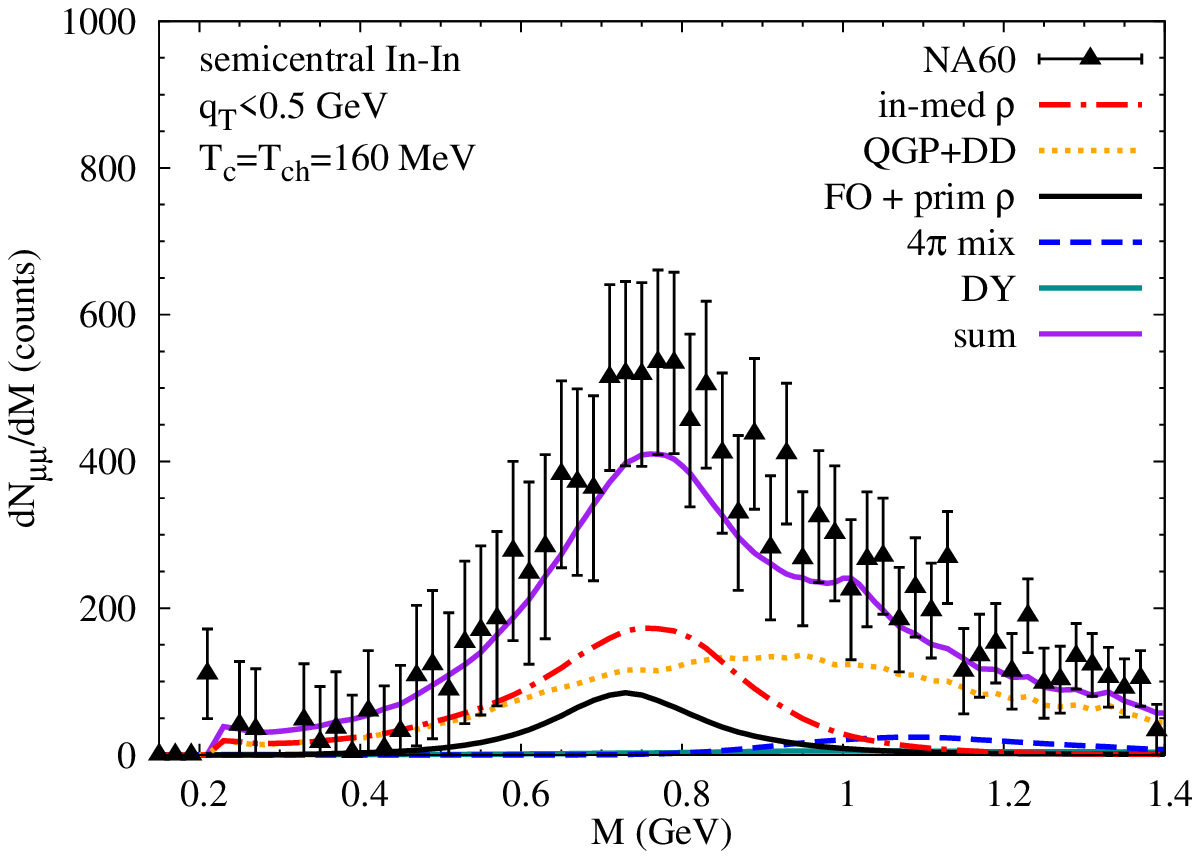}}
\centerline{\includegraphics[width=0.45\textwidth]{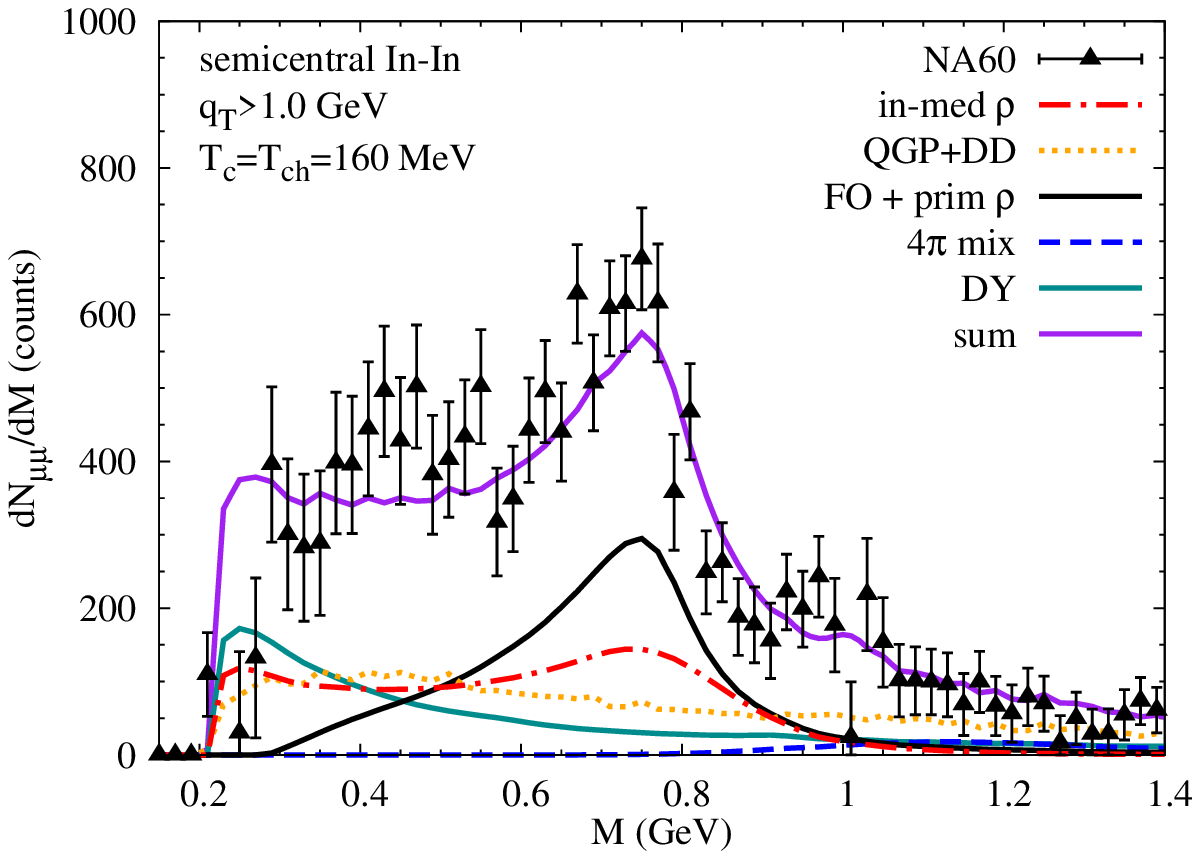}}
\caption{(Color online) The same as Fig.~\ref{fig_Mspecpt}, but for a
  fireball with a critical temperature $T_c=160 \; \text{MeV}$ and a
  chemical-freezeout temperature $T_{\text{ch}}=160 \; \text{MeV}$
  (EoS-B).}
\label{fig_Mspecpt_semi_fb6_Tc160}
\end{figure}
\begin{figure}[!t]
\centerline{\includegraphics[width=0.47\textwidth]{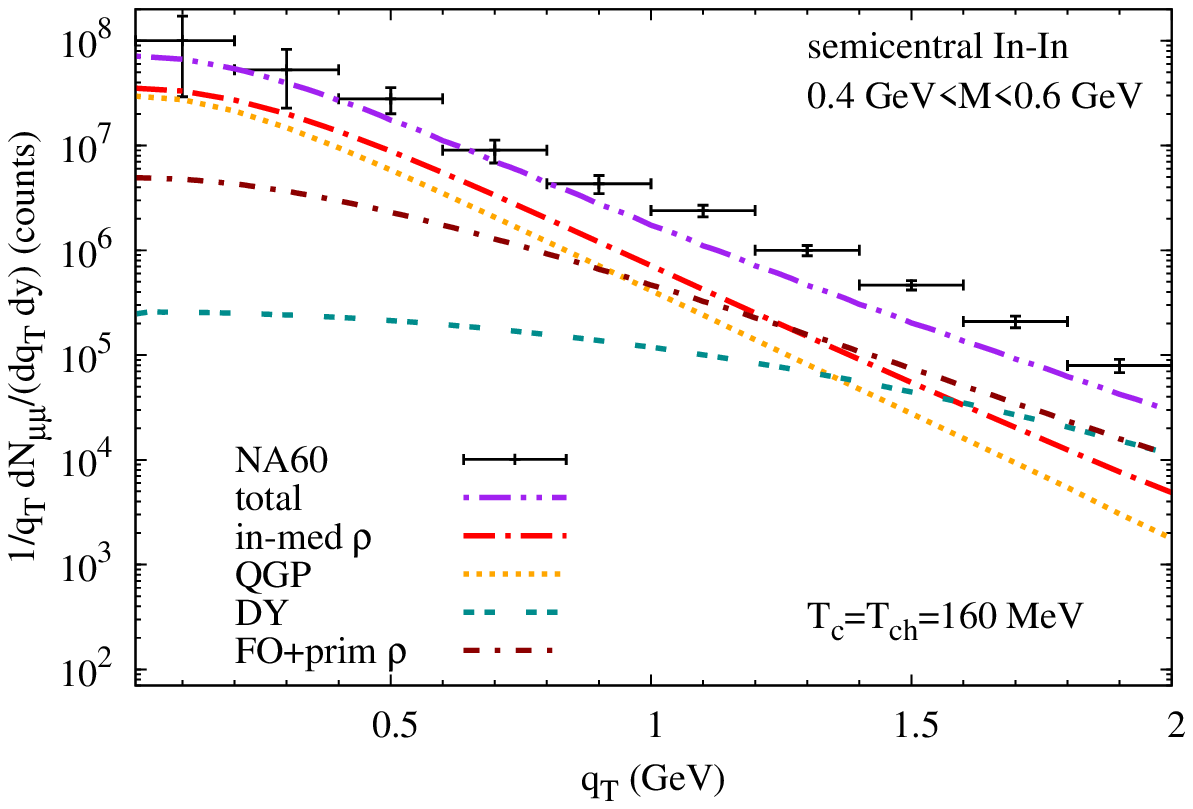}}
\centerline{\includegraphics[width=0.47\textwidth]{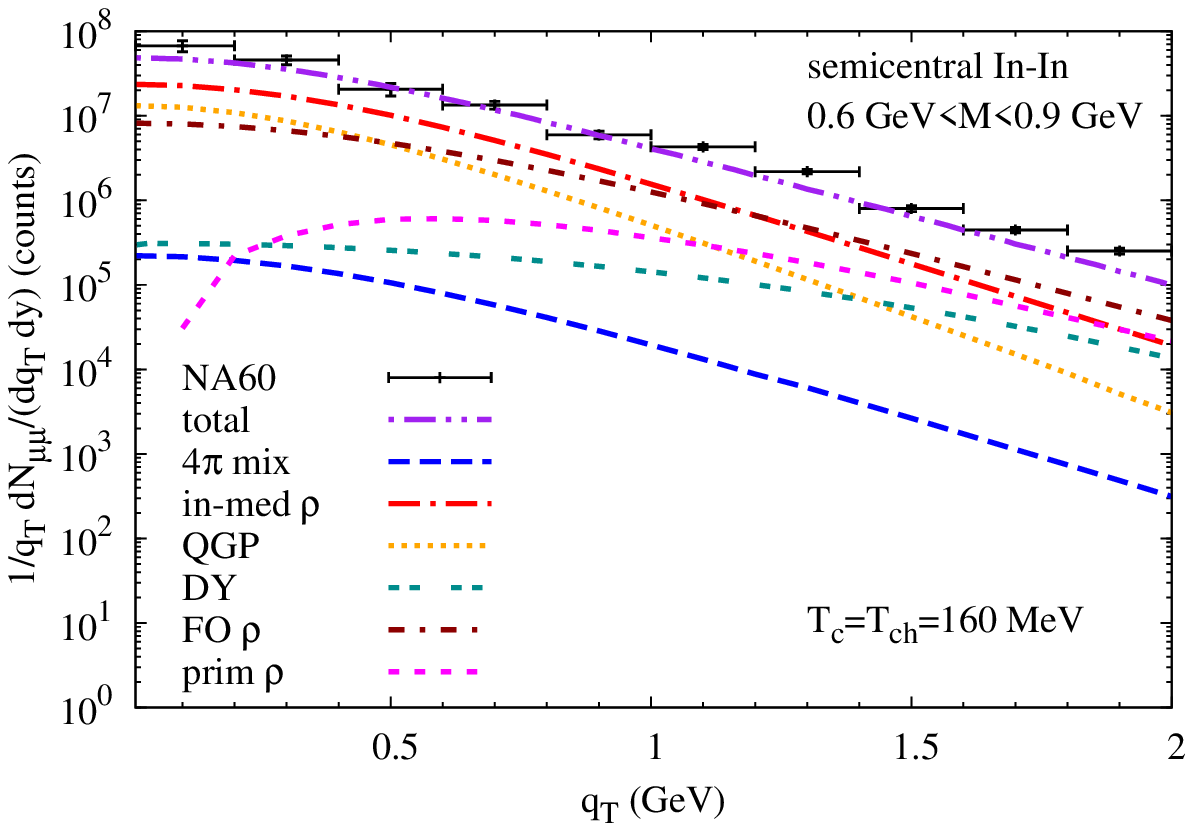}}
\centerline{\includegraphics[width=0.47\textwidth]{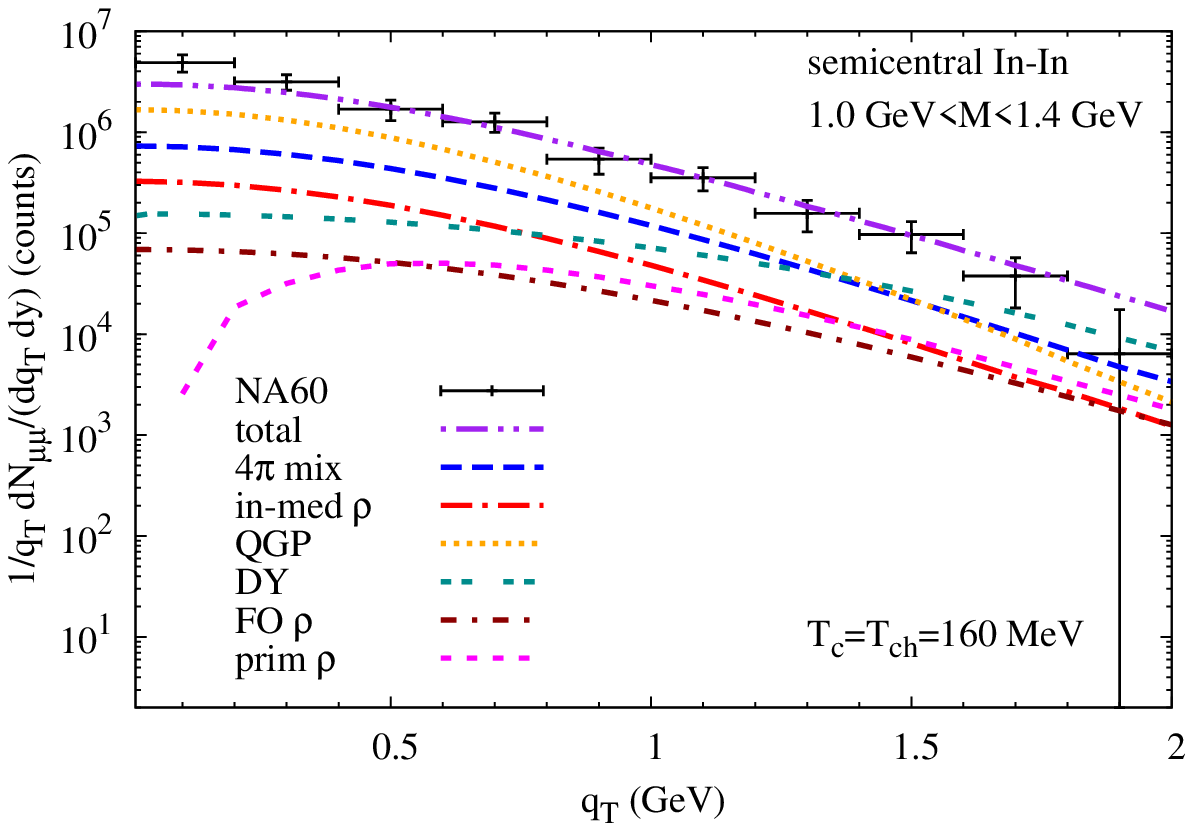}}
\caption{(Color online) The same as Fig.~\ref{fig_ptspec}, but for a
  fireball with a critical temperature $T_c=160 \; \text{MeV}$ and a
  chemical-freezeout temperature $T_{\text{ch}}=160 \; \text{MeV}$
  (EoS-B).}
\label{fig_ptspec_semi_fb6_Tc160}
\end{figure}
\begin{figure}[!t]
\centerline{\includegraphics[width=0.45\textwidth]{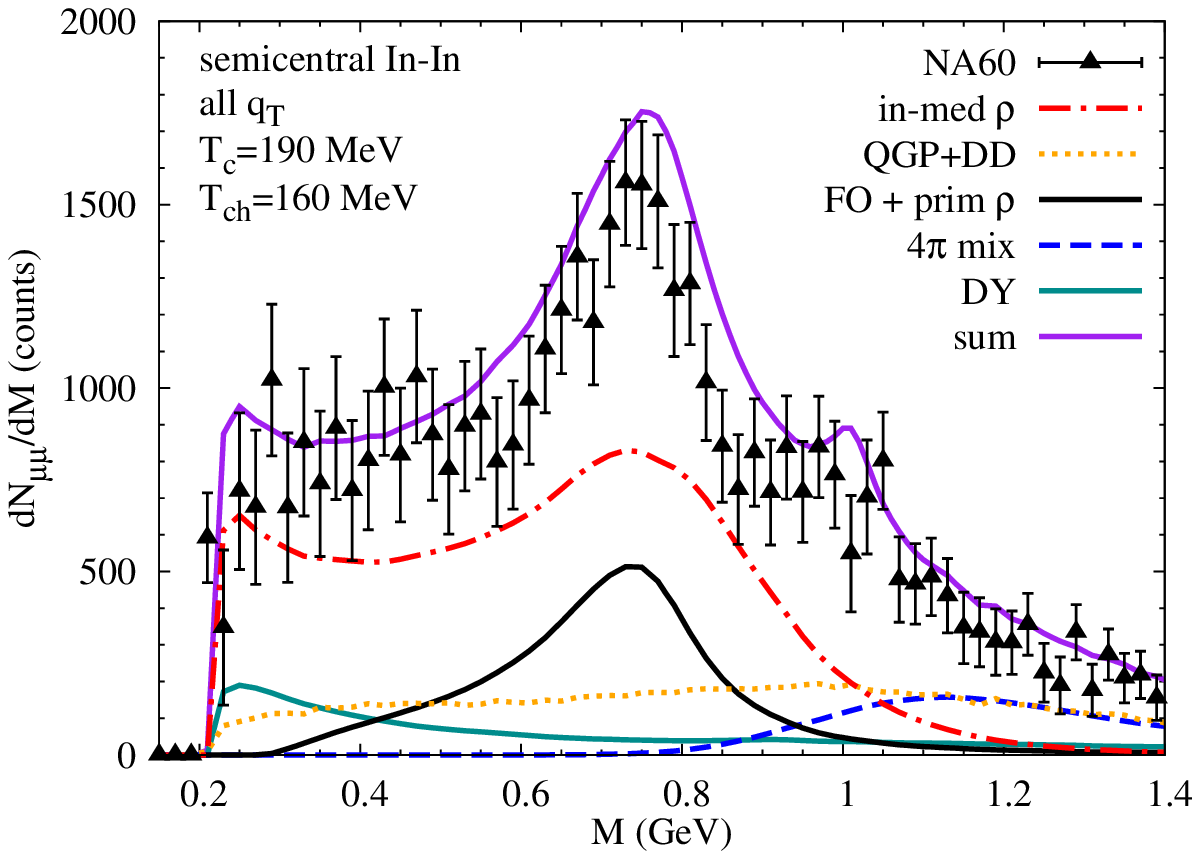}}
\centerline{\includegraphics[width=0.45\textwidth]{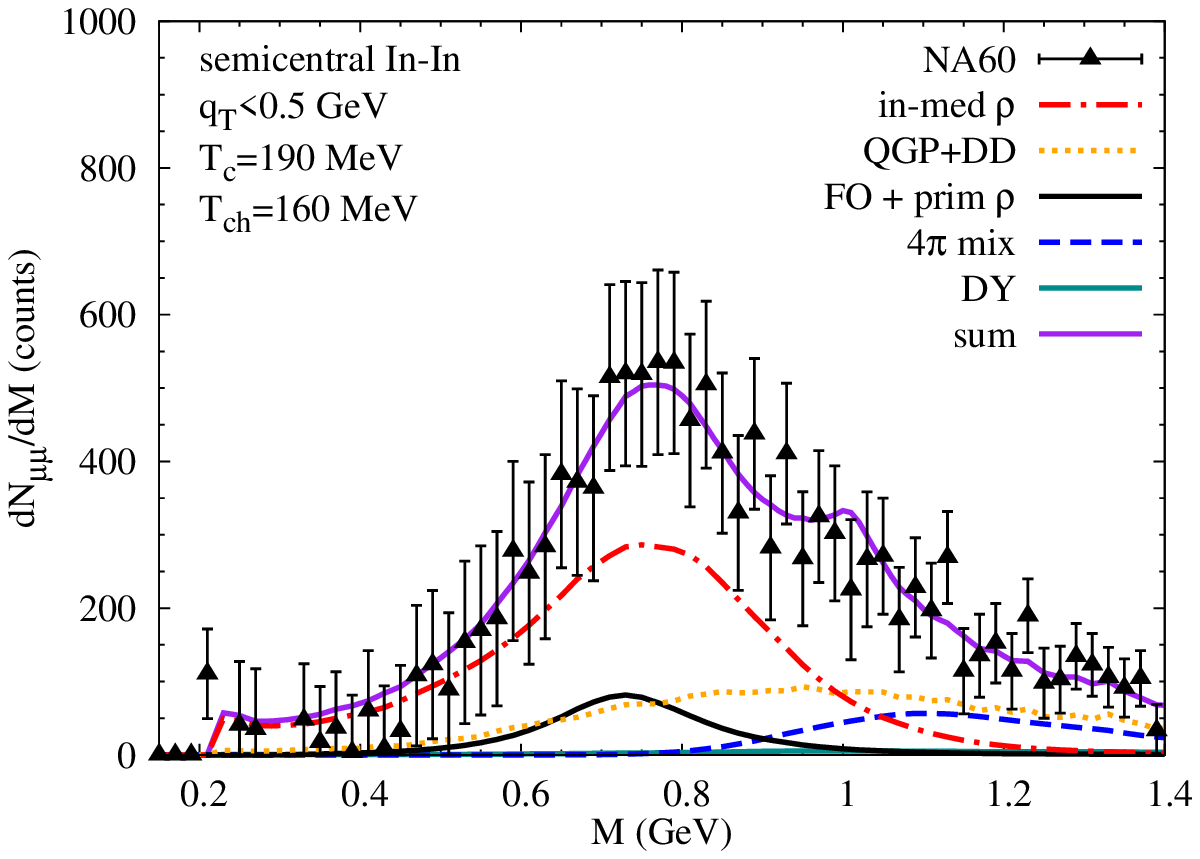}}
\centerline{\includegraphics[width=0.45\textwidth]{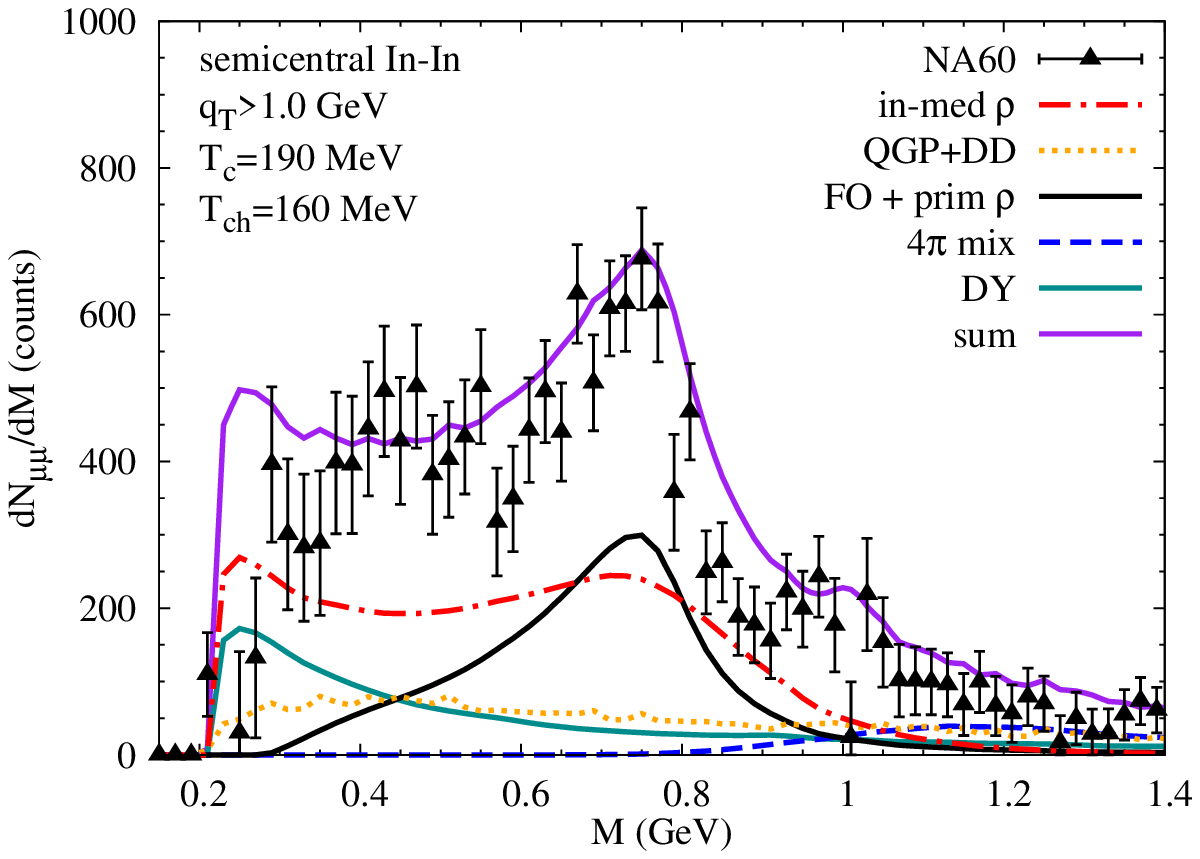}}
\caption{(Color online) The same as Fig.~\ref{fig_Mspecpt}, but for a
  fireball with a critical temperature $T_c=190 \; \text{MeV}$ and a
  chemical-freezeout temperature $T_{\text{ch}}=160 \; \text{MeV}$
  (EoS-C).}
\label{fig_Mspecpt_semi_fb6}
\end{figure}
\begin{figure}[!t]
\centerline{\includegraphics[width=0.47\textwidth]{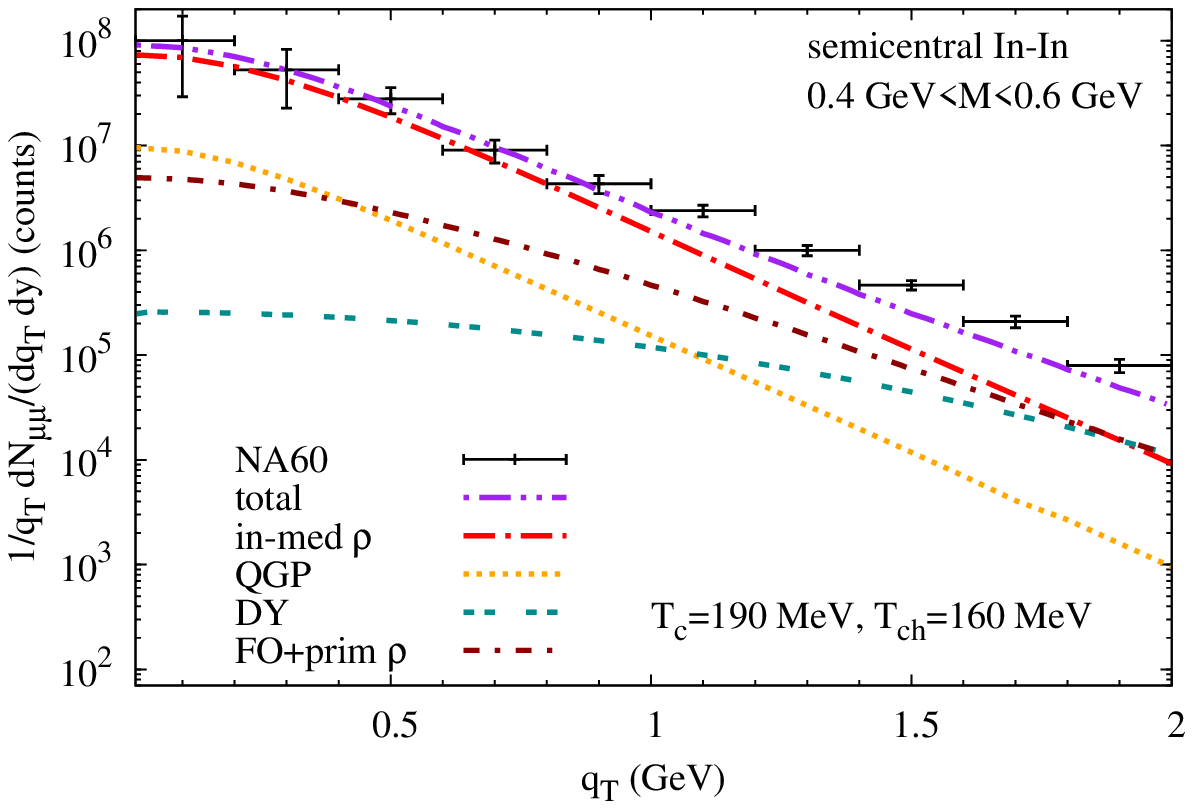}}
\centerline{\includegraphics[width=0.47\textwidth]{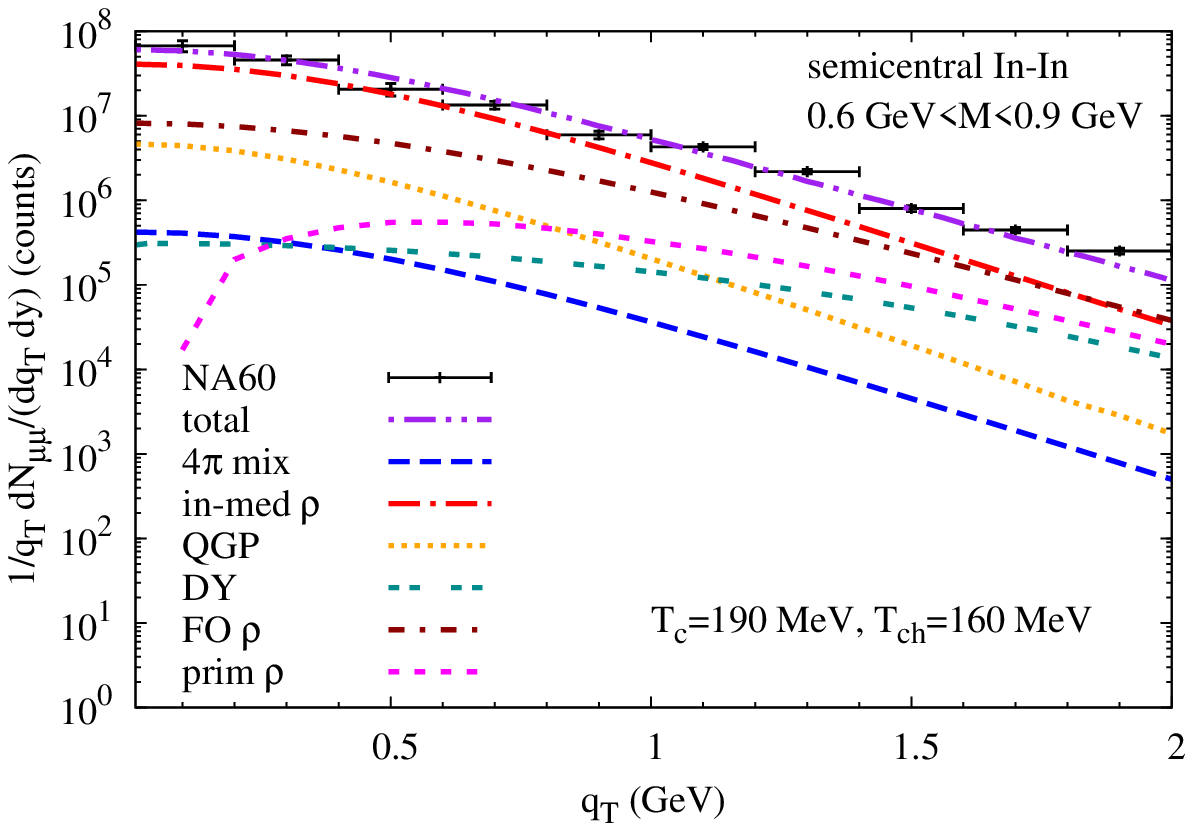}}
\centerline{\includegraphics[width=0.47\textwidth]{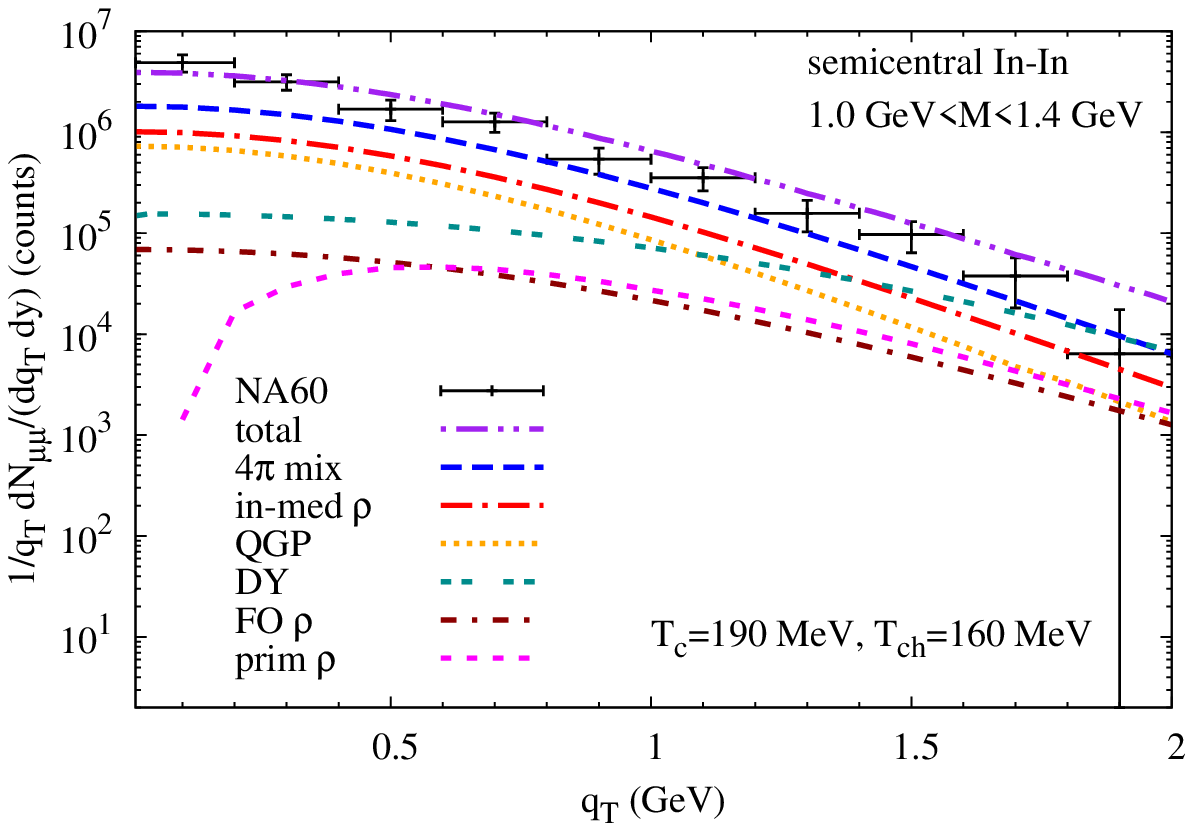}}
\caption{(Color online) The same as Fig.~\ref{fig_ptspec}, but for a
  fireball with a critical temperature $T_c=190 \; \text{MeV}$ and a
  chemical-freezeout temperature $T_{\text{ch}}=160 \; \text{MeV}$
  (EoS-C).}
\label{fig_ptspec_semi_fb6}
\end{figure}

We start our investigation within the ``EoS-B'' scenario, \ie,
$T_c=T_{\rm ch}=160~$MeV. The resulting $M$- and $q_T$-spectra are
compiled in Figs.~\ref{fig_Mspecpt_semi_fb6_Tc160} and
\ref{fig_ptspec_semi_fb6_Tc160}, respectively.  The smaller $T_c$
implies a longer duration of both the QGP and mixed phase (the latent
heat for EoS-B is larger than for EoS-A since entropy density for the
QGP EoS drops slower with $T$ than for the HG EoS), while the duration
of the hadronic phase is accordingly reduced. As noted above, the
thermal-freezeout temperature has increased to $T_{\text{fo}}=136$~MeV
(compared to $T_{\text{fo}}=120$~MeV for EoS-A). As an immediate
consequence, QGP radiation is significantly enhanced, while the hadronic
yield is slightly reduced (both for in-medium $\rho$ in the LMR and
four-pion in the IMR). The overall quality in the description of the
invariant-mass spectra is quite comparable to the EoS-A scenario. The
somewhat harder spectra implied by the larger hadronic temperatures for
EoS-B (most notably for the freezeout $\rho$) lead to a slight increase
of the dilepton yield in the $\rho$ mass region in the $q_T\ge1$~GeV
momentum bin (compare lower panels in Figs.~\ref{fig_Mspecpt} and
\ref{fig_Mspecpt_semi_fb6_Tc160}). However, in the low- ($0.4$~GeV$\le
M\le0.6$~GeV) and intermediate-mass ($0.6$~GeV$\le M\le0.9$~GeV) bins of
the $q_T$ spectra, the improvement at high $q_T$ is rather marginal,
while in the higher mass ($1.0$~GeV$\le M\le1.4$~GeV) there is
essentially no change compared to the EoS-A calculation (recall
Fig.~\ref{fig_ptspec}), despite the fact that the QGP contribution is
now larger than the four-pion one.

Turning to the EoS-C scenario (summarized in
Figs.~\ref{fig_Mspecpt_semi_fb6} and \ref{fig_ptspec_semi_fb6}), the
large value of $T_c=190$~MeV leads to a QGP and mixed phase which is
shorter than for EoS-A, but there is now a ``high-density'' (chemically
equilibrated) hadronic phase down to $T_{\rm ch}=160$~MeV, followed by a
subsequent chemical off-equilibrium evolution which is identical to
EoS-B. Obviously, the QGP emission yield is the smallest of all three
scenarios, while the hadronic yield (both in-medium $\rho$'s and
four-pion annihilation) is the largest, possibly implying a slight
overestimate in the $\rho$-mass region of the inclusive mass spectra
($13\%$ larger than for EoS-A, which could be readjusted by a slightly
reduced lifetime). The high-density hadronic phase helps in the $q_T$
spectra in the low- and intermediate-mass bin, for which, at high $q_T$,
the discrepancy is the smallest for the three scenarios.  In the highest
mass bin, even the rather large ratio of four-pion annihilation over QGP
emission does not upset the agreement with the $q_T$ spectra, indicating
that a large part of the four-pion emission emanates from the
high-density hadronic phase.
\begin{figure}[!t]
\centerline{\includegraphics[width=0.45\textwidth]{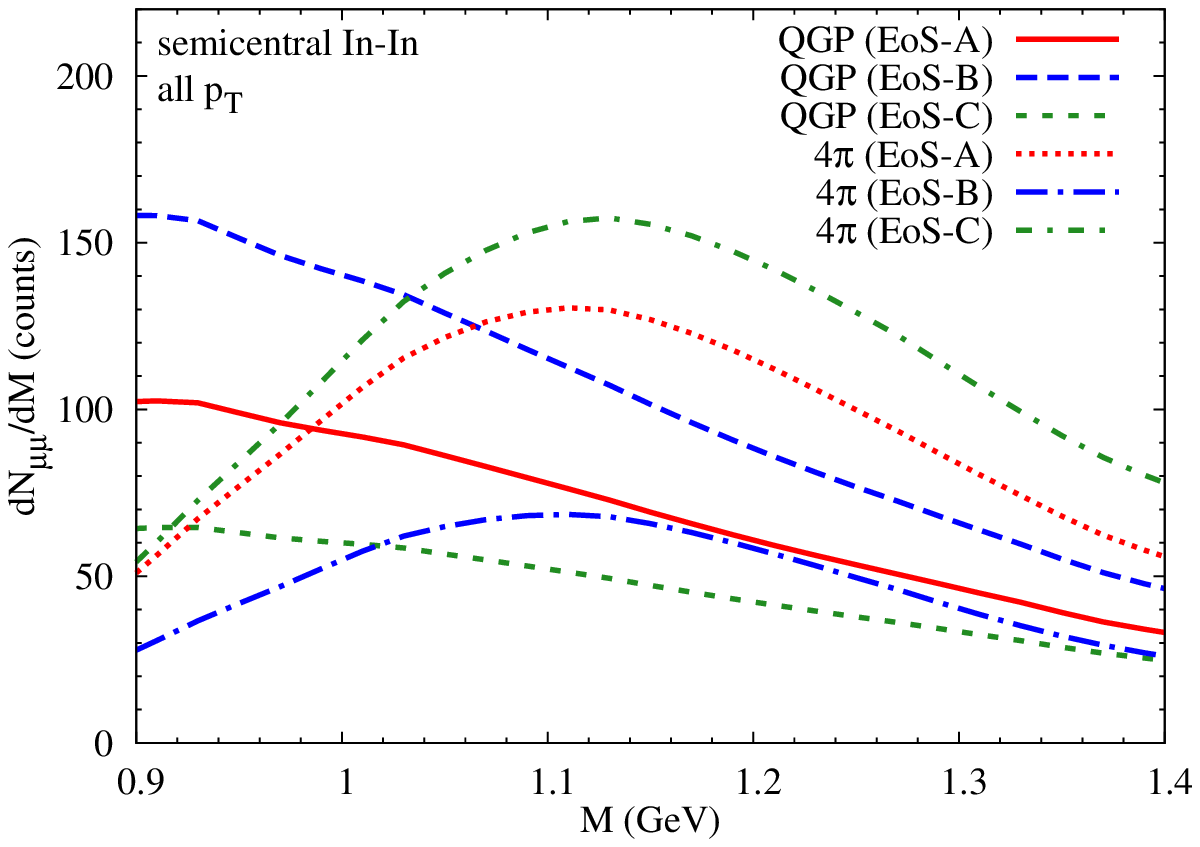}}
\caption{(Color online) Comparison of the QGP and four-pion
  contributions to the dilepton yield in the intermediate mass region
  for the three equations of state, as discussed in the text. 
  The 2-pion part, which is significant up to $M\simeq1.1$~GeV,
  is not shown.}
\label{fig_qgp_4pi_EoS_comp}
\end{figure}

Since the question of four-pion vs. QGP emission in the IMR has drawn
considerable attention in the recent literature~\cite{Specht:2007ez}, we
take a closer look at their interplay in our three scenarios in
Fig.~\ref{fig_qgp_4pi_EoS_comp}. The plot illustrates again that with
EoS-B ($T_c=T_{\rm ch}=160$~MeV) the QGP contribution exceeds the
in-medium four-pion annihilation over the entire IMR considered (note
that the tails of the $\rho$ spectral function are not included). The
opposite trend applies to our default EoS-A scenario, and even more so
for EoS-C. Especially the large difference between EoS-B and EoS-C
clearly demonstrates that the major portion of the four-pion emission
emanates from the high-density hadronic phase (EoS-B and EoS-C have an
identical hadronic evolution below $T_{\rm ch}=160$~MeV); a similar
argument applies to EoS-A, \ie, four-pion emission does not primarily
arise in the late stages (with large flow) despite an enhancement
through large pion-fugacity factors. The basic reason for this can be
understood by inspecting the temperature-mass differential emission
yield which is roughly given by~\cite{Rapp:2004zh}
\begin{equation}
\frac{\dd N_{ll}}{\dd M \dd T} \propto 
\im \Pi_{\rm em}(M,T) \ {\rm e}^{-M/T} \ T^{-5.5} \ , 
\end{equation}
characterized by the standard thermal exponential factor and a power in
temperature resulting from the three-momentum integral over the
Boltzmann factor and, most importantly, the volume expansion. Assuming a
weak temperature dependence of the e.m.~spectral function (medium
effects bias the emission toward higher $T$), a differentiation over $T$
identifies the temperature of maximum emission as $T_{\rm max}\simeq
M/5.5$. In the IMR of interest here, this means a $T_{\text{max}}$ right
around $T_c$ (note that, at $T_c$, the hadronic phase has an extra
``advantage'' of a roughly twice larger volume compared to the QGP due
to the latent heat), consistent with our fireball emission pattern.

We conclude this section by noting that ``reasonable'' variations in
hadro-chemical freezeout and critical temperature have a very moderate
impact on the invariant-mass spectra as seen by NA60. On the one hand,
this implies little sensitivity to the concrete values of $T_c$ and
$T_{\rm ch}$. On the other hand, it means that our results are very
robust with respect to uncertainties in these quantities. The main
reason for this robustness is the ``quark-hadron duality'' of the
underlying (medium-modified) emission rates from the hadronic and QGP
phase in the relevant temperature regime, $T=160$-$190$~MeV, in both LMR
($\rho$ melting) and IMR (including chiral mixing). As such, it provides
additional support to the medium modifications in the employed rates. In
the IMR, this ``duality'' does not allow for a (maybe even academic)
distinction between a high-density hadronic or partonic
source. Partitions with either component dominant are viable in both
mass and transverse momentum spectra. Concerning the high-$q_T$ region
of the lower two mass bins, our calculations indicate a slight
preference for EOS-B and EOS-C, due to their smaller $T_{\rm ch}$,
implying larger temperatures in the hadronic evolution.

\subsection{Slope analysis}
\label{ssec_slope}

\begin{figure}[!t]
\centerline{\includegraphics[width=0.45 \textwidth]{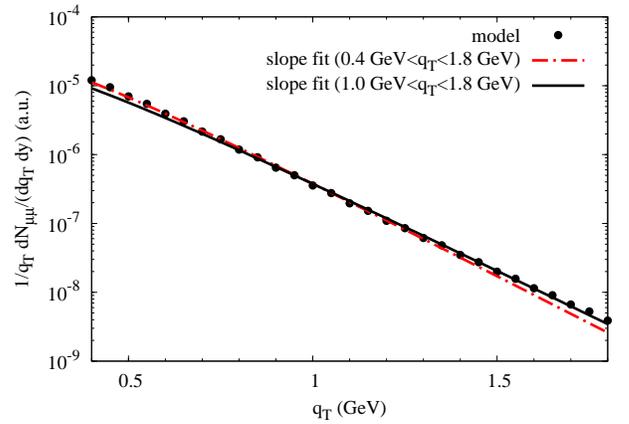}}
\caption{(Color online) Dependence of the effective-slope fits,
  Eq.~(\ref{Teff-fit}), to our theoretical $q_T$ spectra without the
  Drell-Yan contribution (dots) on the fit range in $q_T$ (as indicated
  in the legends) for the mass bin $0.3\;\mathrm{GeV} \leq M \leq
  0.4\;\mathrm{GeV}$.}
\label{fig_Teff_fit_range_dep}
\end{figure}
To complete our analysis of the $q_T$ spectra we perform in this section
a quantitative analysis of effective slope parameters following the
procedure adopted by the NA60 collaboration for the experimental
data~\cite{Damjanovic:2007qm,Specht:2007ez,Arnaldi:2007ru}. The $q_T$
spectra have been divided into several invariant-mass bins, in each of
which the data have been fitted to the function
\begin{equation}
\frac{1}{q_T} \frac{\dd N}{\dd q_T}=\frac{1}{m_T} \frac{\dd N}{\dd
  m_T}=C \exp \left (-\frac{m_T}{T_{\text{eff}}} \right) \ ,
\label{Teff-fit}
\end{equation}
where the fit range is typically taken as $0.4~{\rm GeV}<q_T<1.8$~GeV,
but the extracted slopes are rather insensitive to variations as long as
the lower limit is above $0.4$~GeV. In addition, the experimental $q_T$
spectra exhibit little dependence on centrality (except for peripheral
collisions), and therefore the slope analysis has been performed for
inclusive $q_T$ spectra with $\dd N_{\rm ch}/\dd y>30$. In our
theoretical analyses we focus on semicentral collisions.

We have tried to follow the same procedure (see also
Ref.~\cite{Renk:2006qr}, but as illustrated by the dash-dotted line in
Fig.~\ref{fig_Teff_fit_range_dep}, a fit to our theoretical spectra
(represented by the dots) in the above range ($0.4~{\rm
  GeV}<q_T<1.8$~GeV) slightly underestimates the effective slopes toward
higher $q_T$, especially in the lower mass bins. Part of the problem is
the bias of the fit toward low $q_T$ where the yield is the largest and
thus dominates the total $\chi^2$. To better reproduce the theoretical
spectral shape with the above function we have therefore chosen to (i)
restrict the fit range to $1 \;\text{GeV} <q_T<1.8 \; \text{GeV}$, and
(ii) minimize the $\chi^2$ for the logarithm of the spectra. The
resulting fit (solid line in Fig.~\ref{fig_Teff_fit_range_dep}) indeed
agrees better with the theoretical spectra. This effect becomes more
pronounced when the Drell-Yan component is included (which, in fact,
might indicate that our extrapolation overestimates the DY yield at low
mass).

\begin{figure*}[!t]
\begin{minipage}{0.45 \textwidth}
\centerline{\includegraphics[width=1.0\textwidth]{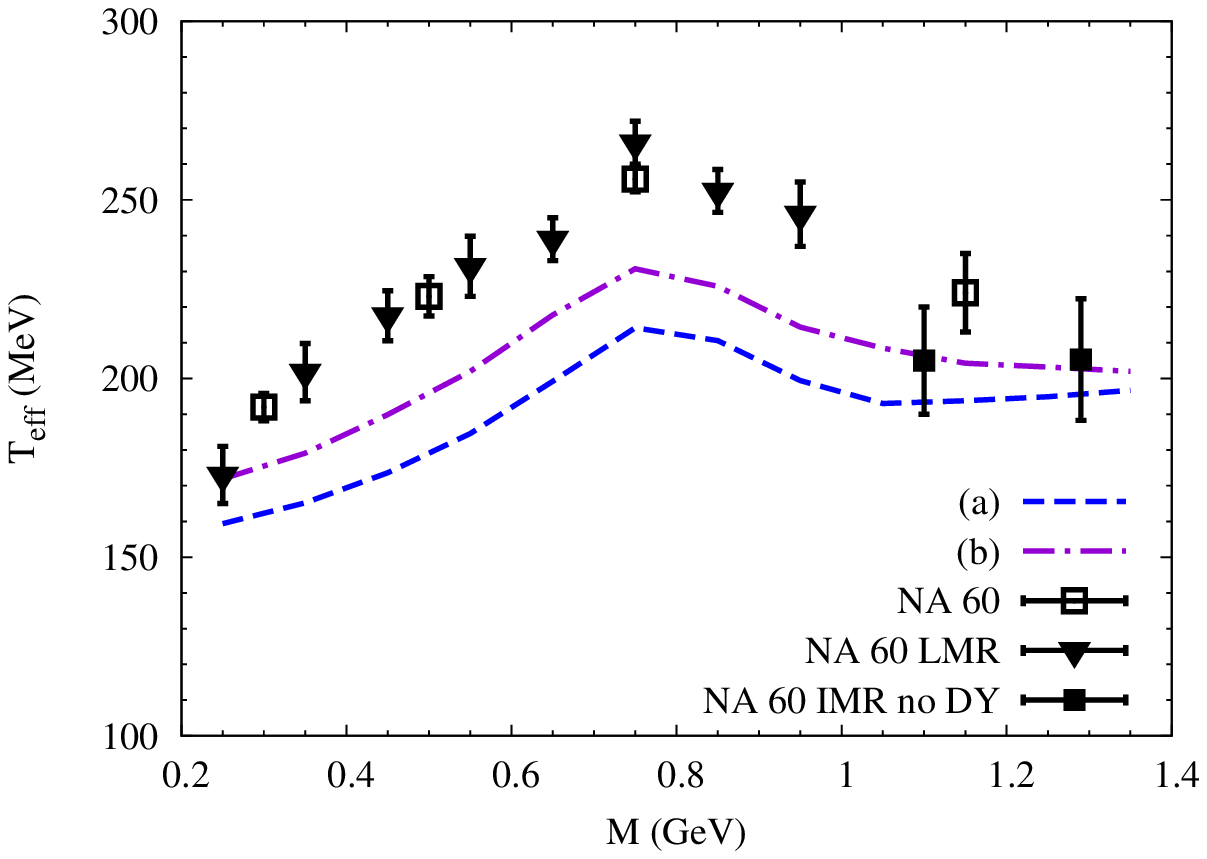}}
\end{minipage}
\begin{minipage}{0.45 \textwidth}
\centerline{\includegraphics[width=1.0\textwidth]{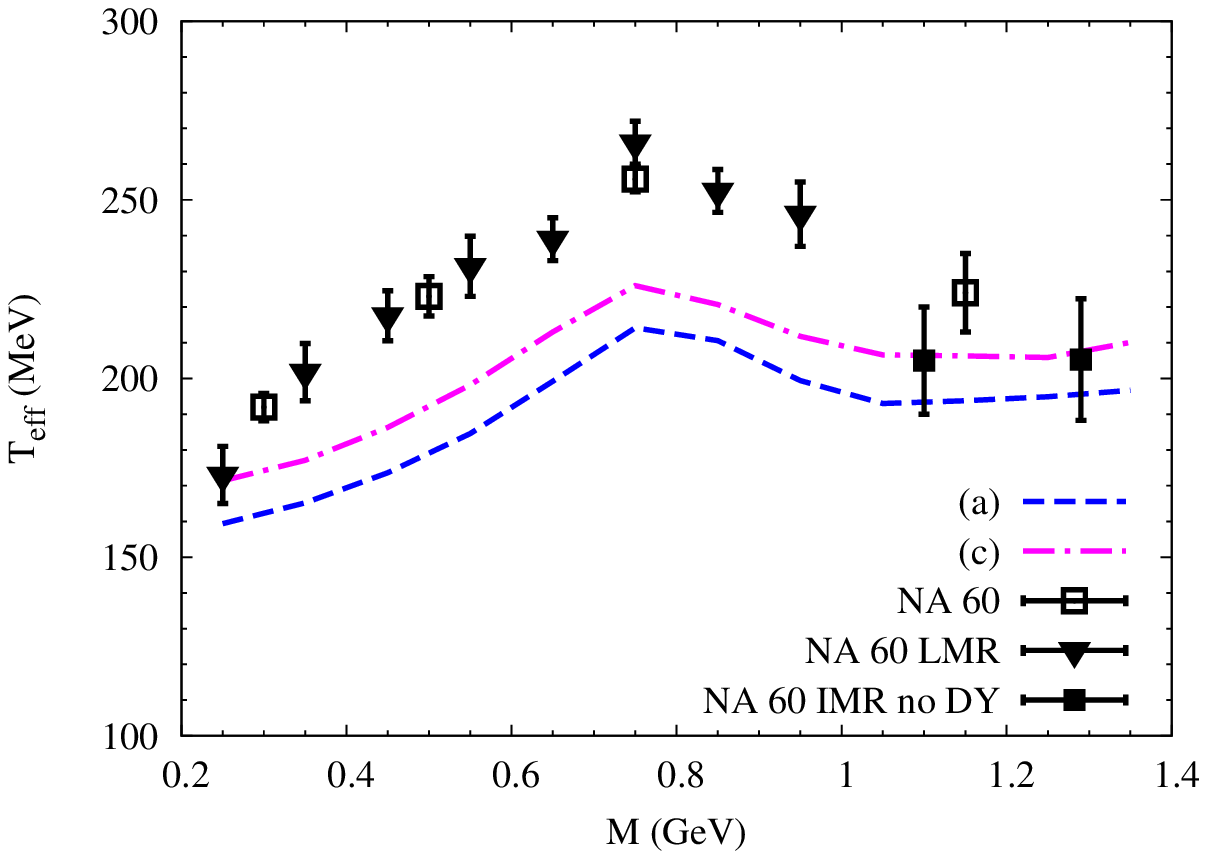}}
\end{minipage}\vspace*{0.5mm}
\begin{minipage}{0.45 \textwidth}
\centerline{\includegraphics[width=1.0\textwidth]{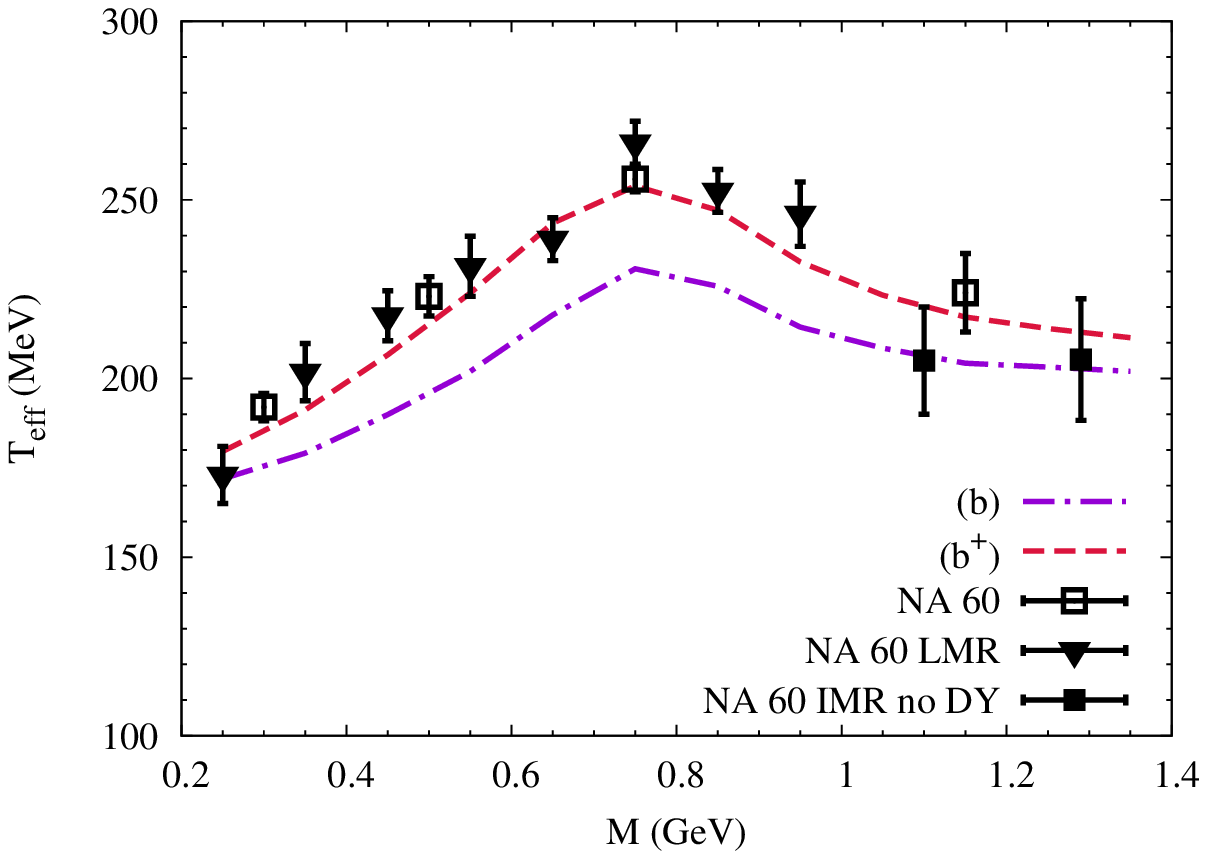}}
\end{minipage}
\begin{minipage}{0.45 \textwidth}
\centerline{\includegraphics[width=1.0\textwidth]{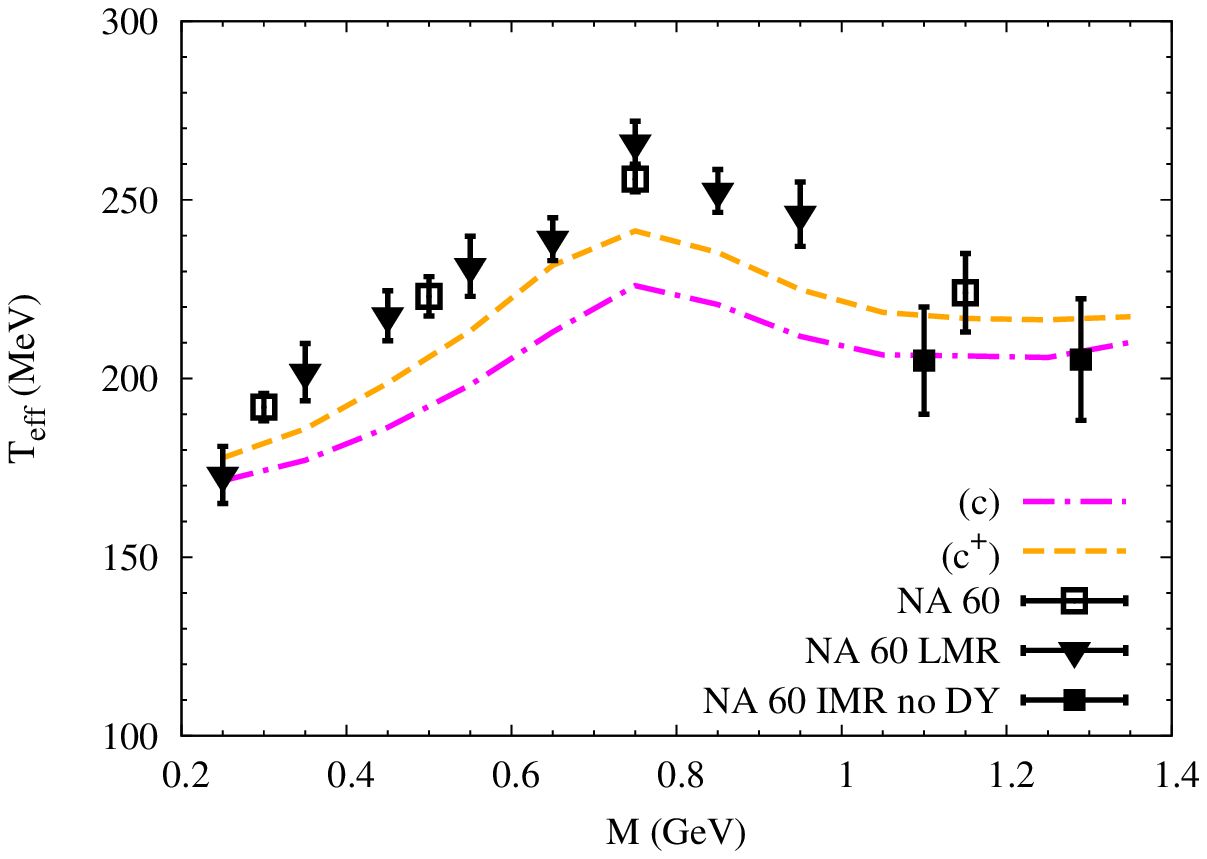}}
\end{minipage}\vspace*{0.5mm}
\begin{minipage}{0.45 \textwidth}
\centerline{\includegraphics[width=1.0\textwidth]{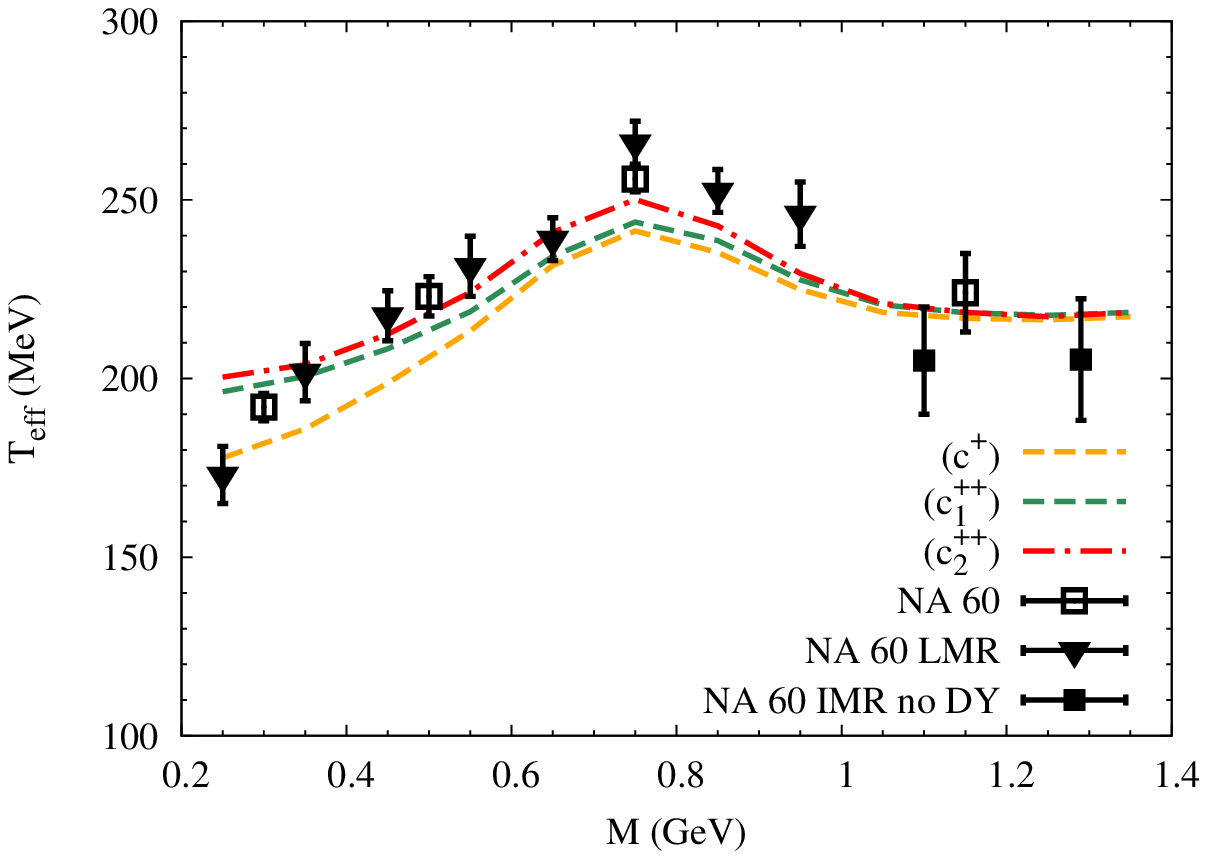}}
\end{minipage}
\begin{minipage}{0.45 \textwidth}
\centerline{\includegraphics[width=1.0\textwidth]{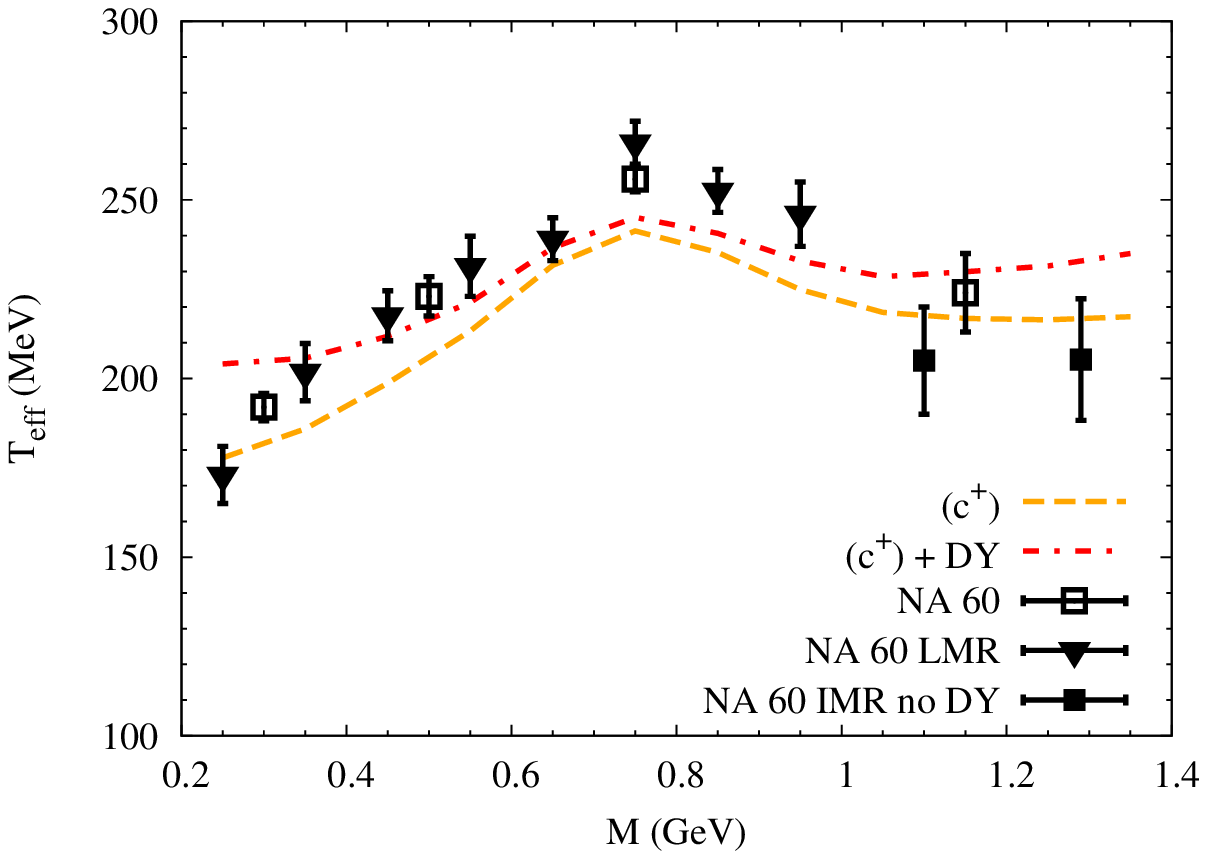}}
\end{minipage}
\caption{(Color online) Effective slope parameters fitted to the theoretical
  $q_T$ spectra using Eq.~(\ref{Teff-fit}) for the following scenarios 
  (by default, DY and $t$-channel meson exchange reactions are not included):
  top left: EoS-A (dashed line) vs. EoS-B (dash-dotted line), both
  with $a_{\perp}=0.085 c^2/\mathrm{fm}$;
  top right: EoS-A (dashed line) vs. EoS-C (dash-dotted line), both
  with $a_{\perp}=0.085 c^2/\mathrm{fm}$;
  middle left: EoS-B with either $a_{\perp}=0.085c^2/\mathrm{fm}$ (dash-dotted
  line) or $a_{\perp}=0.1c^2/\mathrm{fm}$ (dashed line);
  middle right: EoS-C with either $a_{\perp}=0.085c^2/\mathrm{fm}$ (dash-dotted
  line) or $a_{\perp}=0.1c^2/\mathrm{fm}$ (dashed line);
  bottom left: EOS-C with $a_{\perp}=0.1c^2/\mathrm{fm}$ (lower dashed line),
  additionally including  meson $t$-channel
  contributions with free (dash-dotted line) or in-medium (middle dashed line)
  $\rho$ propagator;
  bottom right: EOS-C with $a_{\perp}=0.1c^2/\mathrm{fm}$ (dashed line),
  additionally including Drell-Yan annihilation (dash-dotted line).
  \label{fig_slopes}}
\end{figure*}

Based on the slightly revised procedure, our slope analysis is carried
out for the total sum of all calculated $q_T$ spectra components in
twelve $\Delta M=0.1 \;\text{GeV}$ mass bins in the range
$0.2\;\text{GeV}<M<1.4 \; \text{GeV}$ (corresponding to the experimental
ones) for semicentral In-In; by default, the Drell-Yan and meson
$t$-channel contributions are not included, but studied in separate fits
by adding them to the totals. As in the experimental analysis,
correlated open-charm decays are removed altogether. Our systematic
study addresses variations in the EoS as well as the possibility of
another $\sim$$15\%$ increase in the transverse acceleration of the
fireball expansion, \ie, $a_{\perp}=0.1c^2$/fm (which is, in fact, the
value we have used in our calculations for heavy-flavor observables at
RHIC~\cite{vanHees:2005wb}, as extracted from the hydrodynamic model of
Ref.~\cite{Kolb:2000sd}).  Specifically, we evaluate the following
scenarios, as summarized in the six panels of Fig.~\ref{fig_slopes}:
\begin{enumerate}
\item[(a)] the default fireball expansion with EoS-A
  ($T_c=T_{\text{ch}}=175$~MeV), characterized by a thermal freezeout
  with temperature and radial flow surface velocity of
  $(T_{\text{fo}},\beta_{\perp,{\rm fo}}^s)=(120~{\rm MeV},0.57c)$;
\item[(b)] the default fireball expansion with EoS-B
  ($T_c=T_{\mathrm{ch}}=160$~MeV) and $(T_{\text{fo}},\beta_{\perp,{\rm
      fo}}^s)=(136~{\rm MeV},0.57c)$;
\item[(c)] the default fireball expansion with EoS-C ($T_c=190$~MeV,
  $T_{\mathrm{ch}}=160$~MeV) and $(T_{\text{fo}},\beta_{\perp,{\rm
      fo}}^s)=(136~{\rm MeV},0.57c)$;
\item[(b$^+$)] the same as in (b), but with a transverse fireball
  acceleration $a_{\perp}=0.1 \, c^2/\mathrm{fm}$ yielding
  $(T_{\text{fo}},\beta_{\perp,{\rm fo}}^s)=(135~{\rm MeV},0.65c)$;
\item[(c$^+$)] the same as in (c), but with a transverse fireball
  acceleration $a_{\perp}=0.1 \, c^2/\mathrm{fm}$ yielding 
  $(T_{\text{fo}},\beta_{\perp,{\rm fo}}^s)=(135~{\rm MeV},0.65c)$;
\item[(c$_1^{++}$)] the same as in (c$^+$), but additionally including
  the contribution from meson-$t$-channel exchange with in-medium $\rho$
  propagator (cf.~Sec.~\ref{ssec_omt});
\item[(c$_2^{++}$)] the same as in (c$_1^{++}$), but using the vacuum
  $\rho$ propagator in the meson-$t$-channel exchange contribution.
\end{enumerate}

The upper two panels of Fig.~\ref{fig_slopes} compare the impact of the
three different EoS using our baseline acceleration of
$a_{\perp}=0.085c^2$/fm. The default scenario EoS-A (a) falls short of
the empirical slopes by a significant margin of up to $50$-$60$~MeV in
the LMR, while reaching the lower end of the data in the IMR. The higher
hadronic temperatures of scenarios EoS-B (b) and EoS-C (c) improve the
situation by about $15$-$20$~MeV in the LMR, approximately reflecting
the increase in the hadronic fireball temperatures (e.g.,
$T_{\text{fo}}=136$~MeV compared to $T_{\text{fo}}=120$~MeV at thermal
freezeout).  Scenario EoS-B (b) additionally improves around the $\rho$
peak due to a larger weight of the relatively hard components from
decays of freezeout and primordial $\rho$, since the overall thermal
hadronic emission is smaller than for EoS-A and EoS-C (compare
Fig.~\ref{fig_ptspec_semi_fb6_Tc160} to Figs.~\ref{fig_med} and
\ref{fig_Mspecpt_semi_fb6}). However, for both EoS-B and EoS-C the
slopes in the LMR are still below the data, even though the shape is not
too bad.

In an attempt to improve on the slopes, we investigate a further
increase of the transverse acceleration to values representative for
RHIC energies~\cite{Kolb:2000sd,vanHees:2005wb}, from $0.085 \;
c^2/\text{fm}$ to $0.1 \; c^2/\text{fm}$ for EoS-B and EoS-C (scenarios
(b$^+$) and (c$^+$), shown in the middle panels of
Fig.~\ref{fig_slopes}).  While this increases the effective slopes at
all emission stages, it is more efficient for the later hadronic sources
(in accordance with the blue-shift pocket formulae given in
Sec.~\ref{ssec_qtspec}).  Consequently, the LMR exhibits further
improvement, and, importantly, the consistency with the data in the IMR
is not spoiled due to the fact that the main contributions there arise
from earlier stages with higher temperatures and less flow (dominated
either by the contributions from the QGP phase for EoS-B or four-pion
annihilation in the high-density hadronic phase for EoS-C,
cf.~Fig.~\ref{fig_qgp_4pi_EoS_comp}). At this point, EoS-B provides
somewhat better agreement and one might be tempted to take this as
evidence for a prevalently partonic emission source, as suggested in
Refs.~\cite{Specht:2007ez,Arnaldi:2007ru}.  However, remaining
uncertainties preclude an unambiguous interpretation, as we will now
show.
 
In the two bottom panels of Fig.~\ref{fig_slopes} we focus on the EoS-C
scenario with $a_{\perp}=0.1c^2$/fm.  The left panel illustrates the
sensitivity of the slopes to the additional inclusion of $t$-channel
meson exchange processes. As discussed in the previous section, their
spectra alone carry a slope very similar to the data; not surprisingly,
their inclusion increases the slopes in the LMR with no impact on the
IMR slopes.  When implementing the $t$-channel meson exchanges with the
vacuum $\rho$ propagator, the effect on the slopes is somewhat larger
compared to using the in-medium $D_\rho$, especially in the free $\rho$
mass region.  A single $\pi\rho$ scattering with unmodified $\rho$
mesons would most closely represent interactions of $\rho$ mesons in a
kinetic regime (rescattering without thermalization), and thus provide a
missing link between the two extremes of thermal radiation and
primordial $\rho$'s surviving jet quenching (no reinteraction) in our
baseline calculations (note that there is no contribution from
``quenched'' $\rho$'s).  The overall level of agreement with the data is
now very similar to EoS-B.  Also recall that $T_c=190$~MeV maximizes the
ratio of hadronic to QGP emission; for $T_c=175$~MeV four-pion emission
still outshines the QGP in the IMR, while the slope parameters are
closer to the EoS-B scenario.  The effect of the Drell-Yan contribution
on the slopes is shown in the bottom-right panel of
Fig.~\ref{fig_slopes}. In the IMR the change in slope is consistent with
the experimental results (full vs. open squares), while in the lowest
mass bins the uncertainties in the extrapolation of the DY are large. We
emphasize again that more stringent constraints can be enforced once an
accurate knowledge of the absolute normalization in the experimental
mass and momentum spectra becomes available.

There are uncertainties in the experimental extraction of the slope
parameters beyond the statistical error bars shown in the figures. In
Ref.~\cite{Arnaldi:2007ru} a systematic error of $\sim$7~MeV is
estimated due to the subtraction of background and decay (cocktail)
sources. For the LMR, it has been estimated that the subtraction of a
potential Drell-Yan contribution could lower the effective slopes by
5-10~MeV. One should also be aware of that the applicability of
hydrodynamic (or thermal fireball) descriptions is limited in transverse
momentum, typically to $p_T\le2$~GeV for pion spectra at the SPS
(consistent with where the spectra reach the collision-scaling regime in
Fig.~\ref{fig_raa}), and to significantly smaller $p_T$ for elliptic
flow. Fireball models might compensate for this to some extent by
implementing a larger acceleration, which may be (part of) the reason
that the dilepton slopes prefer $a_{\perp}=0.1c^2$/fm.  To further
scrutinize the $q_T$ spectra, it would be very illuminating to reiterate
earlier implementations of the dilepton rates from hadronic many-body
theory into hydrodynamic~\cite{Huovinen:1998ze} (with a more realistic
hadro-chemsitry) or transport~\cite{Cassing:1997jz} simulations.  In
this context, a measurement of dilepton elliptic flow could provide
further insights in the detailed emission history of the medium at SPS
energies~\cite{Heinz:2006qy}.

\subsection{Update of Comparison to CERES/NA45 Data}
\label{ssec_ceres}

After the refinements in our theoretical approach for SPS dilepton
spectra in the context of the NA60 data, we revisit in this section the
consequences for the comparison with CERES/NA45 in semicentral and
central Pb-Au collisions, which showed good agreement in our earlier
works~\cite{Rapp:1997fs,Rapp:1999us,Rapp:2002tw}. In the following, we
constrain ourselves to the above default scenario for the fireball
evolution with a transverse acceleration of
$a_{\perp}=0.085\,c^2/\text{fm}$. As in the In-In case, the increased
expansion rate entails shorter fireball lifetimes; we choose the latter
as to obtain the same thermal freezeout temperature for central ($T_{\rm
  fo}\simeq106$~MeV) and semicentral ($T_{\rm fo}\simeq112$~MeV) Pb-Au
collisions as in our previous work~\cite{Rapp:1999us,Rapp:1999zw},
implying a reduction by $2$~fm/$c$ for both centralities ($13.5\to
11.5$~fm/$c$ and $12\to 10$~fm/$c$, respectively). Also, since for the
larger system sizes (\ie, lifetimes) the freezeout- and cocktail-$\rho$
contributions~\cite{Agakichiev:2005ai} are significantly suppressed
relative to thermal radiation, and since we are not interested in
(high-) $q_t$ spectra here, we adopt the (old) simplified treatment for
the freezeout $\rho$ by running the fireball an additional duration of
$1$~fm/$c$~\cite{Rapp:1999us}. The larger lifetimes relative to In-In
collisions furthermore imply that the nonrelativistic formula for the
surface expansion velocity, $v_\perp^s=a_\perp t$, approaches values
uncomfortably close to $c$ in the late stages of the collision. To
implement a relativistically covariant acceleration in a simple way, we
model the acceleration of the fireball in analogy to the relativistic
motion for a charged particle in a homogeneous electric field. The
radial surface velocity and fireball radius then take the form
\begin{equation}
\begin{split}
  \beta_\perp^s&=\frac{a_{\perp} t/c}{\sqrt{1+a_{\perp}^2 t^2/c^2}},\\
  r&=r_0+\frac{c^2}{a_{\perp}} \left (\sqrt{1+a_{\perp}^2 t^2/c^2}-1
  \right) \ ,
\end{split}
\end{equation}
which tames the acceleration at late times and matches the
non-relativistic case for early times.

\begin{figure}[!t]
\vspace*{2mm}
\centerline{\includegraphics[width=0.42\textwidth]{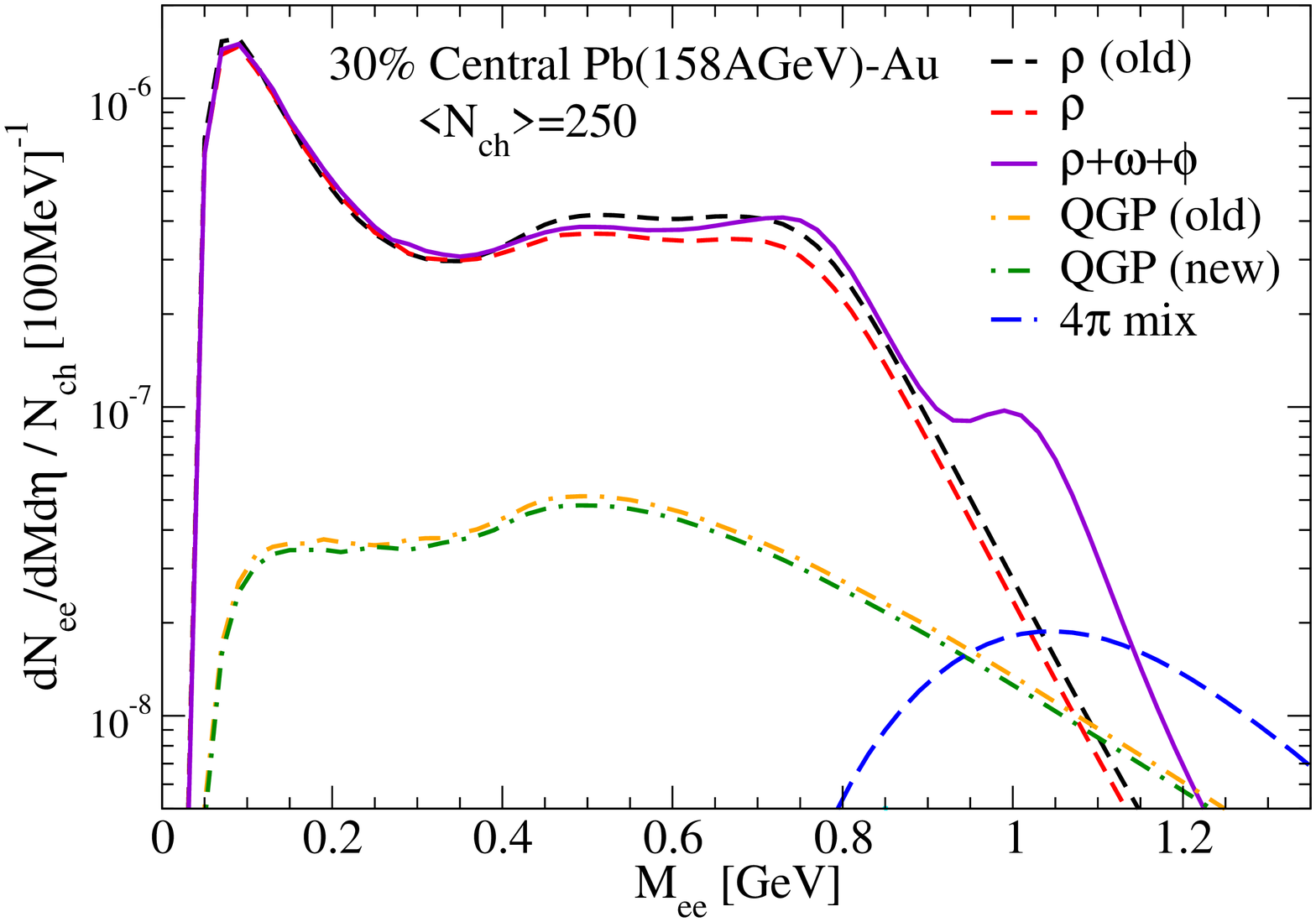}}
\vspace*{2mm}
\centerline{\includegraphics[width=0.45\textwidth]{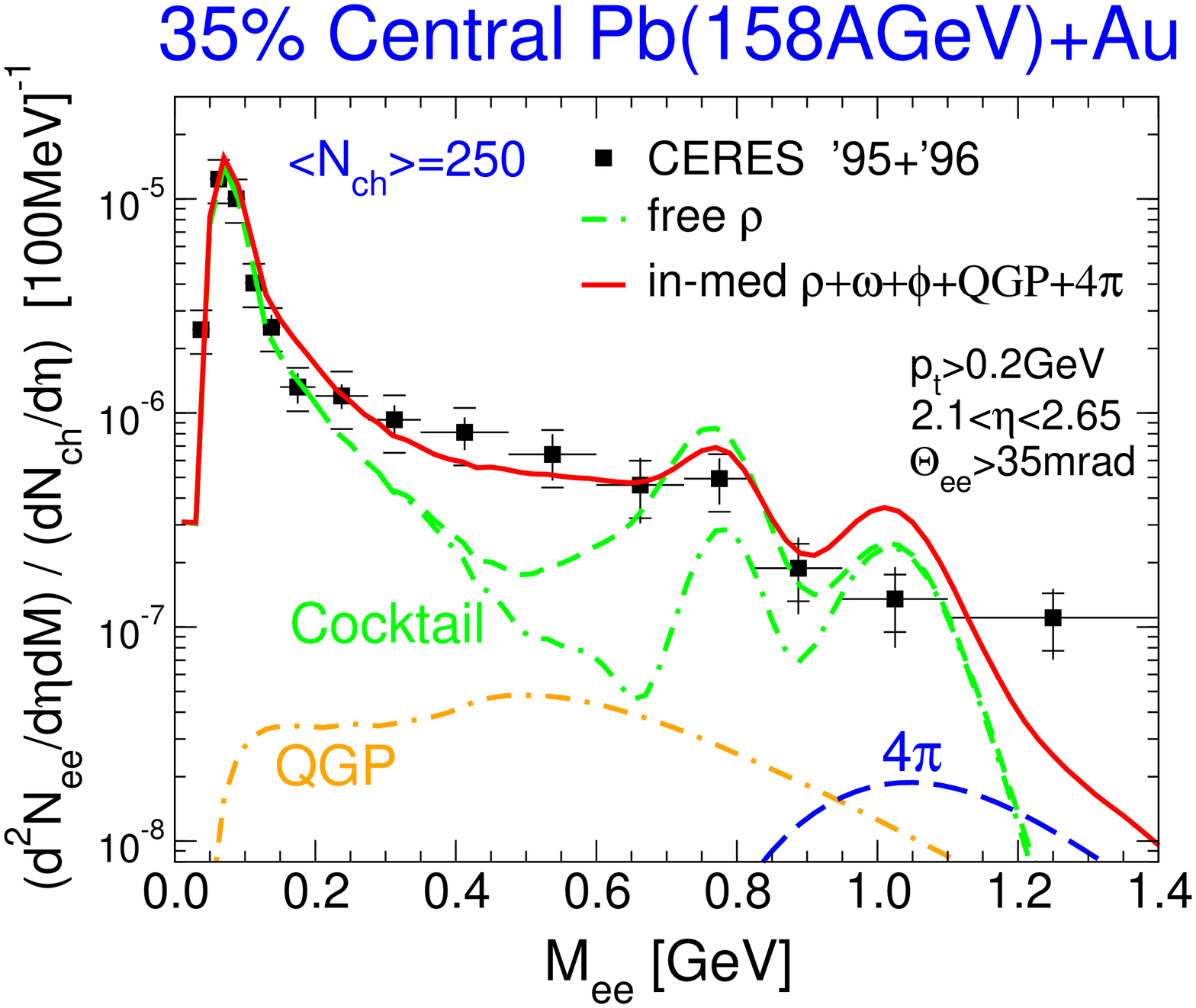}}
\caption{(Color online) Upper panel: comparison of thermal dilepton
  spectra in semicentral Pb-Au collisions (including experimental
  acceptance) for $\rho$ (short-dashed lines) and QGP emission
  (dash-dotted lines) using our previous (labeled ``old'') and updated
  fireball evolution.  The solid line is the sum of in-medium $\rho$,
  $\omega$ and $\phi$ contributions, while the long-dashed line results
  from four-pion annihilation including chiral mixing as described in
  Sec.~\ref{ssec_intmass}. The Drell-Yan contribution is below the
  displayed range; correlated charm decays are neglected. Lower panel:
  The solid line is the total in-medium thermal emission + hadronic
  cocktail (dash-dotted line, adopted from the experimental
  evaluation~\cite{Agakichiev:2005ai}), compared to combined 1995/1996
  CERES/NA45 data~\cite{Agakichiev:2005ai}. The short-dashed line is the
  sum of thermal emission with a vacuum $\rho$ spectral function and the
  cocktail.  }
\label{fig_ceres-semi}
\end{figure}
In the upper panel of Fig.~\ref{fig_ceres-semi} the updated calculations
for dielectron spectra in semicentral Pb-Au collisions are compared to
our previous results (using the same hadro-chemistry and $\rho$ spectral
function), including experimental acceptance cuts as defined by
CERES/NA45. For masses $M\gsim0.6$~GeV, the $\rho$ spectral function
yield is reduced by $15$-$20\%$, directly reflecting the reduction in
the lifetime of the hadronic phase in the fireball. At lower masses,
however, the difference is smaller and eventually disappears for
$M\lsim0.35$~GeV.  The reason for this is the interplay of the
single-electron cuts and the increase in transverse flow, where the
latter gives larger momentum to both electrons which in turn have a
larger probability to make it into the acceptance. For the QGP
contribution, the difference is smaller (less than $10\%$) and weakly
dependent on mass, since the increased transverse flow plays little role
in the early phases (we recall that, for the canonical choice of the
formation time, $\tau_0=1$~fm/$c$, the initial temperature for $30\%$
central Pb-Au amounts to $T_0=203$~MeV, and $210$~MeV for $7\%$
central). The addition of in-medium $\omega$ and $\phi$ decays affects
the total yield mostly around the respective free masses, more
significantly in the $\phi$ region. The in-medium four-pion contribution
becomes relevant for masses $M\gsim1$~GeV, with a relative strength
compared to the other sources that closely resembles the In-In case.
The Drell-Yan contribution does not show up on the scale of
Fig.~\ref{fig_ceres-semi} (and has been restricted to masses above
$\sim$1~GeV). Correlated charm decays are not accounted for. The lower
panel of Fig.~\ref{fig_ceres-semi} summarizes the comparison of the
theoretical calculations with the combined 1996/1996 CERES/NA45
data~\cite{Agakichiev:2005ai}, including the hadron decay cocktail. The
level of agreement is very similar to the one with our earlier
results~\cite{Rapp:1999us,Rapp:2002tw}.

\begin{figure}[!t]
\vspace*{2mm}
\centerline{\includegraphics[width=0.45\textwidth]{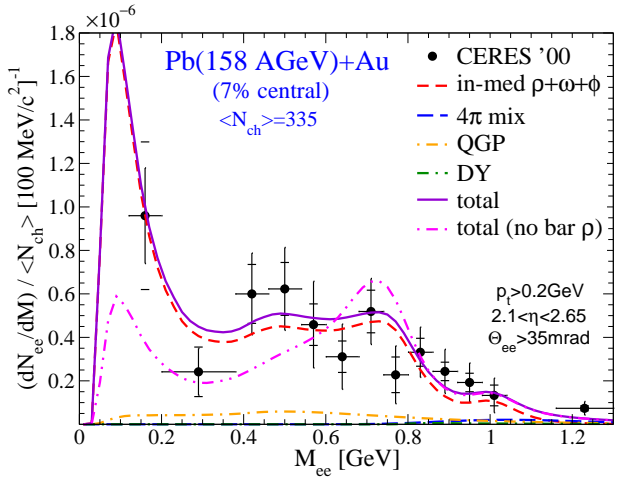}}
\caption{(Color online) Comparison of theoretical calculations of
  dilepton excess spectra in central Pb-Au collision, compared to
  CERES/NA45 data~\cite{Adamova:2006nu}; for the line labeled ``total
  (no bar $\rho$)'' the full in-medium $\rho$ spectral function has been
  replaced by the one which only includes medium effects in a meson
  gas.}
\label{fig_ceres-centr}
\end{figure}
Finally, we turn to the most recent CERES data for central
Pb(158~AGeV)-Au collisions~\cite{Adamova:2006nu}, which are shown in
Fig.~\ref{fig_ceres-centr} in a form similar to the NA60 data in the
previous sections, \ie, on a linear scale and with the cocktail
subtracted. Again, our calculations employing the updated fireball
describe these data fairly well. Variations of $\pm 20\%$ in the
fireball lifetime (affecting the total dilepton yields at the same level
with little change in the spectral shape), would still lead to
reasonable agreement. The importance of baryon-induced medium
modifications at SPS energies is reiterated by the dash-double-dotted
curve where only meson-gas effects are included in the $\rho$ spectral
function; this scenario does not properly reproduce the experimental
spectra. When comparing to the NA60 dimuon mass spectra, the four-pion
and QGP contributions in the IMR appear smaller relative to the signal
in the mass region below the free $\rho$ mass. This is essentially due
to the experimental acceptance, which for CERES/NA45 is significantly
larger in the low-mass region than for NA60, especially for low-momentum
lepton pairs.

A particularly striking feature of the theoretical predictions is the
large enhancement below the two-pion threshold, which is closely
connected with the approach to (and constraints from) the photon
point. The very-low-mass excess is rather sensitive to the
baryon-induced medium effects as well, as illustrated by the curve where
the latter are switched off. Experimentally, this regime is only
accessible with dielectrons (and after subtraction of the $\pi^0$ Dalitz
decay), and the first hint from data is consistent with the theoretical
yield. Clearly, it would be very worthwhile to further explore this
region.

\section{Summary and Conclusions} 
\label{sec_concl}

In the present paper, we have conducted a quantitative study of dilepton
spectra in semi-/central heavy-ion collisions at full SPS energy
($\sqrt{s}=17.3$~GeV). We have supplemented our earlier calculations of
thermal radiation (which reproduce available mass spectra) with sources
of non-thermal origin (expected to become relevant at high transverse
momentum and for small systems), and we have scrutinized the results
with respect to hadro-chemical and flow properties of the underlying
thermal fireball expansion.

The key quantity to describe thermal dilepton rates is the (imaginary
part of the) electromagnetic current correlator. In hadronic matter, we
have evaluated its medium modifications in terms of many-body spectral
functions for the light vector mesons at low mass, and chiral
vector-axialvector mixing for the continuum part at intermediate
mass. Partonic emission above $T_c$ has been approximated by
perturbative quark-antiquark annihilation. An appealing feature of this
description is a smooth merging of the hadronic rates with the partonic
ones at temperatures around the expected phase transition,
$T_c=160$-$190$~MeV. As a new ingredient we have computed thermal
dilepton rates induced by $t$-channel meson exchange in $\pi\rho\to\pi
l^+l^-$ reactions, which we found to become relevant at momenta above
$\sim$$1.5$~GeV. We have augmented our approach by non-thermal sources,
\ie, Drell-Yan annihilation, as well as decays of $\rho$ mesons from
primordial production (subject to jet quenching) and from decoupling at
thermal freezeout. We have emphasized that these $\rho$ decays carry an
extra kinematic Lorentz-$\gamma$ factor relative to thermal radiation,
resulting in harder dilepton-$q_T$ spectra.

All sources have been implemented into a space-time evolution for
$A$-$A$ collisions at SPS, for which we employed an isentropic fireball
model with QGP, mixed and hadron-gas phases and updated expansion
parameters to better match the absolute dilepton yields measured by
NA60. Hadrochemical freezeout and subsequent evolution are constructed
in line with measured hadron multiplicities. The calculated dimuon
invariant-mass spectra agree well with NA60 data in central and
semicentral In-In collisions, including $q_T$ bins below $0.5$~GeV and
above $1.0$~GeV, thus confirming our earlier results for thermal
emission and validating the predicted in-medium vector spectral
functions (dominated, but not exhausted, by the $\rho$). When comparing
to $q_T$ spectra, the new sources contribute to provide good agreement
for central collisions, while discrepancies persist with semicentral
data at $q_T>1~\text{GeV}$ for masses around and below the free $\rho$
mass.  In particular, effective slope parameters for
$q_T\ge1~\text{GeV}$ amount to $T_{\rm eff}\simeq 160$-$210$~MeV,
$\sim$$25\%$ short of the empirically extracted values in the LMR.

We have investigated the sensitivity of our results to variations in the
critical and chemical-freezeout temperature of the fireball.  The
spectral shape of the invariant-mass spectra turned out to be
insensitive for the range $T_c=160-190$~MeV. The reason is the
``duality'' of the thermal emission rates around $T_c$, rendering the
``melting'' of the $\rho$ meson a robust signal. A low value of
$T_c=160$~MeV entails that QGP radiation dominates over hadronic
emission in the IMR. A low value of $T_{\rm ch}=160$~MeV implies smaller
pion chemical potentials (and thus higher temperatures) in the hadronic
evolution, increasing the effective slope parameters by about 15-20~MeV,
still $\sim$30~MeV short of the data in the LMR.

In an attempt to resolve this discrepancy, we have implemented a
$\sim$$15\%$ increase in the transverse fireball acceleration as
previously employed for central Au-Au collisions at RHIC energy.  This
elevates the slope to $\sim$$250$~MeV around the free $\rho$ mass (and
properly reproduces the increase below), while still being consistent
with the (upper) experimental value of $205\pm20$~MeV in the IMR where
the blue shift effect is less pronounced due to the prevalently early
emission. In particular, we have demonstrated that, within the current
theoretical and experimental uncertainties, \emph{both} scenarios for a
predominant emission source in the IMR -- hadronic (four-pion) or
partonic -- are viable. In either case, the radiation mostly emanates
from matter at temperatures around $T_c\simeq160-190$~MeV.  An
unambiguous distinction between a partonic and hadronic source in the
IMR therefore appears difficult at SPS energies.  The larger transverse
flow developed in the partonic stage at RHIC energies could facilitate
this task~\cite{Xu:2007},
although additional complications arise due to the much larger
open-charm contribution and its in-medium modifications.

Finally we have checked that our refined assessment of dilepton sources,
together with the improvements in the fireball expansion, preserves our
previously found agreement with the CERES/NA45 dielectron spectra in
Pb-Au collisions.  The larger system size leads to an appreciable
increase in thermal radiation, thus reducing the uncertainties
associated with the modeling of the break-up stage. In addition,
dielectrons allow access to the mass region below the two-pion
threshold, where the medium effects on the $\rho$-spectral function,
augmented by the thermal Bose factor (and photon propagator), predict a
large thermal dilepton signal. First data in this region support this
signal, and further corroborate the importance of baryonic interactions
for the in-medium $\rho$ spectral function.

Future efforts should be pursued along several directions: first and
foremost, the theoretically calculated vector spectral functions must be
extended to their chiral partners (most notably the axialvector ($a_1$)
channel as the partner of the $\rho$), to establish direct connections
to chiral symmetry restoration. The evaluation of in-medium Weinberg sum
rules~\cite{Kapusta:1993hq}, in connection with constraints from lattice
QCD, will play an important role in this
enterprise~\cite{David:2006sr,Rapp:2007bm}. Second, from the
phenomenological side, the implementation of emission rates into both
hydrodynamic and transport simulations should be reiterated, to compute
$q_T$ spectra and check whether the rather large acceleration of the
fireball model can be justified for the In-In system at SPS.  The study
of dilepton elliptic flow~\cite{Heinz:2006qy} could shed further light
on the time profile of the radiation. At RHIC energies, partonic
collectivity increases significantly; an unexpectedly large dilepton
excess in the LMR recently observed by PHENIX needs to be
understood~\cite{Afanasiev:2007xw}.  Work in several of these directions
is in progress.

\subsection*{Acknowledgments}

We are grateful to S.~Damjanovic and H.~J.~Specht for discussion and
information on the NA60 acceptance, and to A. Marin for providing us
with the new CERES data. This work was supported in part by a
U.S. National Science Foundation CAREER award under grant
no. PHY-0449489.


\end{document}